\documentclass[twoside]{article}
\usepackage{amsmath,amsfonts,amssymb}
\usepackage{qic,epsfig}
\usepackage{cite}
\usepackage{color,colortbl}
\usepackage{graphicx,booktabs}
\usepackage{array,latexsym}
\usepackage{textcomp,booktabs}
\usepackage[usenames,dvipsnames]{xcolor}
\usepackage{graphicx}
\definecolor{classic}{cmyk}{1.0,0.6,0.1,0.35}
\definecolor{classic1}{cmyk}{1.0,0,1.0,0.55}
\definecolor{classic2}{cmyk}{0.8,0.5,0.5,0.15}

\renewcommand{\theequation}{\arabic{section}.\arabic{equation}}

\newtheorem{thm}{Theorem}
\newtheorem{cor}{Corollary}

\newtheorem{defn}{Definition}

\renewenvironment{proof}{\par\noindent{\bf Proof.}}{$\quad\Box$\par}
\newcommand{\ket}[1]{| #1 \rangle}
\newcommand{\bra}[1]{\langle #1 |}

\begin{document}
\setlength{\textheight}{8.0truein}    

\runninghead{Quotient Algebra Partition and Cartan Decomposition
for $su(N)$ II}
            {Zheng-Yao Su and Ming-Chung Tsai}

\normalsize\textlineskip \thispagestyle{empty}
\setcounter{page}{1}

\vspace*{0.88truein}

\alphfootnote

\fpage{1}

\centerline{\bf Quotient Algebra Partition and Cartan
Decomposition for $su(N)$ II} \vspace*{0.035truein}
\centerline{\footnotesize Zheng-Yao Su\footnote{Email:
zsu@nchc.narl.org.tw }}
\centerline{\footnotesize\it National Center for High-Performance
 Computing,}
 \centerline{\footnotesize\it National Applied Research Laboratories,
 Taiwan, R.O.C.}
\vspace*{0.21truein}

\abstracts{This is the sequel exposition following~\cite{Su}.
 The framework quotient algebra partition is rephrased in the language of the {\em s}-representation.
 Thanks to this language, a quotient algebra partition of the simplest form is established under a minimum number of conditions
 governed by a {\em bi-subalgebra} of rank zero, {\em i.e.}, a Cartan subalgebra.
 Within the framework, all Cartan subalgebras of $su(N)$ are classified
 and generated recursively through the process of the {\em subalgebra extension}.
 }{}{}

\vspace{3pt} \vspace*{1pt}\textlineskip

\section{Introduction}\label{secintro}
\renewcommand{\theequation}{\arabic{section}.\arabic{equation}}
\setcounter{equation}{0} \noindent
 In this article as the $2$nd episode of the serial~\cite{Su,SuTsai2,SuTsai3},
 the framework {\em Quotient Algebra Partition (QAP)} unveiled in~\cite{Su}
 is reformulated.
 The concept of {\em bi-subalgebras} and the
 language of the {\em s-representation} are introduced.
 In this language, it admits a recursive classification and
 generation of
 all Cartan subalgebras of the Lie algebra $su(N)$ through the process of the so-called {\em subalgebra extension}.
 Being an equivalence of 1st maximal subgroups in $Z^p_2$,
 1st maximal bi-subalgebras of a Cartan subalgebra in $su(2^p)$
 are expounded in depth.
 The general form of a rank-zero QAP
 is constructed under a minimum number of conditions
 governed by a Cartan subalgebra.

\section{Kinds of Cartan Subalgebras}\label{secclass}
\renewcommand{\theequation}{\arabic{section}.\arabic{equation}}
\setcounter{equation}{0} \noindent
 Notice that the exposition is presented in two views of a QAP structure,
 the quotient algebra and co-quotient algebra.
 To better address quotient algebras, it requires the general forms of quotient algebras and
 Cartan subalgebras. These forms become manifest when being phrased
 in the {\em s-representation}, an expression particularly chosen for generators and
 briefly presented in Appendix~B of~\cite{Su}.
 In this representation, a spinor generator of the Lie algebra $su(2^{p})$ is written as
 \begin{align}\label{defS}
 {\cal S}^{\zeta}_{\alpha}={\cal S}^{\epsilon_{1}\epsilon_{2}\ldots
 \epsilon_{p}}_{a_{1}a_{2}\ldots
 a_{p}}=\bigotimes^{p}_{i=1}\ (\ket{0}\bra{a_{i}}+(-1)^{\epsilon_{i}}\ket{1}\bra{1+a_{i}})\text{.}
 \end{align}
 Here the $p$-digit binary string $\alpha\in Z^p_2$ 
 encodes the {\em binary partitioning} of the generator 
 and the same-length string $\zeta\in Z^p_2$ 
 is interpreted
 as the {\em phase} of the generator.
 It is easy to derive the multiplication of two arbitrary spinors
 ${\cal S}^{\zeta}_{\alpha}$ and ${\cal S}^{\eta}_{\beta}$
 in the form
 \begin{align}\label{eqSproduct}
 {\cal S}^{\zeta}_{\alpha}\cdot{\cal S}^{\eta}_{\beta}
 =(-1)^{\eta\cdot\alpha}{\cal S}^{\zeta+\eta}_{\alpha+\beta}\text{.}
 \end{align}
 There immediately follow the commutation and anti-commutation relations
 \begin{align}\label{eqScommrelate}
 [{\cal S}^{\zeta}_{\alpha},{\cal S}^{\eta}_{\beta}]
 =\{(-1)^{\eta\cdot\alpha}-(-1)^{\zeta\cdot\beta}\}{\cal S}^{\zeta+\eta}_{\alpha+\beta}
 \end{align}
 and
 \begin{align}\label{eqSanticommrelate}
 \{{\cal S}^{\zeta}_{\alpha},{\cal S}^{\eta}_{\beta}\}
 =\{(-1)^{\eta\cdot\alpha}+(-1)^{\zeta\cdot\beta}\}{\cal S}^{\zeta+\eta}_{\alpha+\beta}.
 \end{align}
 A parity rule is thus implied that two generators
 ${\cal S}^{\zeta}_{\alpha}$ and ${\cal S}^{\eta}_{\beta}$ either commute
 if $\eta\cdot\alpha+\zeta\cdot\beta=0$ or anti-commute if
 $\eta\cdot\alpha+\zeta\cdot\beta=1$.
 A great merit of the $s$-representation is to convert the opertion of multiplication
 for spinors into (bitwise-)additions of associated binary strings.
 The whole exposition will proceed in 
 this representation.


 The story begins with a simple lemma.
 \vspace{6pt}
 \begin{lemma}\label{lemScomm}
 For any two commuting spinors ${\cal S}^{\zeta}_{\alpha}$ and
 ${\cal S}^{\eta}_{\beta}\in su(2^p)$,  
 their bi-additive ${\cal S}^{\zeta+\eta}_{\alpha+\beta}$ commutes with both the given.
 \end{lemma}
 \vspace{3pt}
 \begin{proof}
 From Eq.~\ref{eqScommrelate}, there obtains the relation
 $\eta\cdot\alpha=\zeta\cdot\beta$, 
 which leads to the identity
 $\zeta\cdot(\alpha+\beta)=\zeta\cdot\alpha+\zeta\cdot\beta=\zeta\cdot\alpha+\eta\cdot\alpha
 =(\zeta+\eta)\cdot\alpha$
 and thus the vanishing commutator $[{\cal S}^{\zeta}_{\alpha},{\cal S}^{\zeta+\eta}_{\alpha+\beta}]=0$;
 similarly, $[{\cal S}^{\eta}_{\beta},{\cal S}^{\zeta+\eta}_{\alpha+\beta}]=0$.
 \end{proof}
 \vspace{6pt} \noindent
 In spite of its plainness, this lemma is of basic to building abelain subalgebras. The binary-partitioning
 and phase strings associated with spinor generators in a Cartan subalgebra
 are coerced to form an additive subgroup respectively by the lemma.
 For the consistency of subgroups, the identity operator is
 added in the set of generators, which, up to an irrelevant global phase in group actions,
 makes a quotient-algebra structure unchanged.
 The all-zero string ${\mathbf 0}$ is the subgroup identity and 
 signifies the binary partitioning of diagonal operators.

 Redenoted as $\mathfrak{C}_{[{\mathbf 0}]}$
 the intrinsic subalgebra comprising all diagonal operators 
 is the most convenient choice of a Cartan subalgebra 
 and regarded {\em the Cartan subalgebra of the $0$th kind},
 \begin{align}\label{eq0th-kind}
 \mathfrak{C}_{[{\mathbf 0}]}=\{{\cal S}^{\nu_0}_{\mathbf 0}:\forall\hspace{2pt}\nu_0\in Z^{p}_2\}.
 \end{align}
 Including the identity operator, this subalgebra spanned by $2^p$ spinor generators 
 reveals the existence of an isomorphism; the symbol {\em span} is throughout omitted
 from notations of subalgebras.
 Relation~(\ref{eqSproduct}) implies a merit that
 the product of two spinors can be encoded to the additions of two pairs of binary strings.
 The isomorphism arises from this operation.
 \vspace{6pt}
 \begin{thm}\label{thmisoC}
 The set of spinor generators of a Cartan subalgebra $\mathfrak{C}\subset su(2^p)$ forms an
 abelian group isomorphic to $Z^p_2$ under bi-addition $\diamond$:
 $\forall\hspace{2pt}{\cal S}^{\zeta}_{\alpha},{\cal S}^{\eta}_{\beta}\in\mathfrak{C}$,
 ${\cal S}^{\zeta}_{\alpha}\diamond{\cal S}^{\eta}_{\beta}\equiv{\cal S}^{\zeta+\eta}_{\alpha+\beta}\in\mathfrak{C}$.
 \end{thm}
 \vspace{3pt}
 \begin{proof}
 The Cartan subalgebra of the $0$th-kind $\mathfrak{C}_{[\bf 0]}$ is
 by definition isomorphic to $Z^p_2$ according to~(\ref{eq0th-kind}).
 Any other Cartan subalgebra $\mathfrak{C}$ is related to $\mathfrak{C}_{[\bf 0]}$
 by a conjugate transformation.  Explicitly constructed in Section~\ref{sectrans},
 the action $R$ is an option for the transformation, which, up to a plus or minus sign,
 one-to-one maps a spinor generator in $\mathfrak{C}_{[\bf 0]}$ to another in $\mathfrak{C}$.
 Due to relation~(\ref{eqSproduct}), the conjugate transformation preserves the isomorphism.
 \end{proof}
 \vspace{6pt} \noindent
  This isomorphism will be recounted in a varied expression in Section~\ref{secbisubalg}
  that turns into a fundamental concept to formulate quotient algebras. 
  Resulting from the operation, the spinor ${\cal S}^{\zeta+\eta}_{\alpha+\beta}$ is
  termed {\em the bi-additive generator of}
  ${\cal S}^{\zeta}_{\alpha}$ {\em and} ${\cal S}^{\eta}_{\beta}$.
  It will be demonstrated from next section that
  bi-addition is an elementary operation in depicting algebraic structures
  and, compared with the Lie bracket, able
  to provide more {\em refine} information of unitary Lie algebras.

 The next choice, a Cartan subalgebra of the $1$st kind $\mathfrak{C}^{\hspace{2pt}\epsilon}_{[\alpha]}$
 is of the form,
 \begin{align}\label{eq1st-kind}
 \mathfrak{C}^{\hspace{2pt}\epsilon}_{[\alpha]}=\{{\cal S}^{\nu_{1}}_{\mathbf 0},{\cal S}^{\zeta}_{\alpha}
 :\forall\hspace{2pt} \nu_1,\zeta\in Z^p_2,\hspace{5pt} \nu_{1}\cdot\alpha=0\text{ and }\zeta\cdot\alpha=\epsilon\}\text{.}
 \end{align}
 The binary partitioning $\alpha$ is nonzero and $\epsilon$ representing the {\em self parity}
 carries either $0$ or $1$.
 It is apparent that spinors of an identical binary partitioning commute with only
 those of the same self parity.
 As follows is an example of a $1$st-kind Cartan subalgebra in $su(8)$,
 \begin{align}\label{EX1st-kind}
 \mathfrak{C}^{\hspace{0.5pt}1}_{[100]}
 =\{{\cal S}^{000}_{000},{\cal S}^{001}_{000},{\cal S}^{010}_{000},{\cal S}^{011}_{000},
 i{\cal S}^{100}_{100},i{\cal S}^{101}_{100},i{\cal S}^{110}_{100},i{\cal S}^{111}_{100}\}\text{.}
 \end{align}

 Owing to the lemma above, if two binary-partitioning strings 
 $\alpha$ and $\beta$ appear in a Cartan subalgebra,
 there must also exist the $3$rd one 
 $\gamma=\alpha+\beta$, 
 namely the three strings forming an additive subgroup.
 A Cartan subalgebra of the $2$nd kind thus reads as  
 \begin{align}\label{eq2nd-kind}
 \mathfrak{C}^{\hspace{1.5pt}\epsilon_{11}\epsilon_{12}\epsilon_{22}}_{[\alpha,\beta]}
 =\{&{\cal S}^{\nu_{2}}_{\mathbf 0},{\cal S}^{\zeta}_{\alpha},{\cal S}^{\eta}_{\beta},{\cal S}^{\xi}_{\gamma}:
  \forall\hspace{2pt} \nu_2,\zeta,\eta\in Z^p_2,\hspace{5pt}
   \nu_{2}\cdot\alpha=\nu_{2}\cdot\beta=0, \\ \notag
 & \zeta\cdot\alpha=\epsilon_{11},\hspace{1.5pt}
  \eta\cdot\alpha=\zeta\cdot\beta=\epsilon_{12},\hspace{1.5pt}
  \eta\cdot\beta=\epsilon_{22},\hspace{3pt}
  \gamma=\alpha+\beta\text{ and }\xi=\zeta+\eta\}\text{.}
 \end{align}
 The two self parities $\epsilon_{11}$ and $\epsilon_{22}$ and 
 the {\em mutual} parity $\epsilon_{12}$ are independently
 assigned $0$ or $1$; the other three parities $\epsilon_{13}=\xi\cdot\alpha,
 \epsilon_{23}=\xi\cdot\beta$ and $\epsilon_{33}=\xi\cdot\gamma$ are determined
 by the former three, {\em e.g.}, $\epsilon_{33}=\epsilon_{11}+\epsilon_{22}$.
 A $2$nd-kind Cartan subalgebra in $su(8)$ is given for example,
 \begin{align}\label{EX2nd-kind}
 \mathfrak{C}^{\hspace{1pt}101}_{[011,100]}
 =\{{\cal S}^{000}_{000},{\cal S}^{011}_{000},i{\cal S}^{010}_{011},i{\cal S}^{001}_{011},
 i{\cal S}^{111}_{100},i{\cal S}^{100}_{100},{\cal S}^{101}_{111},{\cal S}^{110}_{111}\}\text{.}
 \end{align}

 Based on the commuting condition and the group closure of
 associated strings, a
 Cartan subalgebra of the $k$-th kind in $su(2^p)$ takes the form, $0<k\leq p$,
 \begin{align}\label{eqk-thkind}
 \mathfrak{C}^{\hspace{2pt}\epsilon_{11},\epsilon_{12},\cdots,\epsilon_{kk}}_{[\alpha_1,\alpha_2,\cdots,\alpha_k]}
 =\{&{\cal S}^{\nu_{k}}_{\mathbf 0},{\cal S}^{\zeta_{1}}_{\alpha_{1}},{\cal S}^{\zeta_{2}}_{\alpha_{2}},\cdots,
 {\cal S}^{\zeta_{2^k-1}}_{\alpha_{2^k-1}}:\\ \notag
 &\forall\hspace{2pt} \nu_k,\zeta_i,\zeta_j\in
 Z^p_2,\hspace{3pt}\alpha_i,\alpha_j\in[\alpha]_{k},\hspace{3pt}0<i,j<2^k,\\ \notag
 &\nu_k\cdot\alpha_i=0,\hspace{3pt}\zeta_i\cdot\alpha_j=\zeta_j\cdot\alpha_i=\epsilon_{ij}\}. 
 \end{align}
 There involve in total $2^{k}$ binary-partitioning strings that
 form an additive subgroup of $Z^p_2$.
 Let this subgroup be denoted by $[\alpha_1,\alpha_2,\ldots,\alpha_k]$
 with $\alpha_0\equiv \mathbf 0$ and abbreviated as $[\alpha]_k$ for convenience,
 which is generated by the set $\{\alpha\}_k=\{\alpha_1,\alpha_2,\ldots,\alpha_k\}$ consisting
 of $k$ independent strings.
 Also, the set of any $k$ spinor generators
 $\{{\cal S}^{\zeta_l}_{\alpha_l}:l=1,2,\dots,k\}$ in the subalgebra
 determines all the required parities $\epsilon_{ij},0<i,j< 2^k$,
 as long as $\{{\alpha_l}:l=1,2,\dots,k\}=\{\alpha\}_k$.
 Therefore, the $k$ independent
 binary-partitioning strings and their corresponding parities,
 respectively attached as the subscript and the superscript,
 suffice to characterize a Cartan subalgebra of the
 $k$-th kind.
 The superscript recording a total number $\frac{k(k+1)}{2}$ of parities $\epsilon_{rs}$
 in the order $1\leq r\leq s\leq k$ is marked 
 as $\{\epsilon\}$ whenever no confusion is likely.
 Since imposing an independent condition of the parity is
 equivalent to removing a digit from phase strings, each set of
 $\{{\cal S}^{\nu_k}_{\mathbf 0}\}$ and
 $\{{\cal S}^{\zeta_i}_{\alpha_i}\},0<i<2^k$, contributes the same number $2^{p-k}$
 of elements to the subalgebra.
 An example of a $3$rd-kind Cartan subalgebra in $su(8)$ is given,
 \begin{align}\label{EX3rd-kind}
 \mathfrak{C}^{\hspace{1pt}101011}_{[001,010,100]}=
 \{{\cal S}^{000}_{000},i{\cal S}^{101}_{001},{\cal S}^{100}_{010},i{\cal S}^{001}_{011},i{\cal S}^{111}_{100},
 {\cal S}^{010}_{101},i{\cal S}^{011}_{110},{\cal S}^{110}_{111}\}\text{.}
 \end{align}

 In the following exposition a Cartan subalgebra of $k$-th kind is often
 shortened with the expression
 $\mathfrak{C}^{\hspace{1.pt}\{\epsilon\}}_{[\alpha]_k}
 =\{{\cal S}^{\zeta_i}_{\alpha_i}:0\leq i<2^k,\alpha_i\in [\alpha]_k\text{ and }\zeta_i\in [\zeta]_q \}$,
 where $[\alpha]_k$ and $[\zeta]_q$ denote the subgroups formed by
 binary-partitioning and phase strings respectively;
 the number of generators $q$ for the latter will be explained
 in Section~\ref{secextend} and Appendix~\ref{appcosets}.
 This expression suggests the choice that either subgroup $[\alpha]_k$ or $[\zeta]_q$
 is a legitimate basis for the subalgebra classification,
 although the former is adopted throughout the text.
 In addition,
  attributed to the symmetric form of Eq.~\ref{eqScommrelate}, all Cartan subalgebras
  and quotient algebras are endowed with the so-called {\em bit-phase duality}.
  This duality guarantees the symmetry that, after interchanging the
  binary-partitioning and phase strings of each spinor generator,
  a Cartan subalgebra of the $k$-th kind stays to be a member of the $q$-th kind and
  the structure of a quotient algebra remains valid.

 As to be proved in Section~\ref{secextend},
 Cartan subalgebras of a Lie algebra can be exhaustively searched shell by shell
 through the process of the {\em subalgebra extension} introduced in~\cite{Su}.
 Applied to the Lie algebra $su(2^p)$, the subalgebra extension generates
 all Cartan subalgebras of the $k$-th kind at the $k$-th shell of the extension
 and the process completes at the $p$-th shell.
 It is instructive to have a quick glance of the process in the $s$-representation.  
 The extension starts with
 the quotient algebra 
 given by the intrinsic center subalgebra $\{{\cal Q}(\mathfrak{C}_{[{\mathbf 0}]};2^p-1)\}$
 as outlined in Fig.~\ref{1st-general},
 which is considered the quotient algebra of the $1$st shell.
 \begin{figure}[!ht]
 \begin{center}
 \[\begin{array}{ccc}\label{1st shell}
 &\mathfrak{C}_{[{\mathbf 0}]}=\{{\cal S}^{\nu_0}_{\mathbf 0}:\nu_0\in Z^{p}_2\}&\\

 \hspace{-25pt}
 \begin{array}{c}
 \hspace{15pt}W_{l}=\{{\cal S}^{\zeta}_{\alpha_l}:\forall\hspace{2pt}\zeta\in Z^{p}_2, 
 \hspace{1pt}\zeta\cdot\alpha_l=\epsilon_l\}
 \end{array}
 &&
 \hspace{-5pt}\begin{array}{c}
 \hat{W}_{l}=\{{\cal S}^{\hat{\zeta}}_{\alpha_l}:\forall\hspace{2pt}\hat{\zeta}\in Z^{p}_2, 
 \hspace{1pt}\hat{\zeta}\cdot\alpha_l=\hat{\epsilon}_l\}
 \end{array}
 \end{array}\]\\
 \vspace{8pt}
 \fcaption{The quotient algebra of the $1$st shell given by the Cartan subalgebra $\mathfrak{C}_{[{\mathbf 0}]}$,
           $l=1,2,\dots,2^p-1$.\label{1st-general}}
 \end{center}
 \end{figure}
 Within this quotient algebra, the spinor generators in the two {\em conditioned subspaces}
 $W_l$ and $\hat{W}_{l}$ of a conjugate pair, $1\leq l<2^p$,
 share the identical binary partitioning $\alpha_l$ and yet have opposite self parities,
 {\em i.e.}, $\epsilon_l+\hat{\epsilon}_l=1$.
 For each conjugate pair $\{W_l,\hat{W}_{l}\}$,
 the set of strings $\{\nu_0\}=Z^{p}_2$ is divided into two
 halves $\{\nu_1\}$ and $\{\hat{\nu}_1\}$ such that
 $[{\cal S}^{\zeta}_{\alpha_l},{\cal S}^{\nu_1}_{\mathbf 0}]=0$ and
 $[{\cal S}^{\zeta}_{\alpha_l},{\cal S}^{\hat{\nu}_1}_{\mathbf 0}]\neq 0$
 or $\nu_1\cdot\alpha_l=0$ and $\hat{\nu_1}\cdot\alpha_l=1$;
 however,
 $[{\cal S}^{\hat{\zeta}}_{\alpha_l},{\cal S}^{\nu_1}_{\mathbf 0}]\neq 0$ and
 $[{\cal S}^{\hat{\zeta}}_{\alpha_l},{\cal S}^{\hat{\nu}_1}_{\mathbf 0}]=0$.
 Thanks to the strings $\hat{\nu}_1$, the commutator $[W_l,\mathfrak{C}_{[{\mathbf 0}]}]$
 yields the spinors ${\cal S}^{\zeta+\hat{\nu}_1}_{\alpha_l}$.
 Since $(\zeta+\hat{\nu}_1)\cdot\alpha_l=\zeta\cdot\alpha_l+1$,
 the strings $\hat{\zeta}\equiv\zeta+\hat{\nu}_1$ alter the parity in the conjugate
 subspace $\hat{W}_l$.
 Exhibited in Fig.~\ref{1stshellFig} is an example
 of the quotient algebra for $su(8)$ at the $1$st shell.
 Being the extension from this quotient algebra, a total number $2\times(2^p-1)$ of
 Cartan subalgebras of the $1$st kind
 $\mathfrak{C}^{\hspace{2pt}\epsilon_l}_{[\alpha_l]}
 =\{{\cal S}^{\nu_1}_{\mathbf 0},{\cal S}^{\zeta}_{\alpha_l}:
 \forall \hspace{2pt}\nu_1,\zeta\in Z^p_2,
 \hspace{.1cm}\nu_{1}\cdot\alpha_l=0\text{ and }\zeta\cdot\alpha_l=\epsilon_l\}$
 are produced.

 Once furnished with a Cartan subalgebra extended at the $1$st shell
 $\mathfrak{C}^{\hspace{2pt}\epsilon}_{[\alpha]}\equiv
  \mathfrak{C}^{\hspace{2pt}\epsilon_l}_{[\alpha_l]}$,
  a quotient algebra of the $2$nd shell
 $\{{\cal Q}(\mathfrak{C}^{\hspace{2pt}\epsilon}_{[\alpha]};2^p-1)\}$
 as in Fig.~\ref{2ndshell}, is built by taking
 $\mathfrak{C}^{\hspace{2pt}\epsilon}_{[\alpha]}$ to be the center subalgebra.
  \begin{figure}[!ht]
 \begin{center}
 \[\begin{array}{ccc}\label{2ndshell}
 &\hspace{-50pt}\begin{array}{c}
 \mathfrak{C}^{\hspace{2pt}\epsilon}_{[\alpha]}=
 \{\hspace{3pt}{\cal S}^{\nu_{1}}_{\mathbf 0},{\cal S}^{\zeta}_{\alpha}
 :\forall\hspace{2pt} \nu_1,\zeta\in Z^p_2,\\
 \hspace{53pt}\nu_{1}\cdot\alpha=0\text{ and }\zeta\cdot\alpha=\epsilon\hspace{3pt}\}
 \end{array}&\\
 &&\\
 \hspace{-120pt}W_{\text{old}}&&\hspace{-132pt}\hat{W}_{\text{old}}\\
 \hspace{-20pt}\{\hspace{3pt}{\cal S}^{\hat{\nu}_1}_{\mathbf 0}:\forall\
 \hat{\nu}_1\in Z^p_2,\hspace{3pt}\hat{\nu}_1\cdot\alpha=1\}
 &&
 \hspace{-45pt}\{\hspace{3pt}{\cal S}^{\hat{\zeta}}_{\alpha}:\forall\ \hat{\zeta}\in Z^p_2,\hspace{3pt}\hat{\zeta}\cdot\alpha=\hat{\epsilon}\}\\
 &&\\
 \hspace{-117pt}W_{\text{new}}&&\hspace{-128pt}\hat{W}_{\text{new}}\\
 \hspace{-55pt}
 \begin{array}{c}
 \{\hspace{3pt}{\cal S}^{\eta}_{\beta},{\cal S}^{\xi}_{\gamma}:
 \forall\hspace{2pt} \eta,\xi\in Z^p_2,\\
 \hspace{8pt}\forall\hspace{2pt} \nu_2\in\{\nu_1\}\cap[\breve{\zeta}_0]_{q'},\\
 \forall\hspace{2pt} \tau\in\{\zeta\}\cap[\breve{\zeta}_0]_{q'},\\
 \hspace{27pt}\exists\ \rho\in \{\zeta\}\cap[\breve{\zeta}_0]_{q'},\text{{ \em s.t., }}\\
 \hspace{30pt}\nu_2\cdot\beta=0,\hspace{3pt}\tau\cdot\beta=\eta\cdot\alpha,\\
 \hspace{20pt}\beta+\gamma=\alpha,\hspace{3pt}\eta+\xi=\rho,\\
 \hspace{-28pt}\eta\cdot\beta=\sigma\hspace{3pt}\}
 \end{array}
 &&
 \hspace{-66pt}
 \begin{array}{c}
 \{\hspace{3pt}{\cal S}^{\hat{\eta}}_{\beta},{\cal S}^{\hat{\xi}}_{\gamma}:
 \forall\hspace{2pt} \hat{\eta},\hat{\xi}\in Z^p_2,\\
 \hspace{8pt}\forall\hspace{2pt} \hat{\nu}_2\in\{\nu_1\}\cap[\breve{\zeta}_0]_{q'},\\
 \forall\hspace{2pt} \hat{\tau}\in\{\zeta\}\cap[\breve{\zeta}_0]_{q'},\\
 \hspace{27pt}\exists\ \hat{\rho}\in \{\zeta\}\cap[\breve{\zeta}_0]_{q'},\text{{ \em s.t., }}\\
 \hspace{30pt}\hat{\nu}_2\cdot\beta=0,\hspace{3pt}\hat{\tau}\cdot\beta=\hat{\eta}\cdot\alpha,\\
 \hspace{20pt}\beta+\gamma=\alpha,\hspace{3pt}\hat{\eta}+\hat{\xi}=\hat{\rho},\\
 \hspace{-28pt}\hat{\eta}\cdot\beta=\hat{\sigma}\hspace{3pt}\}
 \end{array}
 \end{array}\]\\
 \fcaption{A quotient algebra of the $2$nd shell given by a Cartan subalgebra of the $1$st kind.\label{2nd-general}}
 \end{center}
 \end{figure}
 In this quotient algebra, 
 there is one ``old" conjugate pair $\{W_{\rm old},\hat{W}_{\rm old}\}$,
 from which only the Cartan subalgebra of the $0$th kind
 $\mathfrak{C}_{[{\mathbf 0}]}=\{{\cal S}^{\nu_1}_{\mathbf 0},{\cal S}^{\hat{\nu}_1}_{\mathbf 0}\}$
 and one 1st-kind
 $\mathfrak{C}^{\hspace{2pt}\hat{\epsilon}}_{[\alpha]}=
 \{\hspace{3pt}{\cal S}^{\nu_{1}}_{\mathbf 0},{\cal S}^{\hat{\zeta}}_{\alpha}\}$
 are recovered, recalling $\epsilon+\hat{\epsilon}=1$.
 By extending from the ``new" conjugate pairs of opposite self parities $\sigma+\hat{\sigma}=1$,
 in total $4\times (2^{p-1}-1)$ Cartan subalgebras of the 2nd kind
 $\mathfrak{C}^{\hspace{1.5pt}\{\epsilon\}}_{[\alpha,\beta]}
 =\{{\cal S}^{\nu_{2}}_{\mathbf 0},{\cal S}^{\tau}_{\alpha},{\cal S}^{\eta}_{\beta},{\cal S}^{\xi}_{\gamma}\}$
 are derived, with the parity string  $\{\epsilon\}=\epsilon\hspace{1.5pt}\epsilon_{12}\hspace{1.5pt}\sigma$
 and $\epsilon_{12}=\tau\cdot\beta=\eta\cdot\alpha$.
 Notice that in a new pair the set $\{\nu_2\}$ is a maximal subgroup of $\{\nu_1\}=Z^p_2$
 and $\{\zeta'\}$ is the coset in the partition of $\{\zeta\}=Z^p_2$ bisected by $\{\nu_2\}$.
 An example of a such quotient algebra of $su(8)$ is given in
 Fig.~\ref{2ndshellFig}.
 In contrast to the old conjugate pair, there require more conditions to describe the new
 pairs. The complexity of these conditions will be rapidly increasing
 when the extension is conducted to higher shells.
 Fortunately, a rank-zero quotient algebra can be
 put into the general form presented in Fig.~\ref{figgform1}
 or even as simple as that in Fig.~\ref{figgform2}. 
 The crux is to discern the {\em group structures} hidden in a unitary Lie algebra.
 The continued exposition is to elaborate upon all these details.


\section{Bi-Subalgebra}\label{secbisubalg}
\renewcommand{\theequation}{\arabic{section}.\arabic{equation}}
\setcounter{equation}{0} \noindent
 To write the general form of a quotient algebra,
 the {\em bi-subalgebra},
 a special form of subalgebra in the Lie algebra $su(2^p)$, 
 need be defined.
\vspace{6pt}
\begin{defn}\label{defbisubalg}
 A set of spinor generators
 $\{{\cal S}^{\zeta}_{\alpha}\}\subset su(2^p)$ forms 
 a bi-subalgebra ${\cal B}$ 
 if its sets of binary-partitioning and phase strings 
 are respectively a subgroup of $Z^p_2$; namely, ${\cal B}$ is a bi-subalgebra of $su(2^p)$ iff\hspace{2pt}
 $\forall\hspace{.1cm}{\cal S}^{\zeta}_{\alpha},{\cal S}^{\eta}_{\beta}\in{\cal B}\subset su(2^p)$,
 ${\cal S}^{\zeta+\eta}_{\alpha+\beta}\in{\cal B}$.
\end{defn}
\vspace{6pt}
 \noindent
 Apparently, $su(2^p)$ is a bi-subalgebra of itself,
 the zeroth {\em maximal}, and
 has numerous others.
 Being an equivalence of a subgroup in $Z^p_2$, a bi-subalgebra
 is {\em a subgroup of} $su(2^p)$ under the operation of bi-addition.
 A Cartan subalgebra $\mathfrak{C}\subset su(2^p)$ is a $p$-th maximal
 in $su(2^p)$, and
 from the zeroth to the $(p-1)$-th maximal bi-subalgebras are all nonabelian.
 Since the abelianness is required to generate quotient algebras
 according to Corollary~1 of~\cite{SuTsai2},
 the attention will be confined to abelian bi-subalgebras
 in this serial of studies.
 Similarly, a Cartan subalgebra of the $k$-th kind
 $\mathfrak{C}^{\hspace{1.pt}\{\epsilon\}}_{[\alpha]_k}$ contains many  bi-subalgebras.
 They can be easily produced by collecting spinor generators of
 binary-partitioning and phase strings forming a subgroup in $[\alpha]_k$
 and $[\zeta]_q$ respectively, {\em cf.} Lemma~\ref{lembisubink-thC} in Section~\ref{secextend}.
 Essential to formulating quotient algebra partitions,
 abelian bi-subalgebras are ordered in {\em maximality}. 
\vspace{6pt}
\begin{defn}\label{defmaxbisubalg}
 A bi-subalgebra $\mathfrak{B}\subset\mathfrak{C}$
 is maximal in a Cartan subalgebra $\mathfrak{C}$ iff
 $\forall\hspace{.1cm}{\cal S}^{\zeta}_{\alpha},{\cal S}^{\eta}_{\beta}\in\mathfrak{B}^c\equiv
  \mathfrak{C}-\mathfrak{B}$,
 ${\cal S}^{\zeta+\eta}_{\alpha+\beta}\in\mathfrak{B}$.
\end{defn}
\vspace{6pt}
 According to this maximality condition,
 a Cartan subalgebra $\mathfrak{C}$ is the {\em 0th maximal} bi-subalgebra of itself
 and a proper maximal bi-subalgebra of $\mathfrak{C}$ is a {\em 1st maximal}
 bi-subalgebra. In the following text unless else specified, the maximal bi-subalgebras
 of $\mathfrak{C}$ embrace both the $0$th and $1$st maximal members.
 For later purposes, siblings of
 succeeding orders, such as $2$nd, $3$rd 
 and maximal bi-subalgebras of other orders will be
 examined in the continued episode~\cite{SuTsai2}.
 In a Cartan subalgebra, 
 a $3$rd member is obtainable from two arbitrary maximal bi-subalgebras. 
\vspace{6pt}
\begin{lemma}\label{lemthe3rdmaxbi}
 Derived from every two maximal bi-subalgebras $\mathfrak{B}_1$ and $\mathfrak{B}_2$ of a
 Cartan subalgebra $\mathfrak{C}$, the subspace 
 $\mathfrak{B}=(\mathfrak{B}_1\cap\mathfrak{B}_2)\cup(\mathfrak{B}^c_1\cap\mathfrak{B}^c_2)$
 is also a maximal bi-subalgebra of $\mathfrak{C}$,
 here $\mathfrak{B}^c_1=\mathfrak{C}-\mathfrak{B}_1$ and $\mathfrak{B}^c_2=\mathfrak{C}-\mathfrak{B}_2$.
 \end{lemma}
\vspace{3pt}
\begin{proof}
 When one of $\mathfrak{B}_1$ and $\mathfrak{B}_2$ is the Cartan subalgebra $\mathfrak{C}$, say
 $\mathfrak{B}_1=\mathfrak{C}$,
 the resulted $\mathfrak{B}$ is identical to the other maximal bi-subalgebra $\mathfrak{B}_2$,
 because of the intersections
 $\mathfrak{B}_1\cap\mathfrak{B}_2=\mathfrak{C}\cap\mathfrak{B}_2=\mathfrak{B}_2$
 and $\mathfrak{B}^c_1\cap\mathfrak{B}^c_2=\{0\}$
 with the null complement $\mathfrak{B}^c_1=\mathfrak{C}^c=\mathfrak{C}-\mathfrak{C}=\{0\}$.

 On the other hand,
 let it be proved first that the subspace $\mathfrak{B}$ is a bi-subalgebra
 when $\mathfrak{B}_1\neq\mathfrak{C}$ and
 $\mathfrak{B}_2\neq\mathfrak{C}$.
 Since both $\mathfrak{B}_1$ and $\mathfrak{B}_2$ are a maximal bi-subalgebra,
 the spinor generator
 ${\cal S}^{\zeta+\eta}_{\alpha+\beta}$ is in $\mathfrak{B}_1\cap\mathfrak{B}_2$ if either
 ${\cal S}^{\zeta}_{\alpha},{\cal S}^{\eta}_{\beta}\in\mathfrak{B}_1\cap\mathfrak{B}_2$ or
 ${\cal S}^{\zeta}_{\alpha},{\cal S}^{\eta}_{\beta}\in\mathfrak{B}^c_1\cap\mathfrak{B}^c_2$;
 ${\cal S}^{\zeta+\eta}_{\alpha+\beta}\in\mathfrak{B}^c_1\cap\mathfrak{B}^c_2$,
 if ${\cal S}^{\zeta}_{\alpha}\in\mathfrak{B}_1\cap\mathfrak{B}_2$ and
 ${\cal S}^{\eta}_{\beta}\in\mathfrak{B}^c_1\cap\mathfrak{B}^c_2$.
 Thus, $\mathfrak{B}$ is a bi-subalgebra.
 To confirm the maximality of $\mathfrak{B}$ in $\mathfrak{C}$,
 inspecting on generators in the subspace
 $\mathfrak{B}^c=\mathfrak{C}-\mathfrak{B}=(\mathfrak{B}_1\cap\mathfrak{B}^c_2)\cup(\mathfrak{B}_1\cap\mathfrak{B}^c_2)$
 is required.
 This subspace is null iff $\mathfrak{B}_1=\mathfrak{B}_2$, which is excluded.
 Once more by the definition of a maximal bi-subalgebra, the generator
 ${\cal S}^{\zeta+\eta}_{\alpha+\beta}$ is in $\mathfrak{B}_1\cap\mathfrak{B}_2\subset\mathfrak{B}$, if either
 ${\cal S}^{\zeta}_{\alpha},{\cal S}^{\eta}_{\beta}\in\mathfrak{B}_1\cap\mathfrak{B}^c_2$ or
 ${\cal S}^{\zeta}_{\alpha},{\cal S}^{\eta}_{\beta}\in\mathfrak{B}^c_1\cap\mathfrak{B}_2$;
 while ${\cal S}^{\zeta+\eta}_{\alpha+\beta}\in\mathfrak{B}^c_1\cap\mathfrak{B}^c_2\subset\mathfrak{B}$,
 if ${\cal S}^{\zeta}_{\alpha}\in\mathfrak{B}_1\cap\mathfrak{B}^c_2$ and
 ${\cal S}^{\eta}_{\beta}\in\mathfrak{B}^c_1\cap\mathfrak{B}_2$.
 The subspace $\mathfrak{B}$ is hence a maximal bi-subalgebra. 
\end{proof}
\vspace{6pt}

 \noindent Importantly, the set of maximal bi-subalgebras of a Cartan subalgebra forms an
 abelian group, which offers another aspect of the isomorphism stated in Theorem~\ref{thmisoC}.
\vspace{6pt}
\begin{thm}\label{thmGofC}
 Given a Cartan subalgebra $\mathfrak{C}\subset su(2^p)$, the set
 $\mathcal{G}(\mathfrak{C})=\{\mathfrak{B}_i;
 \mathfrak{B}_i\ is\ a\ maximal\ bi-subalgebra\ of\ \mathfrak{C}, 0\leq i<2^p\}$
 forms an abelian group isomorphic to $Z^p_2$ under the
 $\sqcap$-operation, that is,
 $\forall\hspace{2pt} \mathfrak{B}_i,\mathfrak{B}_j\in\mathcal{G}(\mathfrak{C})$,
 $\mathfrak{B}_i\sqcap\mathfrak{B}_j\equiv(\mathfrak{B}_i\cap\mathfrak{B}_j)\cup(\mathfrak{B}^c_i\cap\mathfrak{B}^c_j)\in\mathcal{G}(\mathfrak{C})$,
 where $\mathfrak{B}^c_i=\mathfrak{C}-\mathfrak{B}_i$, $\mathfrak{B}^c_j=\mathfrak{C}-\mathfrak{B}_j$
 and $\mathfrak{B}_0=\mathfrak{C}$ is the identity of the group,
 $0\leq i,j< 2^p$.
\end{thm}
\vspace{3pt}
\begin{proof}
 By Lemma~\ref{lemthe3rdmaxbi}, obviously the Cartan subalgebra $\mathfrak{C}$ is the identity of the group and every
 $\mathfrak{B}_i\in\mathcal{G}(\mathfrak{C})$ is the inverse of itself.
 The abelianness and associativity of the group are self evident,
 and the closure is already asserted in last lemma.
 The rest to prove is the isomorphism between $\mathcal{G}(\mathfrak{C})$ and $Z^p_2$.

 Based on the apparent isomorphism of $\mathfrak{C}_{[\bf 0]}$ and $Z^p_2$,
 each maximal bi-subalgebra of the $0$th-kind Cartan subalgebra $\mathfrak{C}_{[\bf 0]}$
 is uniquely associated to one $p$-digit string and has the form
 $\mathfrak{B}_{\alpha}=\{{\cal S}^{\nu}_{{\bf 0}}:\forall\hspace{2pt}\nu\in{Z^p_2},\hspace{2pt}\nu\cdot\alpha=0\}$,
 $\hspace{2pt}\alpha\in{Z^p_2}$ and $\mathfrak{B}_{\bf 0}=\mathfrak{C}$.
 Given any two members
 $\mathfrak{B}_{\alpha}$ and $\mathfrak{B}_{\beta}\in{\cal G}(\mathfrak{C}_{[\bf 0]})$,
 $\alpha,\beta\in{Z^p_2}$, it is easy to derive the relation
 $\mathfrak{B}_{\alpha}\sqcap\mathfrak{B}_{\beta}
 =\mathfrak{B}_{\alpha+\beta}$; an explicit example is given in Fig.~\ref{1stshellFig}. 
 Let ${\cal S}^{\xi}_{{\bf 0}}$ be an arbitrary spinor in $\mathfrak{B}_{\alpha+\beta}$.
 The identity $\xi\cdot(\alpha+\beta)=0$ implies the two cases
 that $\xi\cdot\alpha=\xi\cdot\beta=0$ or
 $\xi\cdot\alpha=\xi\cdot\beta=1$, and therefore ${\cal S}^{\xi}_{{\bf 0}}$ also belongs
 to either $\mathfrak{B}_{\alpha}\cap\mathfrak{B}_{\beta}$
 or $\mathfrak{B}^c_{\alpha}\cap\mathfrak{B}^c_{\beta}$.
 Similarly, since the identity $\xi\cdot(\alpha+\beta)=0$ results from either of the former two
 $\xi\cdot\alpha=\xi\cdot\beta=0$ and
 $\xi\cdot\alpha=\xi\cdot\beta=1$,
 the other direction that ${\cal S}^{\xi}_{{\bf 0}}\in\mathfrak{B}_{\alpha+\beta}$
 if ${\cal S}^{\xi}_{{\bf 0}}\in
 (\mathfrak{B}_{\alpha}\cap\mathfrak{B}_{\beta})\cup(\mathfrak{B}^c_{\alpha}\cap\mathfrak{B}^c_{\beta})$
 is affirmed.
 Therefore, the isomorphism between $\mathcal{G}(\mathfrak{C}_{[\bf 0]})$ and $Z^p_2$
 is validated.

 Furthermore, another arbitrary Cartan subalgebra $\mathfrak{C}$ is connected to
 $\mathfrak{C}_{[\bf 0]}$ by a conjugate transformation $R$,
 which however does not affect
 the relations of intersection, union and complement among subspaces of the Cartan subalgebras.
 In other words, for any given subspaces $V,W\subset\mathfrak{C}_{[\bf 0]}$,
 $V',W'\subset\mathfrak{C}$, $RVR^{\dag}=V'$ and $RWR^{\dag}=W'$,
 the relations keep unchanged in the two subalgebras such that
 $R(V\cap W)R^{\dag}=V'\cap W'$, $R(V\cup W)R^{\dag}=V'\cup W'$ and
 $R(\mathfrak{C}_{[\bf 0]}-V)R^{\dag}=\mathfrak{C}-V'$.
 Accordingly, the isomorphism is preserved under the conjugate transformation.
\end{proof}
\vspace{6pt}
 The number of maximal bi-subalgebras of a Cartan subalgebra
 is an immediate result of the isomorphism.
\vspace{6pt}
\begin{cor}\label{lemnummaxbi}
 A Cartan subalgebra $\mathfrak{C}\subset su(2^p)$ has a total number $2^p$ of
 maximal bi-subalgebras.
\end{cor}
\vspace{6pt}

 The role of maximal bi-subalgebras is to be unveiled in the
 following series of lemmas.
 It is crucial that a spinor generator commutes with a unique maximal bi-subalgebra.
\vspace{6pt}
\begin{lemma}\label{lemunimaxBcomm}
 For every spinor generator ${\cal S}^{\zeta}_{\alpha}\in su(2^p)$, $p>1$,
 there exists a unique maximal bi-subalgebra $\mathfrak{B}\in{\cal G}(\mathfrak{C})$
 of a Cartan subalgebra $\mathfrak{C}\subset{su(2^p)}$ 
 holding the commutation and anti-commutation relations
 $[{\cal S}^{\zeta}_{\alpha},\mathfrak{B}]=0$ and $\{{\cal S}^{\zeta}_{\alpha},\mathfrak{B}^c\}=0$,
 here $\mathfrak{B}^c=\mathfrak{C}-\mathfrak{B}$.
\end{lemma}
\vspace{3pt}
\begin{proof}
 Obviously as ${\cal S}^{\zeta}_{\alpha}\in\mathfrak{C}$,
 a such unique maximal bi-subalgebra is the Cartan
 subalgebra $\mathfrak{B}=\mathfrak{C}$.
 On the other hand, for a generator ${\cal S}^{\zeta}_{\alpha}\notin\mathfrak{C}\subset su(2^p)$, $p>1$,
 there has at least one generator in $\mathfrak{C}$ commuting with ${\cal S}^{\zeta}_{\alpha}$
 or contradictions occur otherwise. If both
 ${\cal S}^{\eta}_{\beta}\text{ and }{\cal S}^{\xi}_{\gamma}\in\mathfrak{C}$ commute
 with ${\cal S}^{\zeta}_{\alpha}$, so does ${\cal S}^{\eta+\xi}_{\beta+\gamma}$.
 Thus the subset ${\cal B}\neq\mathfrak{C}$ collecting all elements in $\mathfrak{C}$ that commute
 with ${\cal S}^{\zeta}_{\alpha}$ is a bi-subalgebra, noting that
 ${\cal S}^{\zeta}_{\alpha}\in\mathfrak{C}$ if ${\cal B}=\mathfrak{C}$.
 Meanwhile, for any two generators
 ${\cal S}^{\eta'}_{\beta'}\text{ and }{\cal S}^{\xi'}_{\gamma'}\in\mathfrak{C}-{\cal B}$
 not commuting (or anti-commuting) with ${\cal S}^{\zeta}_{\alpha}$, the commutator
 $[{\cal S}^{\eta'+\xi'}_{\beta'+\gamma'},{\cal S}^{\zeta}_{\alpha}]$ vanishes, {\em i.e.},
 ${\cal S}^{\eta'+\xi'}_{\beta'+\gamma'}\in{\cal B}$.
 Accordingly, $\mathfrak{B}\equiv{\cal B}$ is a $1$st maximal bi-subalgebra of $\mathfrak{C}$.
 Now let the uniqueness of $\mathfrak{B}$ be shown by contradiction.
 Suppose there have two $1$st maximal bi-subalgebras $\mathfrak{B}_1$ and $\mathfrak{B}_2$
 both commuting with ${\cal S}^{\zeta}_{\alpha}$ and there is at least one generator
 ${\cal S}^{\eta_1}_{\beta_1}\in\mathfrak{B}_1$
 and ${\cal S}^{\eta_1}_{\beta_1}\in\mathfrak{C}-\mathfrak{B}_2$.
 Apparently this leads to the contradiction that
 $[{\cal S}^{\zeta}_{\alpha},{\cal S}^{\eta_1}_{\beta_1}]=
 \{{\cal S}^{\zeta}_{\alpha},{\cal S}^{\eta_1}_{\beta_1}\}= 0$,
 which ends the proof.
\end{proof}
\vspace{6pt}
 An alternative proof by construction for the unique correspondence 
 is given in Appendix~\ref{appBconstruct}. 
 Directly resulting from the above, the following lemma is of use.
\vspace{6pt}
\begin{lemma}\label{lemSinmaxB}
 Given a spinor generator ${\cal S}^{\zeta}_{\alpha}\in{su(2^p)}$ commuting with
 a maximal bi-subalgebra $\mathfrak{B}\in{\cal G}(\mathfrak{C})$ of a Cartan subalgebra $\mathfrak{C}\subset{su(2^p)}$,
 i.e., $[{\cal S}^{\zeta}_{\alpha},\mathfrak{B}]=0$,
 any generator ${\cal S}^{\eta}_{\beta}\in\mathfrak{C}$
 commuting with ${\cal S}^{\zeta}_{\alpha}$
 must be in $\mathfrak{B}$.
\end{lemma}
\vspace{3pt}
\begin{proof}
 The inclusion ${\cal S}^{\eta}_{\beta}\in\mathfrak{B}$ holds when
 $\mathfrak{B}=\mathfrak{C}$ by default.
 While as $\mathfrak{B}\neq\mathfrak{C}$,
 the vanishing commutator $[{\cal S}^{\zeta}_{\alpha},{\cal S}^{\eta}_{\beta}]=0$ violates Lemma~\ref{lemunimaxBcomm} if ${\cal S}^{\eta}_{\beta}\in\mathfrak{C}-\mathfrak{B}$.
\end{proof}
\vspace{6pt}
 As a consequence of this correspondence, every maximal bi-subalgebra is entitled
 to claim its share in the algebra.
\vspace{6pt}
\begin{lemma}\label{lemdisjointconpar}
 Two subspaces ${\cal W}_1$ and ${\cal W}_2\subset su(2^p)$ respectively determined by
 two maximal bi-subalgebras $\mathfrak{B}_1$ and $\mathfrak{B}_2\in{\cal G}(\mathfrak{C})$
 of a Cartan subalgebra $\mathfrak{C}\subset su(2^p)$,
 i.e., $[{\cal W}_1,\mathfrak{B}_1]=0$ and $[{\cal W}_2,\mathfrak{B}_2]=0$,
 share the null intersection ${\cal W}_1\cap {\cal W}_2=\{0\}$.
\end{lemma}
\vspace{3pt}
\begin{proof}
 It is easy to deduce the contradiction that both the two
 maximal bi-subalgebras $\mathfrak{B}_1$ and $\mathfrak{B}_2$ commute with
 ${\cal S}^{\xi}_{\gamma}$ with the assumption that there exists
 a spinor ${\cal S}^{\xi}_{\gamma}\in {\cal W}_1\cap {\cal W}_2$. 
\end{proof}
\vspace{6pt}

 Lemmas~\ref{lemunimaxBcomm} and~\ref{lemdisjointconpar} reveal a truth of fundamental importance that,
 summarized as the following theorem,
 the group ${\cal G}(\mathfrak{C})$ of maximal bi-subalgebras of a
 Cartan subalgebra $\mathfrak{C}$ can generate a partition of $su(2^p)$.
 For the Lie algebra $su(N)$ of dimension $N$ not being a power of $2$, supposing $2^{p-1}<N<2^p$,
 its corresponding Cartan subalgebras, bi-subalgebras and partitioned subspaces are acquirable
 by applying the removing process to those of $su(2^p)$~\cite{Su}.
 Let this partition generally allowed in Lie algebras be called
 a {\em commutator partition}.
\vspace{6pt}
\begin{thm}\label{thmBSparorderp}
 The abelian group ${\cal G}(\mathfrak{C})$ comprising all
 maximal bi-subalgebras of a Cartan subalgebra $\mathfrak{C}$ in
 the Lie algebra $su(N)$ decides a commutator partition of order p of $su(N)$,
 here $2<2^{p-1}<N\leq 2^p$.
\end{thm}
\vspace{6pt}
 As will be exposed in a later episode~\cite{SuTsai3},
 the commutator partition of the above is exactly the {\em bi-subalgebra partition
 of order $p$} generated by $\mathfrak{C}$.
 Equivalent to a partition of a group given by its subgroup,
 a bi-subalgebra partition of order $l$ in $su(N)$ is generated by
 an $l$-th maximal bi-subalgebra ${\cal B}\subset su(N)$
 by employing the operation of bi-addition
 over spinors of the algebra, where the bi-subalgebra ${\cal B}$
 is no necessarily abelian.
 Refer to~\cite{SuTsai3} for detailed discussions.
 However, only partitions given by abelian bi-subalgebras are in current interest,
 recalling that a Cartan subalgebra is a $p$-th maximal bi-subalgebra
 of $su(N)$ as $2^{p-1}<N\leq 2^p$.
 Notably,
 the commutator partition and bi-subalgebra
 partition generated by a same bi-subalgebra in general are not identical,
 but hold a duality relation~\cite{SuTsai3}.

\begin{figure}[ht]
 \vspace{0.3cm} \hspace{-0.1cm}
\begin{tabular}{m{1.4cm}m{1.4cm}m{1.6cm}m{1.4cm}m{1.2cm}b{5.0cm}}
  &$\mathfrak{C}$&${\cal W}_1$&${\cal W}_2$&${\cal W}_3$&$~~~\cdots\cdots$$~$$2^{p}-1$ {\footnotesize conjugate pairs}\\
\end{tabular}

\hspace{-0.1cm} $\mathfrak{su}(2^p)$
\begin{tabular}{| p{0.6cm} | p{0.6cm} | >{\columncolor{PineGreen}}p{1.5cm} | >{\columncolor{Tan}}p{1.5cm} | >{\columncolor{CadetBlue}}p{1.5cm} | p{4.2cm}|}

\hline
\multicolumn{1}{>{\columncolor{Tan}}l}{}& &  &  &  & \raisebox{-4.50ex}[0cm][0cm]{~~~~~$\cdots\cdots$}\\
\multicolumn{1}{>{\columncolor{Tan}}l}{\vspace{-0.07cm}$\mathfrak{B}_2$}& &  &  &  & \\
\cline{1-2}
\multicolumn{1}{>{\columncolor{CadetBlue}}l}{~~~~~~}&\multicolumn{1}{>{\columncolor{PineGreen}}l}{}  &  &  &  & \\
\multicolumn{1}{>{\columncolor{CadetBlue}}l}{}&\multicolumn{1}{>{\columncolor{PineGreen}}l}{\hspace{-0.22cm}$\mathfrak{B}_1$}  &  &  &  & \\
\hline

\end{tabular}
\\

\vspace{0.5cm}

$~~~~~~~~~~~~~~~~~$
\begin{tabular}{|>{\columncolor{PineGreen}}p{0.5cm}>{\columncolor{PineGreen}}p{0.5cm}|p{2cm}|>{\columncolor{Tan}}p{0.5cm}|p{2.925cm}|p{0.5cm}|}
\cline{1-2} \cline{4-4} \cline{6-6}
&&&&&\\
&\hspace{-0.23cm}$\mathfrak{B}_1$&&$\mathfrak{B}_2$&&$~~~~~~~~~~~~~~~$$\mathfrak{B}_3$\\
\cline{1-2} \cline{6-6}
\end{tabular}\\

\vspace{-0.120cm} $~~~~~~~~~~~~~~~~~~~~~~~~~~~~~~$
\hspace{-0.112cm}
\begin{tabular}{p{0.5cm}p{0.5cm}p{0.5cm}|>{\columncolor{Tan}}p{0.5cm}|p{2cm}|>{\columncolor{CadetBlue}}p{0.5cm}|}
\cline{6-6}
&&&&&\\
&&&&&\\
\cline{4-4} \cline{6-6}
\end{tabular}
\\
\\
\\
\fcaption{A schematic diagram of the commutator partition
 of order $p$ decided by the abelian group
 ${\mathcal{G}}({\mathfrak{C}})$ comprising all maximal bi-subalgebras
 of a Cartan subalgebra ${\mathfrak{C}}\subset su(N)$; here ${\mathcal{W}}_{i}$ is the
 conjugate-pair subspace determined by a maximal bi-subalgebra
 ${\mathfrak{B}}_{i}\in {\mathcal{G}}({\mathfrak{C}})$
 according to the condition
 $[{\cal{W}}_{i}, {\mathfrak{B}}_{i}]= 0$, $1\leq i<2^p$,
 the maximal bi-subalgebras of ${\mathfrak{C}}$
 form the abelian group ${\mathcal{G}}({\mathfrak{C}})$
 with the $\sqcap$-operation, and
 the conjugate-pair subspaces are closed under the commutator,
 $e.g.$, $[{\cal{W}}_{1}, {\mathcal{W}}_{2}]={\cal{W}}_{3}$
 iff ${\mathfrak{B}}_{1}\sqcap{\mathfrak{B}}_{2}={\mathfrak{B}}_{3}$.
 \label{figBiSubAlParC}}
\end{figure}

 This partition conveys  
 a significant implication that,
 guided by the commutator rule $[{\cal W},\mathfrak{B}]=0$,
 a maximal bi-subalgebra $\mathfrak{B}\in{\cal G}(\mathfrak{C})$
 of a Cartan subalgebra $\mathfrak{C}\subset su(N)$
 uniquely {\em defines} one partitioned subspace ${\cal W}$ in the algebra.
 A such subspace ${\cal W}$ is termed
 {\em the commutator subspace} or, specifically,
 {\em the conjugate-pair subspace} or simply {\em the conjugate pair}, {\em determined by}
 $\mathfrak{B}$;
 the Cartan subalgebra $\mathfrak{C}$, determiend by itself, is considered a {\em degrade}
 conjugate pair.
 With an abelian group structure,
 the set of all conjugate-pair subspaces
 is closed under the bi-addition and the commutator operations.
 A schematic diagram of a commutator partition
 is illustrated in
 Fig.~\ref{figBiSubAlParC}.
\vspace{6pt}
\begin{lemma}\label{lemconjpaircl}
 For two conjugate-pair subspaces
 ${\cal W}_1$ and ${\cal W}_2$
 respectively determined by maximal bi-subalgebras
 $\mathfrak{B}_1$ and $\mathfrak{B}_2\in{\cal G}(\mathfrak{C})$
 of a Cartan subalgebra $\mathfrak{C}$,
 the closure holds that,
 $\forall\hspace{2pt}{\cal S}^{\zeta}_{\alpha}\in {\cal W}_1$ and ${\cal S}^{\eta}_{\beta}\in {\cal W}_2$,
 the bi-additive generator ${\cal S}^{\zeta+\eta}_{\alpha+\beta}$ belongs to the conjugate-pair
 subspace ${\cal W}=[{\cal W}_1,{\cal W}_2]$ determined by the maximal bi-subalgebra
 $\mathfrak{B}_1\sqcap\mathfrak{B}_2$, i.e.,
 $[{\cal S}^{\zeta+\eta}_{\alpha+\beta},\hspace{1pt}\mathfrak{B}_1\sqcap\mathfrak{B}_2]=0$
 and thus $[{\cal W},\mathfrak{B}_1\sqcap\mathfrak{B}_2]=0$.
\end{lemma}
\vspace{3pt}
\begin{proof}
 For any ${\cal S}^{\xi}_{\gamma}\in\mathfrak{B}_1\sqcap\mathfrak{B}_2$,
 there have the two identities either
 $\zeta\cdot\gamma+\xi\cdot\alpha=0$ and $\eta\cdot\gamma+\xi\cdot\beta=0$
 if ${\cal S}^{\xi}_{\gamma}\in\mathfrak{B}_1\cap\mathfrak{B}_2$ or,
 by Lemma~\ref{lemthe3rdmaxbi},
 $\zeta\cdot\gamma+\xi\cdot\alpha=1$ and $\eta\cdot\gamma+\xi\cdot\beta=1$
 if ${\cal S}^{\xi}_{\gamma}\in\mathfrak{B}^c_1\cap\mathfrak{B}^c_2$;
 only the former pair of identities remain
 when one of $\mathfrak{B}_1$ and $\mathfrak{B}_2$ is identical to $\mathfrak{C}$.
 The relation is then valid for both the cases that
 $(\zeta+\eta)\cdot\gamma=\xi\cdot(\alpha+\beta)$ and therefore
 $[{\cal S}^{\zeta+\eta}_{\alpha+\beta},\hspace{1pt}\mathfrak{B}_1\sqcap\mathfrak{B}_2]=0$.
\end{proof}
\vspace{6pt}

 In a commutator partition, 
 each conjugate pair also corresponds to a coset subspace via the operation of bi-addition.
\vspace{6pt}
\begin{lemma}\label{lemCWcoset}
 The conjugate-pair subspace ${\cal W}$ determined by a maximal
 bi-subalgebra $\mathfrak{B}$ of a Cartan subalgebra $\mathfrak{C}\subset{su(N)}$
 is a coset of $\mathfrak{C}$ under the bi-addition.
\end{lemma}
\vspace{2pt}
\begin{proof}
 It is easy to confirm the lemma by asserting the inclusions
 ${\cal S}^{\zeta+\eta}_{\alpha+\beta}\in\mathfrak{C}$
 and
 ${\cal S}^{\zeta+\xi}_{\alpha+\gamma}\in\mathcal{W}$
 for all ${\cal S}^{\zeta}_{\alpha},{\cal S}^{\eta}_{\beta}\in\mathcal{W}$
 and
 ${\cal S}^{\xi}_{\gamma}\in\mathfrak{C}$.
 The part ${\cal S}^{\zeta+\eta}_{\alpha+\beta}\in\mathfrak{C}$
 is a direct implication of Lemma~\ref{lemconjpaircl} letting
 $\mathfrak{B}_1=\mathfrak{B}_2$ and
 ${\cal W}_1={\cal W}_2$.
 Through the anti-commutator
 $\{{\cal S}^{\zeta}_{\alpha},\mathfrak{B}^c\}=0$
 by Lemma~\ref{lemunimaxBcomm} and the commutator
 $[{\cal S}^{\xi}_{\gamma},\mathfrak{B}^c]=0$
 for $\mathfrak{B}^c=\mathfrak{C}-\mathfrak{B}$,
 there acquires 
 $\{{\cal S}^{\zeta+\xi}_{\alpha+\gamma},\mathfrak{B}^c\}=0$.
 Further based on Lemma~\ref{lemunimaxBcomm} with
 the commuting of
 ${\cal S}^{\zeta+\xi}_{\alpha+\gamma}$ and $\mathfrak{B}$ due to
 $[{\cal S}^{\zeta}_{\alpha},\mathfrak{B}]=[{\cal S}^{\xi}_{\gamma},\mathfrak{B}]=0$,
 the coverage of  ${\cal S}^{\zeta+\xi}_{\alpha+\gamma}$ in ${\cal W}$
 is concluded.
\end{proof}
\vspace{6pt}

 Explicated in Lemma~\ref{lemunimaxBcomm}, every spinor
 commutes with a unique maximal bi-subalgebra $\mathfrak{B}$ of $\mathfrak{C}$
 but anti-commutes with $\mathfrak{C}-\mathfrak{B}$.
 A similar rule is applicable to conjugate pairs.
\vspace{6pt}
\begin{lemma}\label{lemconparscomm}
 Given two conjugate pairs 
 ${\cal W}_1$ and ${\cal W}_2$
 respectively determined by maximal bi-subalgebras
 $\mathfrak{B}_1$ and $\mathfrak{B}_2\in{\cal G}(\mathfrak{C})$
 of a Cartan subalgebra $\mathfrak{C}$,
 the subspace ${\cal W}_1={W}_{12}\cup\hat{W}_{12}$ can divide into two subspaces
 such that every spinor in ${\cal W}_2$ commutes with ${W}_{12}$ 
 and anti-commutes with $\hat{W}_{12}$, 
 here ${W}_{12}$ and $\hat{W}_{12}$ fulfilling the condition
 ${\cal S}^{\zeta+\eta}_{\alpha+\beta}$ and ${\cal S}^{\zeta'+\eta'}_{\alpha'+\beta'}\in\mathfrak{B}_2$
 as well as
 ${\cal S}^{\zeta'+\eta}_{\alpha'+\beta}$ and ${\cal S}^{\zeta+\eta'}_{\alpha+\beta'}
 \in\mathfrak{B}^c_2=\mathfrak{C}-\mathfrak{B}_2$
 for all
 ${\cal S}^{\zeta}_{\alpha},{\cal S}^{\eta}_{\beta}\in{W}_{12}$
 and
 ${\cal S}^{\zeta'}_{\alpha'},{\cal S}^{\eta'}_{\beta'}\in\hat{W}_{12}$.
\end{lemma}
\vspace{2pt}
\begin{proof}
 Remind Lemma~\ref{lemconjpaircl} that each subspace ${\cal W}_i$ is a
 coset of $\mathfrak{C}$ under the bi-addition, $i=1,2$.
 This feature suggests a division of the subspace ${\cal W}_1={W}_{12}\cup\hat{W}_{12}$
 into two subspaces of the forms
 ${W}_{12}=\{{\cal S}^{\zeta_0+\xi}_{\alpha_0+\gamma}: 
 {\cal S}^{\xi}_{\gamma}\in\mathfrak{B}_2 \}$
 and
 $\hat{W}_{12}=\{{\cal S}^{\zeta_0+\xi'}_{\alpha_0+\gamma'}: 
 {\cal S}^{\xi'}_{\gamma'}\in\mathfrak{B}^c_2 \}$
 with some spinor ${\cal S}^{\zeta_0}_{\alpha_0}\in{\cal W}_1$.
 As a result, an arbitrary spinor ${\cal S}^{\nu}_{\tau}\in{\cal W}_2$
 commutes with ${W}_{12}$ (or $\hat{W}_{12}$)
 and anti-commutes with $\hat{W}_{12}$ (or ${W}_{12}$) if $[{\cal S}^{\nu}_{\tau},{\cal S}^{\zeta_0}_{\alpha_0}]=0$
 (or if $\{{\cal S}^{\nu}_{\tau},{\cal S}^{\zeta_0}_{\alpha_0}\}=0$).
\end{proof}
\vspace{6pt}

 As $\mathfrak{B}_1=\mathfrak{B}_2$ in Lemma~\ref{lemconjpaircl}, the
 bi-additive generator ${\cal S}^{\zeta+\eta}_{\alpha+\beta}$ of
 ${\cal S}^{\zeta}_{\alpha}$ and ${\cal S}^{\eta}_{\beta}$ in the same conjugate-pair subspace
 ${\cal W}_1={\cal W}_2$ is contained in the Cartan subalgebra
 $\mathfrak{B}_1\sqcap\mathfrak{B}_2=\mathfrak{C}$,
 which is the degrade conjugate pair ${\cal W}_0=\mathfrak{C}$ determined by $\mathfrak{C}$.
 Note that this property can be rephrased as
 ``$\forall\hspace{2pt}g,g'\in {\cal W}$, $[g,g']\in\mathfrak{C}$"
 when the dimension of the Lie algebra is not a power of $2$.
 An isomorphism relating group structures embedded
 in $ su(2^p)$ is hence concluded.
\vspace{6pt}
\begin{cor}\label{coroisomsu2p}
 Given a Cartan subalgebra $\mathfrak{C}\subset su(2^p)$, there exists the isomorphism relation
 $\{{\cal W}:{\cal W}\subset su(2^p)
 \text{ and }[{\cal W},\mathfrak{B}]=0,
 \mathfrak{B}\in{\cal G}(\mathfrak{C})\}\cong
 {\cal G}(\mathfrak{C})\cong
 \{{\cal S}^{\zeta}_{\alpha}:{\cal S}^{\zeta}_{\alpha}\in\mathfrak{C}\}\cong Z^p_2$
 for the sets of all conjugate pairs in $su(2^p)$,
 of all maximal bi-subalgebras, of all spinor
 generators in $\mathfrak{C}$ and of all $p$-digit strings under
 the commutator, the $\sqcap$, the bi-addition $\diamond$ operations
 and bit-wise addition respectively.
\end{cor}
\vspace{6pt}

 The isomorphism indicates the structure of a hypercube
 topology respectively held by these four sets in common. 
 In particular, based on the one-to-one correspondence of
 subspaces ${\cal W}$ and $\mathfrak{B}$ governed by
 the commutator rule $[{\cal W},\mathfrak{B}]=0$, it is fair to interpret
 the topological structures of the first two groups that share the same
 group identity $\mathfrak{C}$ as {\em the dual image} of each other.
 For a Lie algebra of dimension $N$, $2^{p-1}<N\leq 2^p$, all the required
 Cartan subalgebras, bi-subalgebras and conjugate pairs can be obtained from those for $su(2^p)$
 with the help of removing process. The isomorphism is thus extended to
 algebras of arbitrary dimensions.
\vspace{6pt}
\begin{cor}\label{coroisomsuN}
 Given a Cartan subalgebra $\mathfrak{C}\subset su(N)$, $2^{p-1}<N\leq 2^p$,
 there exists the isomorphism relation
 $\{{\cal W}:{\cal W}\subset su(N)
 \text{ and }[{\cal W},\mathfrak{B}]=0,
 \mathfrak{B}\in{\cal G}(\mathfrak{C})\}\cong
 {\cal G}(\mathfrak{C})\cong Z^p_2$
 for the sets of all conjugate pairs in $su(N)$,
 of all maximal bi-subalgebras in $\mathfrak{C}$ and of all $p$-digit strings under
 the commutator, the $\sqcap$ operations
 and bit-wise addition respectively.
\end{cor}
\vspace{6pt}
 It is naive to reckon that this isomorphism has ``trivialized" a unitary Lie algebra to
 a set of binary strings of a certain length.
 For still the quaternion characters of the algebra are well
 hidden in each conjugate pair. Specifically, unlike that of a
 standard hypercube, every node of the hypercube-topology structure for $su(N)$
 has internals and is composed of a pair of quaternion units.
 More features of this structure will be detailed in next section.

\section{General Form of Quotient Algebra}\label{secformqa}
\renewcommand{\theequation}{\arabic{section}.\arabic{equation}}
\setcounter{equation}{0} \noindent
 Recall Lemma~\ref{lemconjpaircl} asserting the closure of spinors
 among conjugate pairs under the bi-addition.
 Given two  spinors
 ${\cal S}^{\zeta}_{\alpha}\text{ and }{\cal S}^{\eta}_{\beta}$ of the same
 conjugate-pair subspace
 $\cal{W}$ determined by a maximal bi-subalgebra $\mathfrak{B}\subset\mathfrak{C}$,
 this lemma assures their bi-additive
 ${\cal S}^{\zeta+\eta}_{\alpha+\beta}$ to be either in
 $\mathfrak{B}$ or in $\mathfrak{B}^c=\mathfrak{C}-\mathfrak{B}$.
 The so-called {\em coset rule} can further divide 
 $\cal{W}$ into
 two subspaces by placing the bi-additive spinor
 ${\cal S}^{\zeta+\eta}_{\alpha+\beta}$ in $\mathfrak{B}$ only for every pair
 ${\cal S}^{\zeta}_{\alpha}\text{ and }{\cal S}^{\eta}_{\beta}\in\cal{W}$.
 The conjugate pair $\cal{W}$ thereby splits into two subspaces,
 each of which respects the commutator and the coset rules as
 in the following definition.
\vspace{6pt}
\begin{defn}\label{defcondsub}
 A vector subspace $W\subset su(2^p)$ is a conditioned subspace of
 a maximal bi-subalgebra $\mathfrak{B}\in{\cal G}(\mathfrak{C})$ of a
 Cartan subalgebra $\mathfrak{C}\subset su(2^p)$ 
 if two rules are satisfied:
 the commutator rule $[W,\mathfrak{B}]=0$ and the coset rule
 ${\cal S}^{\zeta+\eta}_{\alpha+\beta}\in\mathfrak{B}$, 
 $\forall\hspace{2pt}{\cal S}^{\zeta}_{\alpha},{\cal S}^{\eta}_{\beta}\in W$.
\end{defn}
\vspace{6pt}
 According to this definition, there formulate two {\em degrade} conditioned subspaces
 the Cartan subalgebra $\mathfrak{C}$ and the null space when taking
 $\mathfrak{B}=\mathfrak{C}$.
 Namely, the degrade conjugate pair $\mathfrak{C}$ is ``bisected" into itself and $\{0\}$.
 While every conjugate pair determined by $\mathfrak{B}\neq\mathfrak{C}$
 is bisected into two non-null conditioned subspaces.
\vspace{6pt}
\begin{lemma}\label{lemcosetru}
 Given the conjugate pair 
 ${\cal W}$ determined by a maximal
 bi-subalgebra $\mathfrak{B}\in{\cal G}(\mathfrak{C})$ of a Cartan subalgebra $\mathfrak{C}\subset{su(2^p)}$,
 either ${\cal W}$ itself is a conditioned subspace of
 $\mathfrak{B}=\mathfrak{C}$, or 
 there exist two non-null conditioned subspaces $W\subset{\cal W}$ and
 $\hat{W}={\cal W}-W$ of $\mathfrak{B}\neq\mathfrak{C}$ 
 such that
 ${\cal S}^{\zeta+\eta}_{\alpha+\beta}\in\mathfrak{B}^c=\mathfrak{C}-\mathfrak{B}$
 for every pair
 ${\cal S}^{\zeta}_{\alpha}\in W$ and ${\cal S}^{\eta}_{\beta}\in\hat{W}$.
\end{lemma}
\vspace{3pt}
\begin{proof}
 The degrade conjugate pair $\mathfrak{C}$ is also a conditioned subspace of itself is self-evident,
 so only the conjugate pair determined by $\mathfrak{B}\neq\mathfrak{C}$ is considered here.
 By Lemma~\ref{lemconjpaircl}, any {bi-additive} spinor ${\cal S}^{\zeta_1+\zeta_2}_{\alpha_1+\alpha_2}$
 of 
 ${\cal S}^{\zeta_1}_{\alpha_1}$ and ${\cal S}^{\zeta_2}_{\alpha_2}\in\cal{W}$ must be in
 either $\mathfrak{B}$ or $\mathfrak{B}^c$.
 Suppose
 ${\cal S}^{\zeta_1+\zeta_2}_{\alpha_1+\alpha_2}$ is in $\mathfrak{B}$.
 There exists
 at least one non-null conditioned subspace of $\mathfrak{B}$ and let it be $W$, that is,
 ${W}\neq\{0\}$ and ${\cal S}^{\zeta_1}_{\alpha_1},{\cal S}^{\zeta_2}_{\alpha_2}\in{W}$.
 Moreover, let every ${\cal S}^{\zeta}_{\alpha}\in {\cal W}$ be included in $W$ as long as
 ${\cal S}^{\zeta_1+\zeta}_{\alpha_1+\alpha}\in\mathfrak{B}$.
 The subspace $\hat{W}$ must be non-null too, for $\mathfrak{B}=\mathfrak{C}$ if $\hat{W}=\{0\}$.
 Given any ${\cal S}^{\zeta}_{\alpha}\in W$ and ${\cal S}^{\eta}_{\beta}\in\hat{W}$,
 the inclusion holds 
 ${\cal S}^{\zeta+\eta}_{\alpha+\beta}\in\mathfrak{B}^c$.
 Because it contradicts the assumption of $W$ being a conditioned subspace of $\mathfrak{B}$ if otherwise.
 Since ${\cal S}^{\zeta+\eta_1}_{\alpha+\beta_1}$ and ${\cal S}^{\zeta+\eta_2}_{\alpha+\beta_2}\in\mathfrak{B}^c$
 for any ${\cal S}^{\zeta}_{\alpha}\in W$ and ${\cal S}^{\eta_1}_{\beta_1},{\cal S}^{\eta_2}_{\beta_2}\in\hat{W}$
 and $\mathfrak{B}$ is a maximal bi-subalgebra of $\mathfrak{C}$,
 another inclusion is reached that ${\cal S}^{\eta_1+\eta_2}_{\beta_1+\beta_2}\in\mathfrak{B}$.
 Consequently, it affirms that $\hat{W}$ is a conditioned subspace of $\mathfrak{B}$ as well.
 The same assertions are acquirable when supposing
 ${\cal S}^{\zeta_1+\zeta_2}_{\alpha_1+\alpha_2}\in\mathfrak{B}^c\neq\{0\}$ in the beginning.
\end{proof}
\vspace{6pt}
  This lemma endorses the bisection format to denote a conjugate pair 
 ${\cal W}=\{{W},\hat{W}\}$.
 A subspace decided by Definition~\ref{defcondsub} is literally known as
 a conditioned subspace {\em of rank zero} and let a partition of a unitary Lie algebra
 consisting of such subspaces be called {\em a quotient-algebra partition of rank zero}
 and denoted as $\{\mathcal{P}_{\mathcal{Q}}(\mathfrak{C})\}$,
 to differ from those of higher ranks discussed in the sequels~\cite{SuTsai2,SuTsai3}.
 Yet, the rank specification is ignored often without confusion.
 The non-commuting feature between the two non-null conditioned subspaces of a pair
 is noted. 
\vspace{6pt}
\begin{lemma}\label{lemconsubcommrank0}
 The two conditioned subspaces $W$ and $\hat{W}$ of 
 a maximal bi-subalgebra $\mathfrak{B}\neq\mathfrak{C}$
 of a Cartan subalgebra $\mathfrak{C}$ anti-commute,
 namely
 $\{{\cal S}^{\zeta}_{\alpha},{\cal S}^{\eta}_{\beta}\}=0$
 for every pair
 ${\cal S}^{\zeta}_{\alpha}\in W$ and ${\cal S}^{\eta}_{\beta}\in\hat{W}$.
\end{lemma}
\vspace{3pt}
\begin{proof}
 As known, two arbitrary spinors either commute or anti-commute.
 Assume that there is a commuting pair of spinors
 ${\cal S}^{\zeta}_{\alpha}\in W$ and ${\cal S}^{\eta}_{\beta}\in\hat{W}$.
 Then, it deduces that the bi-additive ${\cal S}^{\zeta+\eta}_{\alpha+\beta}\in\mathfrak{C}$
 commutes with both ${\cal S}^{\zeta}_{\alpha}$ and ${\cal S}^{\eta}_{\beta}$
 by Lemmas~\ref{lemconjpaircl} and~\ref{lemScomm}, and thus belongs to $\mathfrak{B}$ by
 Lemma~\ref{lemSinmaxB}. 
 This outcome contradicts the fact
 ${\cal S}^{\zeta+\eta}_{\alpha+\beta}\in\mathfrak{B}^c=\mathfrak{C}-\mathfrak{B}$
 asserted in Lemma~\ref{lemcosetru}.
 The assumption is hence denied.
\end{proof}
\vspace{6pt}
 The property of the conjugate partition
 is an immediate implication of the bisection over a conjugate pair.
\vspace{6pt}
\begin{lemma}\label{lemconparrank0}
 The relation of the conjugate partition 
 $[W,\mathfrak{C}]\subset\hat{W}$, $[\hat{W},\mathfrak{C}]\subset{W}$ and
 $[W,\hat{W}]\subset\mathfrak{C}$ is fulfilled for
 the pair of conditioned subspaces $W$ and $\hat{W}$ of a 
 maximal bi-subalgebra $\mathfrak{B}\in{\cal G}(\mathfrak{C})$ of a Cartan subalgebra $\mathfrak{C}$.
\end{lemma}
\vspace{3pt}
\begin{proof}
 This property is essentially a restatement of Lemma~\ref{lemconjpaircl}
 with the imposition of the coset rule on the same conjugate-pair subspace.
 The relation trivially holds for $\mathfrak{B}=\mathfrak{C}$.

 For any ${\cal S}^{\zeta}_{\alpha}\in W$ and
 ${\cal S}^{\xi}_{\gamma}\in\mathfrak{B}^c=\mathfrak{C}-\mathfrak{B}$ as $\mathfrak{B}\neq\mathfrak{C}$,
 the commutator $[{\cal S}^{\zeta}_{\alpha},{\cal S}^{\xi}_{\gamma}]\neq 0$
 produces  ${\cal S}^{\zeta+\xi}_{\alpha+\gamma}\notin W$
 and $\notin\mathfrak{C}$ either.
 Still the bi-additive generator ${\cal S}^{\zeta+\xi}_{\alpha+\gamma}$ belongs
 to the conjugate pair ${\cal W}=\{W,\hat{W}\}$
 determined by the maximal bi-subalgebra $\mathfrak{B}$,
 for $[{\cal S}^{\zeta+\xi}_{\alpha+\gamma},\mathfrak{B}]=0$
 owing to the vanishing of 
 $[{\cal S}^{\zeta}_{\alpha},\mathfrak{B}]$ and
 $[{\cal S}^{\xi}_{\gamma},\mathfrak{B}]$.
 Then by Lemma~\ref{lemcosetru}, the inclusion is asserted that
 ${\cal S}^{\zeta+\xi}_{\alpha+\gamma}\in{\cal W}-W=\hat{W}$
 and thus
 $[W,\mathfrak{C}]\subset\hat{W}$; similarly $[\hat{W},\mathfrak{C}]\subset{W}$.
 The $3$rd inclusion $[W,\hat{W}]\subset\mathfrak{C}$
 simply restates Lemmas~\ref{lemconjpaircl} and~\ref{lemcosetru}.
\end{proof}
\vspace{6pt}

 The relation of the conjugate partition faithfully reflects the quaternion feature
 hidden in each node of the aforesaid configuration of hypercube topology.
 Then the structure of a quotient algebra is prepared to appear.
\vspace{6pt}
\begin{lemma}\label{lemabel}
 Every conditioned subspace of rank zero is abelian.
\end{lemma}
 \vspace{3pt}
\begin{proof}
 By the coset and the commutator rules, any two spinors ${\cal S}^{\zeta}_{\alpha}$
 and ${\cal S}^{\eta}_{\beta}$ in a conditioned subspace commute
 with ${\cal S}^{\zeta+\eta}_{\alpha+\beta}$.
 Two equivalent identities are thus obtained
 $\zeta\cdot(\alpha+\beta)=(\zeta+\eta)\cdot\alpha$ and
 $\eta\cdot(\alpha+\beta)=(\zeta+\eta)\cdot\beta$, which lead to 
 the parity identity
 $\zeta\cdot\beta=\eta\cdot\alpha$. 
\end{proof}
\vspace{6pt}
  Alternatively, a conditioned subspace 
  can be obtained via the operation of bi-addition.
\vspace{6pt}
\begin{lemma}\label{lemWcosetofB}
 Each of the two conditioned subspaces $W$ and $\hat{W}$
 determined by a maximal bi-subalgebra $\mathfrak{B}$ 
 of a Cartan subalgebra $\mathfrak{C}\subset{su(N)}$ 
 is a coset of $\mathfrak{B}$ under the bi-addition.
\end{lemma}
 \vspace{3pt}
\begin{proof}
 The proof concerns only the case for $W$ by showing the inclusions
 ${\cal S}^{\zeta+\eta}_{\alpha+\beta}\in\mathfrak{B}$ and
 ${\cal S}^{\zeta+\xi}_{\alpha+\gamma}\in{W}$ for all
 ${\cal S}^{\zeta}_{\alpha},{\cal S}^{\eta}_{\beta}\in{W}$
 and ${\cal S}^{\xi}_{\gamma}\in\mathfrak{B}$;
 a similar argument is applicable to $\hat{W}$.
 The fact ${\cal S}^{\zeta+\eta}_{\alpha+\beta}\in\mathfrak{B}$
 is by Definition~\ref{defcondsub}.
 Since $W$ is abelian by Lemma~\ref{lemabel}
 and $[{\cal S}^{\xi}_{\gamma},W]=0$
 by Definition~\ref{defcondsub},
 the bi-additive ${\cal S}^{\zeta+\xi}_{\alpha+\gamma}$ commutes with $W$.
 Then, owing to the coverage
 ${\cal S}^{\zeta+\xi}_{\alpha+\gamma}\in{\cal W}=W\cup\hat{W}$ of Lemma~\ref{lemCWcoset}
 and the anti-commutating $\{W,\hat{W}\}=0$ from Lemma~\ref{lemconsubcommrank0},
 the subspace $W$ embraces the bi-additive ${\cal S}^{\zeta+\xi}_{\alpha+\gamma}$.
\end{proof}
\vspace{6pt}
 The following lemma paves the way for the success of the
 subalgebra extension.
\vspace{6pt}
\begin{lemma}\label{lemCunionBW}
 Given the two conditioned subspaces $W$ and $\hat{W}$ determined by
 a maximal bi-subalgebra $\mathfrak{B}$ of a Cartan subalgebra
 $\mathfrak{C}\subset{su(N)}$,
 the unions $\mathfrak{B}\cup{W}$ and $\mathfrak{B}\cup\hat{W}$
 form a Cartan subalgebra in $su(N)$ respectively.
\end{lemma}
 \vspace{3pt}
\begin{proof}
 Here consider the case of the abelian subspace
 ${\cal V}=\mathfrak{B}\cup{W}$ only and
 the same conclusion is reachable for $\mathfrak{B}\cup\hat{W}$
 via a similar argument.
 To affirm that ${\cal V}$ is a bi-subalgebra, 
 the inclusion is required that ${\cal S}^{\zeta+\eta}_{\alpha+\beta}\in{\cal V}$
 for all ${\cal S}^{\zeta}_{\alpha}$
 and ${\cal S}^{\eta}_{\beta}\in{\cal V}$.
 This is obviously true
 if ${\cal S}^{\zeta}_{\alpha}$
 and
 ${\cal S}^{\eta}_{\beta}\in\mathfrak{B}$.
 Since the subspace $W$ is a coset of $\mathfrak{B}$
 under the bi-addition as asserted in Lemma~\ref{lemWcosetofB},
 the bi-additive ${\cal S}^{\zeta+\eta}_{\alpha+\beta}$
 belongs to $\mathfrak{B}$
 if ${\cal S}^{\zeta}_{\alpha}$
 and
 ${\cal S}^{\eta}_{\beta}\in{W}$,
 or to $W$
 if ${\cal S}^{\zeta}_{\alpha}\in\mathfrak{B}$
 and ${\cal S}^{\eta}_{\beta}\in{W}$.
 In one word, the subspace ${\cal V}$ encloses the bi-additive
 ${\cal S}^{\zeta+\eta}_{\alpha+\beta}$
 and thus is a bi-subalgebra. 

 To further show that ${\cal V}=\mathfrak{B}\cup{W}$ is a maximal abelian subalgebra of
 $su(N)$, it needs to verify the nonvanishing commutator
 $[{\cal S}^{\xi}_{\gamma},{\cal V}]\neq 0$
 for all ${\cal S}^{\xi}_{\gamma}\in{su(N)-{\cal V}}$.
 The consideration divides into for the three instances
 ${\cal S}^{\xi}_{\gamma}\in\mathfrak{B}^c=\mathfrak{C}-\mathfrak{B}$,
 ${\cal S}^{\xi}_{\gamma}\in\hat{W}$
 and
 ${\cal S}^{\xi}_{\gamma}\in{su(N)}-\mathfrak{C}\cup{\cal W}$
 with ${\cal W}=W\cup\hat{W}$.
 The spinor ${\cal S}^{\xi}_{\gamma}$ anti-commutes with $W$
 both in the 1st instance
 ${\cal S}^{\xi}_{\gamma}\in\mathfrak{B}^c$
 by Lemma~\ref{lemunimaxBcomm}
 and the 2nd ${\cal S}^{\xi}_{\gamma}\in\hat{W}$
 by Lemma~\ref{lemconsubcommrank0}.
 Lastly, the spinor ${\cal S}^{\xi}_{\gamma}$, as in ${su(N)}-\mathfrak{C}\cup{\cal W}$,
 and the subalgebra $\mathfrak{B}$ do not commute,
 because ${\cal S}^{\xi}_{\gamma}$ uniquely commutes with another maximal
 bi-subalgebra of $\mathfrak{C}$ according to Lemma~\ref{lemunimaxBcomm}.
 The proof completes.
\end{proof}
\vspace{6pt}

\vspace{6pt}
\begin{lemma}\label{lemcommrulforWrank0}
 The commutator vanishes 
 $[[W_1,W_2],\hspace{1pt}\mathfrak{B}_1\sqcap\mathfrak{B}_2]=0$
 for $W_1$ and $W_2$
 respectively being a conditioned subspace of maximal bi-subalgebras
 $\mathfrak{B}_1$ and $\mathfrak{B}_2\in{\cal G}(\mathfrak{C})$
 of a Cartan subalgebra $\mathfrak{C}\subset{su(2^p)}$.
\end{lemma}
 \vspace{3pt}
\begin{proof}
 This is an immediate result of Lemma~\ref{lemconjpaircl}.
\end{proof}

\vspace{6pt}
\begin{lemma}\label{lemcosetrulforWrank0}
 Given 
 ${\cal S}^{\zeta_1}_{\alpha_1},{\cal S}^{\zeta_2}_{\alpha_2}\in W_1$ and
 ${\cal S}^{\eta_1}_{\beta_1},{\cal S}^{\eta_2}_{\beta_2}\in W_2$
 where $W_1$ and $W_2$ are respectively a non-null conditioned subspace
 of maximal bi-subalgebras
 $\mathfrak{B}_1$ and $\mathfrak{B}_2\in{\cal G}(\mathfrak{C})$ of a Cartan subalgebra $\mathfrak{C}\subset{su(2^p)}$,
 the inclusion holds
 ${\cal S}^{\zeta_1+\zeta_2+\eta_1+\eta_2}_{\alpha_1+\alpha_2+\beta_1+\beta_2}
 \in\mathfrak{B}_1\sqcap\mathfrak{B}_2$\hspace{1pt}
 if $[{\cal S}^{\zeta_i}_{\alpha_i},{\cal S}^{\eta_i}_{\beta_i}]\neq 0$, $i=1,2$.
\end{lemma}
\vspace{3pt}
 \begin{proof}
 Owing to the coset rule, it is obvious that
 ${\cal S}^{\zeta_1+\zeta_2}_{\alpha_1+\alpha_2}\in\mathfrak{B}_1$ and
 ${\cal S}^{\eta_1+\eta_2}_{\beta_1+\beta_2}\in\mathfrak{B}_2$.
 The inclusion
 ${\cal S}^{\zeta_1+\zeta_2+\eta_1+\eta_2}_{\alpha_1+\alpha_2+\beta_1+\beta_2}
 \in\mathfrak{B}_1\sqcap\mathfrak{B}_2$ apparently holds when 
 $\mathfrak{B}_1$ or $\mathfrak{B}_2$ is identical to $\mathfrak{C}$.
 As $\mathfrak{B}_1\neq\mathfrak{C}$ and $\mathfrak{B}_2\neq\mathfrak{C}$,
 the validness of this lemma rests upon the fact that
 ${\cal S}^{\zeta_1+\zeta_2}_{\alpha_1+\alpha_2}\in\mathfrak{B}_2$ iff
 ${\cal S}^{\eta_1+\eta_2}_{\beta_1+\beta_2}\in\mathfrak{B}_1$ or
 ${\cal S}^{\zeta_1+\zeta_2}_{\alpha_1+\alpha_2}\in\mathfrak{B}^c_2$ iff
 ${\cal S}^{\eta_1+\eta_2}_{\beta_1+\beta_2}\in\mathfrak{B}^c_1$.
 Based on the beginning assumption
 ${\cal S}^{\zeta_1+\zeta_2}_{\alpha_1+\alpha_2}\in\mathfrak{B}_2$,
 the commutator rule offers two identities
 $(\zeta_1+\zeta_2)\cdot\beta_i+\eta_i\cdot (\alpha_1+\alpha_2)=0, i=1,2$.
 Since $[{\cal S}^{\zeta_i}_{\alpha_i},{\cal S}^{\eta_i}_{\beta_i}]\neq 0, i=1,2$, two more
 identities are provided $\zeta_i\cdot\beta_i+\eta_i\cdot\alpha_i=1, i=1,2$.
 Consequently, two parity relations are derived
 $\zeta_1\cdot\beta_2+\eta_2\cdot\alpha_1=1\text{ and }\zeta_2\cdot\beta_1+\eta_1\cdot\alpha_2=1$,
 which lead to the vanishing commutators 
 $[{\cal S}^{\eta_1+\eta_2}_{\beta_1+\beta_2},{\cal S}^{\zeta_i}_{\alpha_i}]=0, i=1,2$, for
 $\zeta_i\cdot (\beta_1+\beta_2)+(\eta_1+\eta_2)\cdot\beta_i=0$.
 Then by Lemma~\ref{lemSinmaxB}, the result
 ${\cal S}^{\eta_1+\eta_2}_{\beta_1+\beta_2}\in\mathfrak{B}_1$
 is drawn.
 It can be similarly affirmed that
 ${\cal S}^{\zeta_1+\zeta_2}_{\alpha_1+\alpha_2}\in\mathfrak{B}_2$ if
 ${\cal S}^{\eta_1+\eta_2}_{\beta_1+\beta_2}\in\mathfrak{B}_1$.
 The proof for the $2$nd case is closely similar to that for the $1$st and
 only some identities are slightly adjusted during the derivation.
 Such as supposing
 ${\cal S}^{\zeta_1+\zeta_2}_{\alpha_1+\alpha_2}\in\mathfrak{B}^c_2$,
 by Lemma~\ref{lemthe3rdmaxbi}
 the initial two identities change into
 $(\zeta_1+\zeta_2)\cdot\beta_i+\eta_i\cdot (\alpha_1+\alpha_2)=1, i=1,2$.
 Nevertheless, the whole derivation is conducted in parallel, which leads
 to the commutators
 $[{\cal S}^{\eta_1+\eta_2}_{\beta_1+\beta_2},{\cal S}^{\zeta_i}_{\alpha_i}]\neq 0, i=1,2$.
 Since ${\cal S}^{\eta_1+\eta_2}_{\beta_1+\beta_2}\in\mathfrak{B}_2\subset\mathfrak{C}$
 and $[\mathfrak{B}_1,{\cal S}^{\zeta_i}_{\alpha_i}]=0, i=1,2$,
 it deduces that ${\cal S}^{\eta_1+\eta_2}_{\beta_1+\beta_2}\in\mathfrak{B}^c_1$.
 The other direction that
 ${\cal S}^{\zeta_1+\zeta_2}_{\alpha_1+\alpha_2}\in\mathfrak{B}^c_2$ if
 ${\cal S}^{\eta_1+\eta_2}_{\beta_1+\beta_2}\in\mathfrak{B}^c_1$ is similarly asserted.
 Finally, the $1$st case implies that
 ${\cal S}^{\zeta_1+\zeta_2+\eta_1+\eta_2}_{\alpha_1+\alpha_2+\beta_1+\beta_2}\in
 \mathfrak{B}_1\cap\mathfrak{B}_2$ and the $2$nd
 ${\cal S}^{\zeta_1+\zeta_2+\eta_1+\eta_2}_{\alpha_1+\alpha_2+\beta_1+\beta_2}\in
 \mathfrak{B}^c_1\cap\mathfrak{B}^c_2$.
  The proof completes.
 \end{proof}
 \vspace{6pt}
  \noindent
  Respectively confirming the commutator and the coset rules,
  Lemmas~\ref{lemcommrulforWrank0} and~\ref{lemcosetrulforWrank0} validate the condition of closure held
  among conditioned subspaces of two arbitrary maximal bi-subalgebras and of the $3$rd
  bi-subalgebra generated from them with the $\sqcap$-operation.
  The explicit formulation of the condition is further assured as follows.
 \vspace{6pt}
 \begin{lemma}\label{lemcondofcloserank0}
  For two conjugate pairs $\{W_1,\hat{W}_1\}$ and $\{W_2,\hat{W}_2\}$
  respectively determined by maximal bi-subalgebras $\mathfrak{B}_1$ and
  $\mathfrak{B}_2\in{\cal G}(\mathfrak{C})$ of a Cartan subalgebra $\mathfrak{C}$,
  $\mathfrak{B}_1\neq\mathfrak{C}$ and
  $\mathfrak{B}_2\neq\mathfrak{C}$,
  if one of the commutator inclusions is true
  $[W_1,W_2]\subset\hat{W}_3$, $[W_1,\hat{W}_2]\subset{W}_3$, $[\hat{W}_1,W_2]\subset{W}_3$  and
  $[\hat{W}_1,\hat{W}_2]\subset\hat{W}_3$,  so are the rest three,
  here  $\{W_3,\hat{W}_3\}$ being the conjugate pair
  determined by $\mathfrak{B}_1\sqcap\mathfrak{B}_2$.
 \end{lemma}
 \vspace{3pt}
 \begin{proof}
  Only one occasion is proved here that the other three inclusions hold
  if $[W_1,W_2]\subset\hat{W}_3$ is assumed.
  Due to the assumed relation, the nonvanishing commutators
  $[{\cal S}^{\zeta}_{\alpha},{\cal S}^{\eta}_{\beta}]\neq 0$ and
  $[{\cal S}^{\zeta'}_{\alpha'},{\cal S}^{\hat{\eta}}_{\hat{\beta}}]\neq 0$
  produce the generators ${\cal S}^{\zeta+\eta}_{\alpha+\beta}$
  and ${\cal S}^{\zeta'+\hat{\eta}}_{\alpha'+\hat{\beta}}$
  with ${\cal S}^{\zeta}_{\alpha}$, ${\cal S}^{\zeta'}_{\alpha'}\in W_1$,
  ${\cal S}^{\eta}_{\beta}\in W_2$ and ${\cal S}^{\hat{\eta}}_{\hat{\beta}}\in\hat{W}_2$.
  By going through the parity identities given from the facts 
  $[{\cal S}^{\zeta}_{\alpha},{\cal S}^{\zeta'}_{\alpha'}]=0$,
  $[{\cal S}^{\eta}_{\beta},{\cal S}^{\hat{\eta}}_{\hat{\beta}}]\neq 0$,
  ${\cal S}^{\zeta+\zeta'}_{\alpha+\alpha'}$ and
  ${\cal S}^{\eta+\hat{\eta}}_{\beta+\hat{\beta}}\in\mathfrak{C}$, it is easy to verify that
  $[{\cal S}^{\zeta+\eta}_{\alpha+\beta},{\cal S}^{\zeta'+\hat{\eta}}_{\alpha'+\hat{\beta}}]\neq 0$.
  Since both ${\cal S}^{\zeta+\eta}_{\alpha+\beta}$ and
  ${\cal S}^{\zeta'+\hat{\eta}}_{\alpha'+\hat{\beta}}$ are in the conjugate pair determined
  by $\mathfrak{B}_1\sqcap\mathfrak{B}_2$ but do not commute, by Lemma~\ref{lemconsubcommrank0}
  ${\cal S}^{\zeta'+\hat{\eta}}_{\alpha'+\hat{\beta}}$ must be in $W_3$ if
  ${\cal S}^{\zeta+\eta}_{\alpha+\beta}\in\hat{W}_3$.
  The inclusion $[W_1,\hat{W}_2]\subset{W}_3$ is thus asserted.
  The inclusion $[\hat{W}_1,W_2]\subset W_3$ holds similarly.
  Likewise, let the commutators $[{\cal S}^{\zeta}_{\alpha},{\cal S}^{\eta}_{\beta}]\neq 0$
  and $[{\cal S}^{\hat{\zeta}}_{\hat{\alpha}},{\cal S}^{\hat{\eta}}_{\hat{\beta}}]\neq 0$
  produce the generators ${\cal S}^{\zeta+\eta}_{\alpha+\beta}$ and
  ${\cal S}^{\hat{\zeta}+\hat{\eta}}_{\hat{\alpha}+\hat{\beta}}$
  with ${\cal S}^{\zeta}_{\alpha}\in W_1$, ${\cal S}^{\hat{\zeta}}_{\hat{\alpha}}\in \hat{W}_1$,
  ${\cal S}^{\eta}_{\beta}\in W_2$ and ${\cal S}^{\hat{\eta}}_{\hat{\beta}}\in\hat{W}_2$.
  It immediately derives that
  $[{\cal S}^{\zeta+\eta}_{\alpha+\beta},{\cal S}^{\hat{\zeta}+\hat{\eta}}_{\hat{\alpha}+\hat{\beta}}]=0$,
  for $[{\cal S}^{\zeta}_{\alpha},{\cal S}^{\hat{\zeta}}_{\hat{\alpha}}]\neq 0$,
  $[{\cal S}^{\eta}_{\beta},{\cal S}^{\hat{\eta}}_{\hat{\beta}}]\neq 0$,
  ${\cal S}^{\zeta+\hat{\zeta}}_{\alpha+\hat{\alpha}}$ and
  ${\cal S}^{\eta+\hat{\eta}}_{\beta+\hat{\beta}}\in\mathfrak{C}$.
  Since the two generators are in the same conjugate pair and commute,
  by Lemma~\ref{lemconsubcommrank0} again,
  ${\cal S}^{\hat{\zeta}+\hat{\eta}}_{\hat{\alpha}+\hat{\beta}}$ is in $\hat{W}_3$ too
  if ${\cal S}^{\zeta+\eta}_{\alpha+\beta}\in\hat{W}_3$, that is,
  $[\hat{W}_1,\hat{W}_2]\subset\hat{W}_3$.
  It is equally easy to assert the rest three occasions based on the other assumptions.
 \end{proof}
 \vspace{6pt}

  In fact, there is a $2$nd option to phrase the condition of closure
  $[W_1,W_2]\subset{W}_3$, $[W_1,\hat{W}_2]\subset\hat{W}_3$, $[\hat{W}_1,W_2]\subset\hat{W}_3$  and
  $[\hat{W}_1,\hat{W}_2]\subset{W}_3$.
  However, in order to keep consistent with the original formulation in~\cite{Su}, the former expression
  is herein designated and also in the continued work \cite{SuTsai2,SuTsai3}.
  The degrade conjugate pair is thus put as
  $\mathfrak{C}={\cal W}_0=\{{W}_0=\mathfrak{C},\hat{W}_0=\{0\}\}$ such that
  $[{W},\hat{W}]\subset{W}_0=\mathfrak{C}$.
  Moreover, by assigning the associated maximal bi-subalgebra and introducing
  an appropriate parity to each subspace, 
  the two conditioned subspaces of $\mathfrak{B}\in{\cal G}(\mathfrak{C})$
  are redenoted in the forms  $\hat{W}={W}^0(\mathfrak{B})$ and $W={W}^1(\mathfrak{B})$.
  With this new notation, both the conjugate partition of Lemma~\ref{lemconparrank0}
  and the condition of closure of Lemma~\ref{lemcondofcloserank0}
  can be concisely written into one commutator inclusion as in the following theorem.
 \vspace{6pt}
 \begin{thm}\label{thmgenWcommrank0}
  With the abelian group ${\cal G}(\mathfrak{C})$ consisting of all maximal
  bi-subalgebras of a Cartan subalgebra $\mathfrak{C}\subset{su(N)}$,
  $2^{p-1}<N\leq 2^p$,
  the Lie algebra $su(N)$ can be partitioned into $2^{p+1}$ 
  conditioned subspaces following the commutation relation,
  $\forall\hspace{2pt}\epsilon,\sigma\in Z_2$ and $\forall\hspace{2pt}\mathfrak{B},\mathfrak{B}'\in{\cal G}(\mathfrak{C})$,
  \begin{align}\label{eqgenWcommrank0}
  [{W}^{\epsilon}(\mathfrak{B}), {W}^{\sigma}(\mathfrak{B}')]\subset{W}^{\epsilon+\sigma}(\mathfrak{B}\sqcap\mathfrak{B}'),
  \end{align}
  where ${W}^0(\mathfrak{B})=\hat{W}(\mathfrak{B})$ and ${W}^1(\mathfrak{B})={W}(\mathfrak{B})$
  are the two conditioned subspaces of $\mathfrak{B}\neq\mathfrak{C}$, and
  ${W}^0(\mathfrak{B})=\{0\}$ and ${W}^1(\mathfrak{B})=\mathfrak{C}$
  as $\mathfrak{B}=\mathfrak{C}$.
 \end{thm}
 \vspace{6pt}
 The commutation relation of Eq.~\ref{eqgenWcommrank0} is
  a clear and terse representation of the {\em quaternion democracy}
  enjoyed in a quotient-algebra partition of a unitary Lie algebra.
  That is, within this partition, three arbitrary
  conditioned subspaces behave as the three quaternion units
  obeying the rule of a commutator inclusion.
  It is favorable to term this commutation relation 
  {\em the quaternion condition of closure of rank zero}.

  In a quotient algebra partition
  of rank zero $\{\mathcal{P}_{\mathcal{Q}}(\mathfrak{C})\}$,
  there create two types of structures as a direct consequence of Eq.~\ref{eqgenWcommrank0}:
  {\em a quotient algebra of rank zero} when
  a Cartan subalgebra $\mathfrak{C}$ serves as the center subalgebra and
  {\em a co-quotient algebra of rank zero}
  when a conditioned subspace else from $\mathfrak{C}$ 
  plays the role instead.
  The exposition of (co-)quotient algebras of higher ranks is postponed to
  next episode~\cite{SuTsai2}.
\vspace{6pt}
\begin{thm}\label{thmgformQArank0}
 In the quotient algebra partition of rank zero $\{\mathcal{P}_{\mathcal{Q}}(\mathfrak{C})\}$
 generated by a Cartan subalgebra $\mathfrak{C}\subset{su(N)}$,
 $2^{p-1}<N\leq 2^p$, there determine 
 the quotient algebra of rank zero given by $\mathfrak{C}$,
 denoted as $\{{\cal Q}(\mathfrak{C};2^p-1)\}$ and as illustrated in Fig.~\ref{figgform1},
 when $\mathfrak{C}$ is taken as the center subalgebra,
 or a co-quotient algebra of rank zero given by a conditioned
 subspace ${W}(\mathfrak{B})$ (or $\hat{W}(\mathfrak{B})$) of a $1$st maximal bi-subalgebra
 $\mathfrak{B}$ of $\mathfrak{C}$, denoted as $\{{\cal Q}({W}(\mathfrak{B});2^p-1)\}$
 (or $\{{\cal Q}(\hat{W}(\mathfrak{B});2^p-1)\}$) and
 as illustrated in Fig.~\ref{figgformQAtocoQA}, when
 ${W}(\mathfrak{B})$ (or $\hat{W}(\mathfrak{B})$)  as the center subalgebra.
\end{thm}
\vspace{6pt}

 \begin{figure}[!ht]
 \hspace{0pt}
 \begin{center}
 \[\hspace{-50pt}\begin{array}{c}
 \mathfrak{C}
 \end{array}\]
 \[\begin{array}{ccc}
 \hspace{-40pt}{W}(\mathfrak{B})& &\hspace{150pt}\hat{W}(\mathfrak{B})
 \end{array}\]
 \vspace{-15pt}
 \[\begin{array}{ccc}
 \hspace{-15pt}
 \begin{array}{c}
 \{\hspace{2pt}\forall\hspace{2pt} {\cal S}^{\eta}_{\beta},{\cal S}^{\xi}_{\gamma}\notin\mathfrak{C},\\
 \hspace{3pt}[{\cal S}^{\eta}_{\beta},\mathfrak{B}]=0,\\
 \hspace{3pt}{\cal S}^{\eta+\xi}_{\beta+\gamma}\in\mathfrak{B}\hspace{2pt}\}
 \end{array}
 &
 \hspace{0pt}
 &
  \hspace{105pt}
  \begin{array}{c}
 \{\hspace{2pt}\forall\hspace{2pt} {\cal S}^{\hat{\eta}}_{\hat{\beta}},{\cal S}^{\hat{\xi}}_{\hat{\gamma}}\notin\mathfrak{C},\\
 \hspace{3pt}[{\cal S}^{\hat{\eta}}_{\hat{\beta}},\mathfrak{B}]=0,\\
 \hspace{3pt}{\cal S}^{\hat{\eta}+\hat{\xi}}_{\hat{\beta}+\hat{\gamma}}\in\mathfrak{B}\hspace{2pt}\}
 \end{array}
 \end{array}\]\\
 \vspace{6pt}
 \fcaption{The general form of the quotient algebra of rank zero given by a Cartan subalgebra $\mathfrak{C}$, 
 where ${W}(\mathfrak{B})$ and $\hat{W}(\mathfrak{B})$
 are the two conditioned subspaces of a maximal bi-subalgebra $\mathfrak{B}$ of $\mathfrak{C}$.\label{figgform1}}
 \end{center}
 \end{figure}
 \noindent
  Although the algorithm introduced in~\cite{Su} works equally well to
  generate a co-quotient algebra, it is instructive to comment on an
  apparent alternative to produce a such variant structure from its equivalent.
  As the general form shown in Fig.~\ref{figgformQAtocoQA}(b),
  a co-quotient algebra of rank zero is
  given by a conditioned subspace ${W}(\mathfrak{B}_l)$ instead
  of the Cartan subalgebra $\mathfrak{C}$ as in (a). 
  The two structures are equivalent up to a simple rearrangement of conditioned subspaces.
  Due to the change of the center subalgebra in $\{{\cal Q}({W}(\mathfrak{B}_l);2^p-1)\}$,
  every conditioned subspace 
  other than ${W}(\mathfrak{B}_l)$ needs to find its new conjugate partner
  to form a conjugate pair
  via 
  Eq.~\ref{eqgenWcommrank0}
  such that the property of the conjugate partition is preserved. 
  In contrast to $\{{\cal Q}(\mathfrak{C};2^p-1)\}$ having only one degrade 
  pair $\{\mathfrak{C},\{0\}\}$,
  the structure $\{{\cal Q}({W}(\mathfrak{B}_l);2^p-1)\}$ has
  a degrade $\{{W}(\mathfrak{B}_l),\{0\}\}$ and
  an irregular $\{\mathfrak{C},\hat{W}(\mathfrak{B}_l)\}$
  in addition to a number $2^{p}-2$ of regular conjugate pairs.
  It is easily recognized in these structures that a $1$st maximal bi-subalgebra
  $\mathfrak{B}\subset\mathfrak{C}$ has three 
  Cartan subalgebras as its supersets $\mathfrak{C}$,
  $\mathfrak{B}\cup{W}(\mathfrak{B})$ and $\mathfrak{B}\cup\hat{W}(\mathfrak{B})$,
  {\em cf.} Lemma~\ref{lemCunionBW}.
  Both quotient and co-quotient algebras are important to decompositions
  of unitary Lie algebras, which will be examined in depth in continued episodes~\cite{SuTsai2,SuTsai3}.
 \begin{figure}[!ht]
 \hspace{0pt}
 \begin{center}
 \[\begin{array}{ccc}
         \begin{array}{c}
                 \hspace{-150pt}(a)
                 \\
                 \begin{array}{c}
                 \mathfrak{C}
                 \end{array}
                 \\
                 \\
                 \begin{array}{ccc}
                 {W}(\hspace{2pt}\mathfrak{B}_l\hspace{2pt})&&\hat{W}(\hspace{2pt}\mathfrak{B}_l\hspace{2pt})\\
                 &\hspace{20pt}&\\
                 {W}(\mathfrak{B}_m)&&\hat{W}(\mathfrak{B}_m)\\
                 &&\\
                 {W}(\hspace{1pt}\mathfrak{B}_n\hspace{1pt})&&\hat{W}(\hspace{1pt}\mathfrak{B}_n\hspace{1pt})
                 \end{array}
         \end{array}
         &\hspace{20pt}&
         \begin{array}{c}
                 \hspace{-150pt}(b)
                 \\
                 \begin{array}{c}
                 {W}(\hspace{2pt}\mathfrak{B}_l\hspace{2pt})
                 \end{array}
                 \\
                 \\
                 \begin{array}{ccc}
                 \mathfrak{C}&&\hat{W}(\hspace{2pt}\mathfrak{B}_l\hspace{2pt})\\
                 &\hspace{20pt}&\\
                 {W}(\mathfrak{B}_m)&&\hat{W}(\hspace{2pt}\mathfrak{B}_n\hspace{2pt})\\
                 &&\\
                 {W}(\hspace{1pt}\mathfrak{B}_n\hspace{1pt})&&\hat{W}(\hspace{1pt}\mathfrak{B}_m\hspace{1pt})
                 \end{array}
         \end{array}
 \end{array}\]\\
 \vspace{6pt}
 \fcaption{Alternatively producing the co-quotient algebra of rank zero
 $\{{\cal Q}({W}(\mathfrak{B}_l);2^p-1)\}$, as in (b),
 given by a conditioned subspace ${W}(\mathfrak{B}_l)$ of a $1$st maximal bi-subalgebra $\mathfrak{B}_l$ of a Cartan
 subalgebra $\mathfrak{C}\subset{su(2^p)}$ from the quotient algebra of rank zero $\{{\cal Q}(\mathfrak{C};2^p-1)\}$,
 as in (a), via a re-arrangement of abelian subspaces following the conjugate partition;
 here $\mathfrak{B}_l,\mathfrak{B}_m,\mathfrak{B}_n\in{\cal G}(\mathfrak{C})$, $1\leq l,m,n<2^p$
 and $\mathfrak{B}_m\sqcap\mathfrak{B}_n=\mathfrak{B}_l$.\label{figgformQAtocoQA}}
 \end{center}
 \end{figure}

  The general form of a quotient algebra of rank zero 
  admits a further simplification.
  Similar to the proof of Lemma~\ref{lemabel},
  it is plain to show that
  the coset rule has a
  substitute requiring each subspace to be abelian. In other words, a conditioned subspace of a maximal
  bi-subalgebra $\mathfrak{B}$ is an abelian subspace commuting with $\mathfrak{B}$.
\vspace{6pt}
\begin{lemma}\label{lemabelness}
 Based on the commutator rule
 $[W,\mathfrak{B}]=0$ for
 a conditioned subspace $W$ of a maximal
 bi-subalgebra $\mathfrak{B}\in{\cal G}(\mathfrak{C})$ of a Cartan subalgebra
 $\mathfrak{C}\subset{su(2^p)}$, 
 $W$ is abelian if and only if 
 ${\cal S}^{\zeta+\eta}_{\alpha+\beta}\in\mathfrak{B}$
 for every pair
 ${\cal S}^{\zeta}_{\alpha}$ and ${\cal S}^{\eta}_{\beta}\in{W}$.
\end{lemma}
\vspace{3pt}
\begin{proof}
 Suppose that $W$ is abelian and let ${\cal S}^{\zeta}_{\alpha}$
 and ${\cal S}^{\eta}_{\beta}$ be two arbitrary generators in $W$.
 Based on the commutator relation $[W,\mathfrak{B}]=0$ and by Lemma~\ref{lemconjpaircl},
 the generator ${\cal S}^{\zeta+\eta}_{\alpha+\beta}$ must fall in the Cartan subalgebra
 $\mathfrak{C}$. Since commuting with both
 ${\cal S}^{\zeta}_{\alpha}$ and ${\cal S}^{\eta}_{\beta}$, {\em cf.} Lemma~\ref{lemScomm},
 ${\cal S}^{\zeta+\eta}_{\alpha+\beta}$ is further affirmed, by Lemma~\ref{lemSinmaxB},
 to be included in $\mathfrak{B}$.
 While the fact that $W$ is abelian if the coset and the commutator rules
 are satisfied is already asserted in Lemma~\ref{lemabel}.
\end{proof}
\vspace{6pt}
  Recall that the corresponding Cartan subalgebras and maximal bi-subalgebras
  are obtainable through the removing process
  for a Lie algebra of dimension not being a power of $2$.
  Replacing the coset rule by imposing the abelianness in each conditioned subspace,
  this simplified version of the general form shall be widely suited.
\vspace{6pt}
\begin{cor}\label{corogform}
 An alternative general form of a quotient algebra of rank zero  
 $\{{\cal Q}(\mathfrak{C})\}$ applicable to
 unitary Lie algebras of any dimensions is as shown in Fig.~\ref{figgform2}.
\end{cor}
 \begin{figure}[!ht]
 \begin{center}
 \begin{align*}
 \hspace{-25pt}\mathfrak{C}  
 \end{align*}
 \[\begin{array}{ccc}
 \hspace{-50pt}{W}(\mathfrak{B}) &&\hspace{-20pt}\hat{W}(\mathfrak{B}) \\ 
 \hspace{-20pt}\begin{array}{c}
 \{g:
 \forall\hspace{2pt} g,g'\notin\mathfrak{C},\\
 \hspace{-5pt}[\hspace{2pt}g,\mathfrak{B}_{}\hspace{2pt}]=0,\\
 \hspace{-2pt}[\hspace{2pt}g,g'\hspace{2pt}]=0\hspace{2pt}\}
 \end{array}
 &\hspace{100pt}&
 \hspace{10pt}
 \begin{array}{c}
 \{\hat{g}:
 \forall\hspace{2pt} \hat{g},{\hat{g}}'\notin\mathfrak{C},\\
 \hspace{-5pt}[\hspace{2pt}\hat{g},\mathfrak{B}_{}\hspace{2pt}]=0,\\
 \hspace{-2pt}[\hspace{2pt}\hat{g},{\hat{g}}'\hspace{2pt}]=0\hspace{2pt}\}
 \end{array}
 \end{array}\]\\
 \vspace{12pt}
 \fcaption{An alternative general form of a quotient algebra of rank zero 
  admitted in 
  any dimensions.\label{figgform2}}
 \end{center}
 \end{figure}
 \noindent
 Similarly, a co-quotient algebra of rank zero of this form can be constructed from $\{{\cal Q}(\mathfrak{C})\}$
 by rearranging conditioned subspaces following Eq.~\ref{eqgenWcommrank0}.

 \section{Subalgebra Extension}\label{secextend}
 \renewcommand{\theequation}{\arabic{section}.\arabic{equation}}
\setcounter{equation}{0} \noindent
 As will become apparent in continued episodes~\cite{SuTsai2,SuTsai3},
 locating all Cartan subalgebras within a unitary Lie algebra is essential to
 systematic decompositions of the algebra and factorizations of corresponding
 transformations.
 This section is aimed at proving the mathematical truth that all the Cartan subalgebras
 of $su(2^p)$ can be completely searched in $p$ shells of the subalgebra extension.
 Toward this goal, it demands a better detailed understanding of maximal bi-subalgebras. 
 Before proceeding further, some notations in frequent use have to be clarified.
 Given a Cartan subalgebra of the $k$-th kind
 $\mathfrak{C}^{\hspace{1.pt}\{\epsilon\}}_{[\alpha]_k}
 =\{{\cal S}^{\zeta_i}_{\alpha_i}:0\leq i<2^k, {\cal S}^{\zeta_0}_{\alpha_0}\equiv{\cal S}^{\nu_k}_{\mathbf{0}}\}$,
 since subject to the condition $\nu_k\cdot\alpha_i=0$
 as specified in Eq.~\ref{eqk-thkind} for a number $k$ of independent $\alpha_i$,
 the set of the phase strings $\{\nu_k\}$ associated to diagonal spinors
 $\{{\cal S}^{\nu_k}_{\mathbf{0}}\}$ is itself a 
 subgroup of $p-k$ generators and denoted as $[\nu_k]_{p-k}$ or $[\zeta_0]_{p-k}$.
 Being a subgroup of $\{\zeta\}$ that consists of all phase strings in
 $\mathfrak{C}^{\hspace{1.pt}\{\epsilon\}}_{[\alpha]_k}$, $[\zeta_0]_{p-k}$
 can generate a partition in $\{\zeta\}$. 
 Within this partition, each set of the phase strings $\{\zeta_i\}$ associated with
 the binary partitioning $\alpha_i$ corresponds to a coset of $[\zeta_0]_{p-k}$,
 {\em cf.} Fig.~\ref{phasecoset} in Appendix~\ref{appcosets}.
 However, these cosets may have repetitions and vary with
 the integer $h$ which, ranging from $0$ to $k$,
 counts the number of independent cosets in $\{\zeta\}$.
 For instance, as $h=0$ every coset is simply a replica of $[\zeta_0]_{p-k}$, and
 every coset is distinct from each other as $h=k$.
 The subgroup $\{\zeta\}$ is hence granted the notation
 $[\zeta]_q$ with $q=p-k+h$ and $0\leq h\leq k$.
 A Cartan subalgebra of the $k$-th kind is then legitimately abbreviated in the form
$\mathfrak{C}^{\hspace{1.pt}\{\epsilon\}}_{[\alpha]_k}
 =\{{\cal S}^{\zeta_i}_{\alpha_i}:0\leq i<2^k,\alpha_i\in [\alpha]_k\text{ and }\zeta_i\in [\zeta]_q \}$.

 The following subgroup feature as an explicit implication of Lemma~\ref{lemScomm} paves the way for the classification of bi-subalgebras.
\vspace{6pt}
\begin{lemma}\label{lembisubink-thC}
 For a  Cartan subalgebra 
 $\mathfrak{C}^{\hspace{1.pt}\{\epsilon\}}_{[\alpha]_k}
 =\{{\cal S}^{\zeta_i}_{\alpha_i}:0\leq i<2^k,\alpha_i\in [\alpha]_k\text{ and }\zeta_i\in [\zeta]_q\}$,
 all the phase strings associated to a subset of spinor generators
 $\{{\cal S}^{\zeta_j}_{\alpha_j}\}\subset\mathfrak{C}^{\hspace{1.pt}\{\epsilon\}}_{[\alpha]_k}$
 must form a subgroup of $[\zeta]_q$ 
 if the set of binary-partitioning strings $\{\alpha_j\}$ is
 a subgroup of $[\alpha]_k$, 
 i.e., $\{\alpha_j\}=[\alpha]_{k^{'}}$ with $k'\leq k$, and likewise
 the set of binary-partitioning strings must be a subgroup of $[\alpha]_k$ if the set of
 phase strings $\{\zeta_j\}$ is a subgroup $[\zeta]_{q^{'}}\subset[\zeta]_q$ with $q'\leq q$;
 in either case, the set $\{{\cal S}^{\zeta_j}_{\alpha_j}\}$ forms a bi-subalgebra of
 $\mathfrak{C}^{\hspace{1.pt}\{\epsilon\}}_{[\alpha]_k}$.
\end{lemma}
\vspace{6pt}
 Classified by the means of construction, there exist two types of $1$st maximal bi-subalgebras:
 the {\em bit type} constructed according to a maximal
 subgroup in $[\alpha]_k$ and the {\em phase type} according to the like in $[\zeta]_q$.
 The first step of building a bit-type maximal bi-subalgebra is to arbitrarily
 decide a subgroup $[\alpha]_{k-1}$ of $k-1$ generators from $[\alpha]_k$.
 Collecting all the spinor generators with the binary partitioning belonging to $[\alpha]_{k-1}$,
 the set
 $\mathfrak{B}^{\hspace{1.pt}[\zeta_0]_{p-k}}_{[\alpha]_{k-1}}=\{{\cal S}^{\zeta_l}_{\alpha_l}: \forall\hspace{2pt}\alpha_l\in[\alpha]_{k-1}\}$
 is a bit-type maximal bi-subalgebra of
 $\mathfrak{C}^{\hspace{1.pt}\{\epsilon\}}_{[\alpha]_k}$.
 For the latter type, similarly let an arbitrary maximal subgroup $[\zeta_0]_{p-k-1}$ be chosen
 in $[\zeta_0]_{p-k}$ first.
 A maximal subgroup $[\breve{\zeta}_0]_{q'}\subset[\zeta]_q$, $q'=q\text{ or }q-1,$
  is then formed by taking the union
 of $[\zeta_0]_{p-k-1}$ and all its cosets in $[\zeta]_q$, noting that
 every set of phase strings $\{\zeta_i\}$ associated to the spinor
 generators ${\cal S}^{\zeta_i}_{\alpha_i}$, $0<i<2^k$,
 is a coset of $[\zeta_0]_{p-k}$, {\em cf.} Appendix~\ref{appcosets}.
 A phase-type maximal bi-subalgebra
 $\mathfrak{B}^{\hspace{1.pt}[\zeta_0]_{p-k-1}}_{[\alpha]_{k}}=\{{\cal S}^{\zeta_l}_{\alpha_l}: \forall\hspace{2pt}\zeta_l\in[\breve{\zeta}_0]_{q'}\}$
 is acquired by gathering all spinor generators of phase strings included in
 $[\breve{\zeta}_0]_{q'}$.
 Both so created satisfy
 the maximality condition of Definition~\ref{defmaxbisubalg}.
 \vspace{6pt}
 \begin{lemma}\label{lembitphase}
 A bit-type bi-subalgebra
 $\mathfrak{B}^{\hspace{1.pt}[\zeta_0]_{p-k}}_{[\alpha]_{k-1}}=\{{\cal S}^{\zeta_l}_{\alpha_l}: \forall\hspace{2pt}\alpha_l\in[\alpha]_{k-1}\}$,
 constructed according to a maximal subgroup $[\alpha]_{k-1}\subset[\alpha]_{k}$,
 and a phase-type bi-subalgebra
 $\mathfrak{B}^{\hspace{1.pt}[\zeta_0]_{p-k-1}}_{[\alpha]_{k}}=\{{\cal S}^{\zeta_r}_{\alpha_r}: \forall\hspace{2pt}\zeta_r\in[\breve{\zeta}_0]_{q'}\}$,
 according to a maximal subgroup $[\breve{\zeta}_0]_{q'}\subset[\zeta]_q$,
 are both maximal in a given Cartan subalgebra
 $\mathfrak{C}^{\hspace{1.pt}\{\epsilon\}}_{[\alpha]_k}
 =\{{\cal S}^{\zeta_i}_{\alpha_i}:0\leq i<2^k,\alpha_i\in [\alpha]_k\text{ and }\zeta_i\in [\zeta]_q \}$.
 \end{lemma}
 \vspace{6pt} \noindent
 A $1$st maximal bi-subalgebra in a Cartan subalgebra
 $\mathfrak{C}\subset su(2^p)$
 is literally a bisected portion of $\mathfrak{C}$ and, regardless of the type,
 composed of in total $2^{p-1}$ spinor generators.
 Importantly, the union of the bit and the phase types yields the complete set of maximal bi-subalgebras
 for a given Cartan subalgebra.
 In Appendix~\ref{appcosets} more details of their constructions are elaborated. 

  Since a maximal bi-subalgebra is either  bit type or  phase type, 
  two conjugate pairs respectively determined
  by a bit-type and a phase-type 
  are particularly illustrated 
  in the quotient algebra 
  generated by a Cartan subalgebra
  $\mathfrak{C}^{\{\epsilon\}}_{[\alpha]_k}$ in Fig.~\ref{genQA3}.
  According to the alternative proof of Lemma~\ref{lemunimaxBcomm} provided in Appendix~\ref{appBconstruct},
  a spinor ${\cal S}^{\eta_{}}_{\beta_{}}\notin\mathfrak{C}^{\{\epsilon\}}_{[\alpha]_k}$
  is in a conditioned subspace 
  of a bit-type maximal bi-subalgebra $\mathfrak{B}^{[\zeta_0]_{p-k}}_{[\alpha]_{k-1}}$
  as in the first pair if $\beta\in[\alpha]_k$.
  Extended from this pair, the two unions $W_{\rm old}\cup\mathfrak{B}^{[\zeta_0]_{p-k}}_{[\alpha]_{k-1}}$ and
  $\hat{W}_{\rm old}\cup\mathfrak{B}^{[\zeta_0]_{p-k}}_{[\alpha]_{k-1}}$
  form Cartan subalgebras by Lemma~\ref{lemCunionBW}, and
  remain to be a $(k-1)$-th and a $k$-th kinds respectively;
  the conditioned subspaces are hence labeled
  with the subscript ``old."
  While in the $2$nd pair determined by a phase-type maximal bi-subalgebra
  $\mathfrak{B}^{[\zeta_0]_{p-k-1}}_{[\alpha]_{k}}$,
  the addition of two arbitrary binary-partitioning strings 
  $\gamma+\gamma'$,  $\gamma$ and $\gamma'\notin[\alpha]_k$,
  is obliged to be in $[\alpha]_k$ by the coset rule.
  It implies that the union of $[\alpha]_k$ and all binary-partitioning strings $\gamma$ in this pair
  grows into a subgroup $[\alpha]_{k+1}$ of $k+1$ generators.
  Indicated by the subscript ``new," the two Cartan subalgebras hereby extended
  $W_{\rm new}\cup\mathfrak{B}^{[\zeta_0]_{p-k-1}}_{[\alpha]_{k}}$ and
  $\hat{W}_{\rm new}\cup\mathfrak{B}^{[\zeta_0]_{p-k-1}}_{[\alpha]_{k}}$,
  {\em cf.} Lemma~\ref{lemCunionBW},
  are respectively a $(k+1)$-th kind.
  This is the major clue hinting the success of the extension.
  Being crucial to the proof of the algorithm, a structure format
  is endorsed in advance.
 \vspace{6pt}
 \begin{lemma}\label{lemcondsub}
 A bit-type maximal bi-subalgebra
 $\mathfrak{B}^{\hspace{1.pt}[\zeta_0]_{p-k}}_{[\alpha]_{k-1}}
 =\{{\cal S}^{\zeta_l}_{\alpha_l}:\forall\hspace{2pt}\alpha_l\in[\alpha]_{k-1}\subset[\alpha]_k\}$
 and a phase-type 
 $\mathfrak{B}^{\hspace{1.pt}[\zeta_0]_{p-k-1}}_{[\alpha]_k}
 =\{{\cal S}^{\zeta_r}_{\alpha_r}:\forall\hspace{2pt}\zeta_r\in[\breve{\zeta}_0]_{q'}\subset[\zeta]_q\}$
 of a Cartan subalgebra of the $k$-th kind
 $\mathfrak{C}^{\hspace{1.pt}\{\epsilon\}}_{[\alpha]_k}
 =\{{\cal S}^{\zeta_i}_{\alpha_i}:0\leq i<2^k,\hspace{0.pt}\alpha_i\in [\alpha]_k\text{ and }\zeta_i\in [\zeta]_q \}$
 respectively determine
 the conjugate pairs $\{W_{\rm old},\hat{W}_{\rm old}\}$
 and $\{W_{\rm new},\hat{W}_{\rm new}\}$
 of the forms with detailed contents as shown in Fig.~\ref{genQA3}.
 \end{lemma}
 \vspace{3pt}
 \begin{proof}
 At first, let ${\cal W}_{\rm bit}$ and ${\cal W}_{\rm phase}$ denote the conjugate-pair
 subspaces determined by the bit-type and the phase-type maximal bi-subalgebras
 $\mathfrak{B}^{\hspace{1.pt}[\zeta_0]_{p-k}}_{[\alpha]_{k-1}}$ and
 $\mathfrak{B}^{\hspace{1.pt}[\zeta_0]_{p-k-1}}_{[\alpha]_k}$ respectively.
 The inclusion of Lemma~\ref{lemconjpaircl} holds for both the subspaces,
 namely 
 $\forall\hspace{2pt}{\cal S}^{\eta}_{\beta},{\cal S}^{{\eta'}}_{{\beta'}}\in{\cal W}_{\rm bit},\hspace{2pt}
              {\cal S}^{\xi}_{\gamma},{\cal S}^{{\xi'}}_{{\gamma'}}\in{\cal W}_{\rm phase}$,\hspace{2pt}
 ${\cal S}^{\eta+{\eta'}}_{\beta+{\beta'}}\in\mathfrak{C}^{\hspace{1.pt}\{\epsilon\}}_{[\alpha]_k}$
 and
 ${\cal S}^{\xi+{\xi'}}_{\gamma+{\gamma'}}\in\mathfrak{C}^{\hspace{1.pt}\{\epsilon\}}_{[\alpha]_k}$.
 This inclusion leads to the two possibilities that either all the binary partitioning $\beta$
 of ${\cal W}_{\rm bit}$ are in the subgroup $[\alpha]_k$ or all in the coset
 $[\alpha]^c_k=[\alpha]_{k+1}-[\alpha]_k$ belonging to a subgroup of $k+1$ generators
 $[\alpha]_{k+1}\supset [\alpha]_k$;
 the same two possibilities for all the binary partitioning $\gamma$ of ${\cal W}_{\rm phase}$.

 To fix this freedom, the commutator rule comes into play and distinguishes between
 the two conjugate pairs.
 Since $\forall\hspace{2pt}\zeta_0\in [\zeta_0]_{p-k}$,\hspace{1pt}
 $\zeta_0\cdot\beta=0$ owing to the rule
 $[{\cal S}^{\eta}_{\beta},{\cal S}^{\zeta_0}_{\bf 0}]=0$,
 it is discerned that
 $\beta\in[\alpha]_k$ for all ${\cal S}^{\eta}_{\beta}\in{\cal W}_{\rm bit}$.
 On the other hand, since $\forall\hspace{2pt}\zeta'_0\in [\zeta_0]_{p-k-1},\hspace{1pt}
 \zeta''_0\in [\zeta_0]_{p-k}-[\zeta_0]_{p-k-1}$,\hspace{1pt}
 $\zeta'_0\cdot\gamma=0$ owing to the rules
 $[{\cal S}^{\xi}_{\gamma},{\cal S}^{\zeta'_0}_{\bf 0}]=0$
 and $\zeta''_0\cdot\gamma=1$
 by Lemma~\ref{lemSinmaxB} and noting ${\cal S}^{\zeta''_0}_{\bf 0}\in
 \mathfrak{C}-\mathfrak{B}^{\hspace{1.pt}[\zeta_0]_{p-k-1}}_{[\alpha]_k}$,
 it is reached that $\gamma\in [\alpha]_{k+1}-[\alpha]_{k}$
 for all ${\cal S}^{\xi}_{\gamma}\in{\cal W}_{\rm phase}$.

 With the application of the coset rule,
 the subspace ${\cal W}_{\rm bit}$ determined by
 $\mathfrak{B}^{\hspace{1.pt}[\zeta_0]_{p-k}}_{[\alpha]_{k-1}}$
 can be further discriminated into the two conditioned subspaces
 $W_{\rm bit}$ and $\hat{W}_{\rm bit}$.
 Similar to the inclusion above,
 the binary partitioning $\beta$ of $W_{\rm bit}$ must be either all in $[\alpha]_{k-1}$
 or all in $[\alpha]^c_{k-1}=[\alpha]_{k}-[\alpha]_{k-1}$, because
 $\beta+\beta'\in [\alpha]_{k-1}$ 
 for arbitrary ${\cal S}^{\eta}_{\beta}\text{ and }{\cal S}^{{\eta'}}_{{\beta'}}\in W_{\rm bit}$;
 the same  for all the binary partitioning $\hat{\beta}$ of $\hat{W}_{\rm bit}$.
 These two scenarios, attributed to the commutator and the coset rules,
 respectively characterize the binary partitionings of
 the two conditioned subspaces of one conjugate pair.
 Also recall the non-nullness of each conditioned subspace 
 affirmed in Lemma~\ref{lemcosetru}.
 It is therefore recognized that the conjugate pair
 $\{W_{\rm bit},\hat{W}_{\rm bit}\}$ should correspond to
 the pair $\{W_{\rm old},\hat{W}_{\rm old}\}$ of Fig.~\ref{genQA3}.
 Despite the coset rule making no difference in the respective expressions
 of the two conditioned subspaces, solely considering the commutator rule identifies
 ${\cal W}_{\rm phase}$ with the pair $\{W_{\rm new},\hat{W}_{\rm new}\}$.
 \end{proof}

 \begin{figure}[!ht]
 \begin{center}
 \[\hspace{-80pt}\begin{array}{c}
 \mathfrak{C}^{\{\epsilon\}}_{[\alpha]_k}=
 \{\hspace{2pt}{\cal S}^{\zeta_0}_{\mathbf 0},{\cal S}^{\zeta_{1}}_{\alpha_{1}},{\cal S}^{\zeta_{2}}_{\alpha_{2}},\cdots,
 {\cal S}^{\zeta_{2^k-1}}_{\alpha_{2^k-1}}:\\
 \hspace{80pt}\forall\hspace{2pt} \alpha_i\in[\alpha]_k,\hspace{2pt}\zeta_i\in[\zeta]_q,
 \hspace{2pt}\zeta_i\cdot\alpha_i=\epsilon_{ii}\hspace{2pt}\}
 \end{array}\]
 \[\begin{array}{ccc}
 \hspace{-118pt}W_{\text{old}}&&\hspace{-13pt}\hat{W}_{\text{old}}\\
 \hspace{-35pt}\begin{array}{c}
 \{\hspace{2pt}{\cal S}^{\eta_i}_{\beta_i},\cdots:
 \hspace{2pt}\forall\hspace{2pt} {\cal S}^{\eta_i}_{\beta_i},{\cal S}^{\eta_j}_{\beta_j}\notin\mathfrak{C},\\
 \hspace{-3pt}\beta_i,\beta_j\in{[\alpha]}_{k-1}\subset{[\alpha]}_{k},\\
 \hspace{-42pt}1\leq i,j\leq 2^k,\\
 \hspace{-14pt}[\hspace{2pt}{\cal S}^{\eta_i}_{\beta_i},\mathfrak{B}^{[\zeta_0]_{p-k}}_{[\beta]_{k-1}}\hspace{2pt}]=0,\\
 \hspace{-21pt}{\cal S}^{\eta_i+\eta_j}_{\beta_i+\beta_j}\in\mathfrak{B}^{[\zeta_0]_{p-k}}_{[\beta]_{k-1}},\\
 \hspace{-42pt}\eta_i\cdot\beta_i=\sigma_{ii}\hspace{2pt}\}
 \end{array}
 &&
 \hspace{70pt} \begin{array}{c}
 \{\hspace{2pt}{\cal S}^{\hat{\eta}_i}_{\hat{\beta}_i},\cdots:
 \hspace{2pt}\forall\hspace{2pt} {\cal S}^{\hat{\eta}_i}_{\hat{\beta}_i},{\cal S}^{\hat{\eta}_j}_{\hat{\beta}_j}\notin\mathfrak{C},\\
 \hspace{37pt}\hat{\beta}_i,\hat{\beta}_j\in{[\alpha]}^c_{k-1}=[\alpha]_k-{[\alpha]}_{k-1},\\
 \hspace{-40pt}1\leq i,j\leq 2^k,\\
 \hspace{-12pt}[\hspace{2pt}{\cal S}^{\hat{\eta}_i}_{\hat{\beta}_i},\mathfrak{B}^{[\zeta_0]_{p-k}}_{[\beta]_{k-1}}\hspace{2pt}]=0,\\
 \hspace{-21pt}{\cal S}^{\hat{\eta}_i+\hat{\eta}_j}_{\hat{\beta}_i+\hat{\beta}_j}\in\mathfrak{B}^{[\zeta_0]_{p-k}}_{[\beta]_{k-1}},\\
 \hspace{-42pt}\hat{\eta}_i\cdot\hat{\beta}_i=\hat{\sigma}_{ii}\hspace{2pt}\}
 \end{array}\\
 &&\\
 \hspace{-113pt}W_{\text{new}}&&\hspace{-8pt}\hat{W}_{\text{new}}\\
 \hspace{-35pt}\begin{array}{c}
 \{\hspace{2pt} {\cal S}^{\xi_i}_{\gamma_i},\cdots:
 \hspace{2pt}\forall\hspace{2pt} {\cal S}^{\xi_i}_{\gamma_i},{\cal S}^{\xi_j}_{\gamma_j}\notin\mathfrak{C},\\
 \hspace{28pt}\gamma_i,\gamma_j\in{[\alpha]}^c_k=[\alpha]_{k+1}-[\alpha]_{k},\\
 \hspace{-38pt}1\leq i,j\leq 2^k,\\
 \hspace{-3pt}[\hspace{2pt}{\cal S}^{\xi_i}_{\gamma_i},\mathfrak{B}^{[\zeta_0]_{p-k-1}}_{[\alpha]_{k}}\hspace{2pt}]=0,\\
 \hspace{-10pt}{\cal S}^{\xi_i+\xi_j}_{\gamma_i+\gamma_j}\in\mathfrak{B}^{[\zeta_0]_{p-k-1}}_{[\alpha]_{k}},\\
 \hspace{-40pt}\xi_i\cdot\gamma_i=\chi_{ii}\hspace{2pt}\}
 \end{array}
 &&
  \hspace{70pt}\begin{array}{c}
 \{\hspace{2pt}{\cal S}^{\xi_i}_{\gamma_i},\cdots:
 \hspace{2pt}\forall\hspace{2pt} {\cal S}^{\hat{\xi}_i}_{\gamma_i},{\cal S}^{\hat{\xi}_j}_{\gamma_j}\notin\mathfrak{C},\\
 \hspace{28pt}\hat{\gamma}_i,\hat{\gamma}_j\in{[\alpha]}^c_k=[\alpha]_{k+1}-[\alpha]_{k},\\
 \hspace{-38pt}1\leq i,j\leq 2^k,\\
 \hspace{-3pt}[\hspace{2pt}{\cal S}^{\hat{\xi}_i}_{\gamma_i},\mathfrak{B}^{[\zeta_0]_{p-k-1}}_{[\alpha]_{k}}\hspace{2pt}]=0,\\
 \hspace{-10pt}{\cal S}^{\hat{\xi}_i+\hat{\xi}_j}_{\gamma_i+\gamma_j}\in\mathfrak{B}^{[\zeta_0]_{p-k-1}}_{[\alpha]_{k}},\\
 \hspace{-40pt}\hat{\xi}_i\cdot\gamma_i=\hat{\chi}_{ii}\hspace{2pt}\}
 \end{array}
 \end{array}\]\\
 \fcaption{The general form of a quotient algebra of rank zero with details in conjugate pairs.\label{genQA3}}
 \end{center}
 \end{figure}

  \noindent

  Since a Cartan subalgebra of the $k$-th kind
  $\mathfrak{C}_{\hspace{.8pt}k}$ has
  $(2^{p-k}-1)2^k$ phase-type maximal bi-subalgebras, {\em cf}. Appendix~\ref{appcosets},
  there are in total $(2^{p-k}-1)2^{k+1}$ Cartan subalgebras of the
  $(k+1)$-th kind extended from the quotient algebra
  $\{{\cal Q}(\mathfrak{C}_{\hspace{.8pt}k})\}$.
  This is the set of all Cartan subalgebras of the $(k+1)$-th kind that can be
  formed within the quotient algebra given by a Cartan subalgebra of the $k$-th kind.
  Despite having redundancy, the union of these  sets
  must render the complete set of those the $(k+1)$-th kind provided
  they are extended from quotient algebras
  given by the complete set of Cartan subalgebras of the $k$-th kind.
  Along this thinking, the following theorem concludes the extension.
 \vspace{6pt}
 \begin{thm}\label{thmextend}
  All Cartan subalgebras of the Lie algebra $su(2^p)$ can be generated in the first $p$ shells
  through the process of the subalgebra extension.
 \end{thm}
 \vspace{3pt}
 \begin{proof}
 The key to the proof is the assertion that
 for a given Cartan subalgebra of the $(k+1)$-th kind
 $\mathfrak{C}^{\hspace{1.pt}\{\epsilon\}}_{[\alpha]_{k+1}}=
 \{{\cal S}^{\zeta}_{\alpha}:\alpha\in[\alpha]_{k+1}\text{ and }\zeta\in[\zeta]_q\}$,
 $0\leq k<p$, there always exists a Cartan subalgebra of the $k$-th kind
 $\mathfrak{C}_{\hspace{.8pt}k}$ such that
 $\mathfrak{C}^{\hspace{1.pt}\{\epsilon\}}_{[\alpha]_{k+1}}$ is extended from
  the quotient algebra $\{{\cal Q}(\mathfrak{C}_{\hspace{.8pt}k})\}$.
 It is soon seen that a Cartan subalgebra
 $\mathfrak{C}_{\hspace{.8pt}k}$ of this purpose has multiple
 choices, actually as many as the bit-type maximal bi-subalgebras of the given
 $\mathfrak{C}^{\hspace{1.pt}\{\epsilon\}}_{[\alpha]_{k+1}}$.

 A picturized quotient-algebra structure with details 
 is essential to the argument. 
 A representing structure of
 $\{{\cal Q}(\mathfrak{C}^{\hspace{1.pt}\{\epsilon\}}_{[\alpha]_{k+1}})\}$
 is portrayed therein by letting
 every index or parameter $k$ in that of
 Fig.~\ref{genQA3} be replaced by $k+1$.
 According to Lemma~\ref{lemcondsub}, a conditioned subspace ${W}(\mathfrak{B})$
 specifically corresponding to $W_{\rm old}$ in the figure can be formed by fulfilling
 the commutator and the coset rules determined by a bit-type maximal bi-subalgebra
 $\mathfrak{B}$ of $\mathfrak{C}^{\hspace{1.pt}\{\epsilon\}}_{[\alpha]_{k+1}}$.
 That is,
 ${W}(\mathfrak{B})=W_{\text{old}}=\{{\cal S}^{\eta}_{\beta}: 
 \forall\hspace{2pt}{\cal S}^{\eta}_{\beta},{\cal S}^{\eta'}_{\beta'}\notin
 \mathfrak{C}^{\hspace{1.pt}\{\epsilon\}}_{[\alpha]_{k+1}},\hspace{2pt}
 \beta,\beta'\in[\alpha]_k\subset[\alpha]_{k+1},\hspace{2pt}
 [{\cal S}^{\eta}_{\beta},\mathfrak{B}]=0\text{ and }
 {\cal S}^{\eta+\eta'}_{\beta+\beta'}\in\mathfrak{B}\}$ and
 $\mathfrak{B}=\mathfrak{B}^{\hspace{1.pt}[\zeta_0]_{p-k-1}}_{[\alpha]_k}=
 \{{\cal S}^{\zeta'}_{\alpha'}:\forall\hspace{2pt}\alpha'\in[\alpha]_k\subset[\alpha]_{k+1}\}
 \subset\mathfrak{C}^{\hspace{1.pt}\{\epsilon\}}_{[\alpha]_{k+1}}$.
 Simple is the construction that the union  
 $\mathfrak{B}\cup{W}(\mathfrak{B})$
 of an arbitrary choice of a bit-type maximal bi-subalgebra and
 one of its conditioned subspaces in
 $\{{\cal Q}(\mathfrak{C}^{\hspace{1.pt}\{\epsilon\}}_{[\alpha]_{k+1}})\}$
 makes a
 Cartan subalgebra of the $k$-th kind just needed for the extension.

 It is easy to verify that truly $\mathfrak{B}\cup{W}(\mathfrak{B})$ is a Cartan subalgebra,
 {\em cf.} Lemma~\ref{lemCunionBW}.
 Being the union of two commuting abelian subspaces, 
 $\mathfrak{B}\cup{W}(\mathfrak{B})$ is abelian. Due to the partition of the space
 $su(2^p)-\mathfrak{C}^{\hspace{1.pt}\{\epsilon\}}_{[\alpha]_{k+1}}$ prescribed in
 Theorem~\ref{thmBSparorderp}, any element which commutes with $\mathfrak{B}$ but not
 in $\mathfrak{C}^{\hspace{1.pt}\{\epsilon\}}_{[\alpha]_{k+1}}$ must be included
 in the conjugate pair $\{W_{\rm old},\hat{W}_{\rm old}\}$ determined by
 $\mathfrak{B}$.
 Moreover, since $\forall\hspace{2pt}{\cal S}^{\hat{\eta}}_{\hat{\beta}}\in\hat{W}(\mathfrak{B}),
 \hspace{1pt}{\cal S}^{\zeta''}_{\alpha''}\in
 \mathfrak{C}^{\hspace{1.pt}\{\epsilon\}}_{[\alpha]_{k+1}}-\mathfrak{B}$,
 $[{W}(\mathfrak{B}),{\cal S}^{\hat{\eta}}_{\hat{\beta}}]\neq 0$ by Lemma~\ref{lemconsubcommrank0}
 and $[{W}(\mathfrak{B}),{\cal S}^{\zeta''}_{\alpha''}]\neq 0$ by Lemma~\ref{lemunimaxBcomm},
 there exists no commuting element for $\mathfrak{B}\cup{W}(\mathfrak{B})$
 in its complement in $su(2^p)$.
 Carrying all binary partitioning $\alpha'\in[\alpha]_k$, the subspace
 $\mathfrak{B}\cup{W}(\mathfrak{B})\equiv\mathfrak{C}_{\hspace{.8pt}k}$ is
 then confirmed to be a Cartan subalgebra of the $k$-th kind.
 Note that still $\mathfrak{B}$ is a maximal bi-subalgebra of $\mathfrak{C}_{\hspace{.8pt}k}$,
 for all elements in ${W}(\mathfrak{B})$ obeying the coset rule,
 yet being phase type rather than bit type.
 In addition, since all spinor generators in
 $\mathfrak{B}^c=\mathfrak{C}^{\hspace{1.pt}\{\epsilon\}}_{[\alpha]_{k+1}}-\mathfrak{B}
 \subset su(2^p)-\mathfrak{C}_{\hspace{.8pt}k}$ satisfy the commutator and the coset rules,
 the subset $\mathfrak{B}^c$ is a conditioned subspace of $\mathfrak{B}$
 in $\{{\cal Q}(\mathfrak{C}_{\hspace{.8pt}k})\}$.
 Therefore, it asserts that the given Cartan subalgebra
 $\mathfrak{C}^{\hspace{1.pt}\{\epsilon\}}_{[\alpha]_{k+1}}$ can be constructed in
 the quotient algebra $\{{\cal Q}(\mathfrak{C}_{\hspace{.8pt}k})\}$
 by taking the union of a maximal bi-subalgebra
 $\mathfrak{B}\subset\mathfrak{C}_{\hspace{.8pt}k}$ and one of its
 conditioned subspaces $\mathfrak{B}^c$ in $\{{\cal Q}(\mathfrak{C}_{\hspace{.8pt}k})\}$.

 The above argument by induction begins with $k=0$, where the complete set
 of Cartan subalgebras of the $1$st kind, {\em cf.} Eq.~\ref{eq1st-kind}, is extended from
 the intrinsic quotient algebra as shown in Fig.~\ref{1st-general}.
 The Lie algebra $su(2^p)$ has Cartan subalgebras at most up to the $p$-th kind.
 Let the induction proceed from $k=0$ through $k=p-1$, which
 correspond to the first $p$ shells of the subalgebra extension.  The proof completes.
 \end{proof}
 \vspace{6pt} \noindent
%
%

\section{Local Equivalence among Cartan Subalgebras}\label{secLocalCar}
\renewcommand{\theequation}{\arabic{section}.\arabic{equation}}
\setcounter{equation}{0} \noindent
  Any two Cartan subalgebras of a Lie algebra
  as well known are
  related via a conjugate transformation.
  In particular, a Cartan subalgebra of the $k$-th
  kind in $su(N)$, 
  for $2^{p-1}<N\leq 2^p$ and
  $0\leq k<p$, can be mapped to certain of the $p$-th kind by local transformations.
 \vspace{6pt}
 \begin{cor}\label{corolocequiC}
  Every Cartan subalgebra of the $k$-th kind
  $\mathfrak{C}_{\hspace{.8pt}k}$, $0\leq k<p$, is locally equivalent to a Cartan subalgebra of
  the $p$-th kind $\mathfrak{C}_{\hspace{.8pt}p}$ in the Lie algebra $su(2^p)$; namely,
  for a Cartan subalgebra of the $k$-th kind
 \begin{align*}
  \mathfrak{C}_{\hspace{.8pt}k}=\mathfrak{C}^{\hspace{1.pt}\{\epsilon\}}_{[\alpha]_k}
  =\{{\cal S}^{\zeta_i}_{\alpha_i}:0\leq i<2^k,\hspace{2pt}\alpha_i\in[\alpha]_k\text{ and }\zeta_i\in[\zeta]_{q}\},
 \end{align*}
  there exist a local transformation
 \begin{align}\label{eqloctransktop}
  U_{k\rightarrow p}=\prod^{p-k}_{j=1}h^{\hat{\zeta}_j}_{\hat{\alpha}_j}=h^{\hat{\zeta}_1}_{\hat{\alpha}_1}h^{\hat{\zeta}_2}_{\hat{\alpha}_2}\cdots h^{\hat{\zeta}_{p-k}}_{\hat{\alpha}_{p-k}}\in SU(2)^{\otimes p}
 \end{align}
  and
  a Cartan subalgebra of the $p$-th kind
 \begin{align*}
  \hspace{0pt}\mathfrak{C}_{\hspace{.8pt}p}=\mathfrak{C}^{\hspace{1.pt}\{\epsilon'\}}_{[\alpha]_p}=
  \{\hspace{2pt}&{\cal S}^{\zeta_i+\zeta'_i}_{\alpha_i+\alpha'_i}:
  \forall\hspace{2pt}{\cal S}^{\zeta_i}_{\alpha_i}\in\mathfrak{C}_{\hspace{.8pt}k},
  \zeta'_i=\sum^{p-k}_{j=1}{\hat{\zeta}_j}^{(\sigma_{ij})},\hspace{2pt}
  \alpha'_i=\sum^{p-k}_{j=1}{\hat{\alpha}_j}^{(\sigma_{ij})},\hspace{2pt}0\leq i<2^k,\hspace{2pt}1\leq j\leq p-k,\\
  &\hat{\zeta}_j\cdot\alpha_i+\zeta_i\cdot\hat{\alpha}_j=\sigma_{ij}\in{Z_2},
  \hspace{2pt}{\hat{\zeta}_j}^{(0)}={\hat{\alpha}_j}^{(0)}=\mathbf{0},
  \hspace{2pt}{\hat{\zeta}_j}^{(1)}=\hat{\zeta}_j\text{ and }{\hat{\alpha}_j}^{(1)}=\hat{\alpha}_j\hspace{2pt}\}
 \end{align*}
  such that $U_{k\rightarrow p}\hspace{1pt}\mathfrak{C}_{\hspace{.8pt}k} U^{\dag}_{k\rightarrow p}=\mathfrak{C}_{\hspace{.8pt}p}$,
  where
  $h^{\hspace{0.5pt}\hat{\zeta}_j}_{\hat{\alpha}_j}
  =\frac{1}{\sqrt{2}}({\cal S}^{\bf 0}_{\hspace{.01cm}{\bf 0}}+i\cdot(-i)^{\hat{\zeta}_j\cdot\hat{\alpha}_j}{\cal S}^{\hat{\zeta}_j}_{\hat{\alpha}_j})$
  is a one-qubit rotation, 
  $\hat{\zeta}_j\in{Z^p_2}$,
  $\hat{\alpha}_j\in{Z^p_2}-[\alpha]_k$, $\hat{\zeta}_j=\mathbf{0}$ or $\hat{\zeta}_j=\hat{\alpha}_j$,
  $\hat{\alpha}_j+\hat{\alpha}_l\in{Z^p_2}-[\alpha]_k$ as $j\neq l$,
  $1\leq j,l\leq p-k$,
  and each string $\hat{\alpha}_j=a_1 a_2\cdots a_p$ has
  only one single nonzero bit, i.e., $a_s=1$ for some $s$ and $a_t=0$ for all $t\neq s$, $1\leq s,t\leq p$.
 \end{cor}
 \vspace{3pt}
 \begin{proof}
  This lemma can be proved by a straightforward calculation.
  Every action 
  $h^{\hspace{0.5pt}\hat{\zeta}_j}_{\hat{\alpha}_j}
  =I\otimes\cdots I\otimes{R}\otimes I\otimes\cdots \otimes I$ 
  is local which
  makes a $SU(2)$ rotation $R=e^{i\frac{\pi}{4}{\cal S}^{0}_{1}}$
  or $R=e^{\frac{\pi}{4}{\cal S}^{1}_{1}}$ only on one quibit.
 \end{proof}
 \vspace{6pt} \noindent
  For a given $\mathfrak{C}_{\hspace{.8pt}k}$, the choice of $\mathfrak{C}_{\hspace{.8pt}p}$
  locally connected is not unique and parameterized by $p-k$.
  The rotation operator 
  is an example of {\em basic transformations}, whose
  detailed exposition is postponed to Section~\ref{sectrans}.
  Knowing the number of Cartan subalgebras is of interest. 
 \vspace{6pt}
 \begin{cor}\label{coronumC}
  The Lie algebra $su(2^p)$ has a total number
  ${\cal N}(2^p)=\sum^{p}_{k=0}{\cal N}_{k}(2^{p})=\prod^{p}_{i=1}(2^i+1)$ of Cartan subalgebras,
  here
  ${\cal N}_{k}(2^{p})
  =2^{\frac{k(k+1)}{2}}\prod^{k}_{i=1}\frac{2^{p-i+1}-1}{2^{i}-1}$
  being the number of Cartan subalgebras of the $k$-th kind.
 \end{cor}
 \vspace{3pt}
 \begin{proof}
  To construct a Cartan subalgebra of the $k$-th kind
  $\mathfrak{C}^{\hspace{1.pt}\{\epsilon\}}_{[\alpha]_k}$, a subgroup of $k$ generators
  $[\alpha]_k$ is first to be decided from $Z^p_2$ and there are in total
  $\prod^{k}_{i=1}\frac{2^{p-i+1}-1}{2^{i}-1}$ such subgroups.
  Once $[\alpha]_k$ is decided, there have $2^{\frac{k(k+1)}{2}}$ options
  of the parity set $\{\epsilon\}$ for choice.
  Then ${\cal N}_{k}(2^{p})$ is the product of these two numbers.
  Expanding the product in the form
 \begin{align*}
  {\cal N}_{k}(2^{p})=\sum^p_{i_1<i_2<\cdots<i_k,i_1,i_2,\cdots,i_k=1}2^{i_1+i_2+\cdots+i_k}
 \end{align*}
  and summing ${\cal N}_{k}(2^{p})$ from $k=0$ through $k=p$ yields
  the total number $\prod^{p}_{i=1}(2^i+1)$ of Cartan subalgebras.
  It is instructive to read the number ${\cal N}_{k}(2^{p})$
  from the coefficient of the term $x^k$ of the generating function $\prod^{p}_{i=1}(2^ix+1)$.
 \end{proof}
 \vspace{6pt} \noindent
  Despite being more complicated,
  an alternative way to earn the number of Cartan subalgebras 
  is to count those of different local transformations associated to Cartan subalgebras of the $p$-th kind.
  Since the Cartan subalgebras of the Lie algebra $su(N)$, $2^{p-1}<N<2^{p}$,
  are acquirable by applying the removing process to those of $su(2^p)$,
  the number of these Cartan subalgebras ${\cal N}(N)$ should be bounded between 
  ${\cal N}(2^{p-1})$ and ${\cal N}(2^p)$.

  Besides the classification based on their kinds,
  Cartan subalgebras can be classified by specifically reconsidering those
  of the $p$-th kind.
  Recall Eq.~\ref{eqk-thkind} that a Cartan subalgebra of the $k$-th kind
  $\mathfrak{C}^{\hspace{1.pt}\{\epsilon\}}_{[\alpha]_k}
  =\mathfrak{C}^{\hspace{2pt}\epsilon_{11},\epsilon_{12},\cdots,\epsilon_{kk}}_{[\alpha_1,\alpha_2,\cdots,\alpha_k]}
  =\{{\cal S}^{\zeta_l}_{\alpha_l}:0\leq l<2^k\}$
  is characterized by a set of $k$ independent binary-partitioning strings
  $\{\alpha_1,\alpha_2,\cdots,\alpha_k\}$ and the corresponding
  parities, including both self parities $\epsilon_{ii}=\zeta_i\cdot\alpha_i$ and the mutual 
  $\epsilon_{ij}=\zeta_i\cdot\alpha_j=\zeta_j\cdot\alpha_i$ for $1\leq i<j\leq k$.
  Since Cartan subalgebras of the $p$-th kind
  $\mathfrak{C}^{\hspace{1.pt}\{\epsilon\}}_{[\alpha]_p}$
  share the identical set of binary-partitioning strings $[\alpha]_p=Z^p_2$,
  it is sufficient to sort these subalgebras according to their parities.
  In the following assertions,
  a redenoted $p$-th-kind Cartan subalgebra
 \begin{align}\label{eqpthkindnota}
  \mathfrak{C}^{\epsilon_{se},\epsilon_{mu}}_{[\widehat{\alpha}]_p}
  \equiv\mathfrak{C}^{\hspace{1.pt}\{\epsilon\}}_{[\widehat{\alpha}]_p}
  =\mathfrak{C}^{\hspace{2pt}\epsilon_{11},\epsilon_{12},\cdots,\epsilon_{pp}}_{[\widehat{\alpha}_1,\widehat{\alpha}_2,\cdots,\widehat{\alpha}_p]}
 \end{align}
  with the assigning of the set of binary-partitioning strings
  $[\widehat{\alpha}]_p
  =Span\{\widehat{\alpha}_r:\forall\hspace{2pt}\widehat{\alpha}_r=a_1a_2\cdots a_p\in{Z^p_2}\text{ and }a_i
  =\delta_{ir}\text{ for } 1\leq i,r\leq p\}$
  is thus superscribed with two binary strings of
  self parities $\epsilon_{se}=\epsilon_{11}\epsilon_{22}\cdots\epsilon_{pp}\in{Z^p_2}$
  and of mutual parities
  $\epsilon_{mu}=\epsilon_{12}\epsilon_{13}\cdots\epsilon_{p-1p}\in{Z^{\frac{p(p-1)}{2}}_2}$.
 \vspace{6pt}
 \begin{lemma}\label{lempthkindC}
  Given two arbitrary Cartan subalgebras of the $p$-th kind
  $\mathfrak{C}^{\epsilon_{se},\epsilon_{mu}}_{[\widehat{\alpha}]_p}$
  and
  $\mathfrak{C}^{\bar{\epsilon}_{se},\epsilon_{mu}}_{[\widehat{\alpha}]_p}$
  carrying the same mutual-parity string
  $\epsilon_{mu}\in{Z^{\frac{p(p-1)}{2}}_2}$,
  the subalgebra $\mathfrak{C}^{\epsilon_{se},\epsilon_{mu}}_{[\widehat{\alpha}]_p}$
  is mapped to
  $\mathfrak{C}^{\bar{\epsilon}_{se},\epsilon_{mu}}_{[\widehat{\alpha}]_p}$
  by the local transformation
  \begin{align}\label{eqpthCsamemuU}
  U_{se}=\prod^{p}_{i=1}h^{\eta_i}_{\mathbf{0}}=h^{\eta_1}_{\mathbf{0}}h^{\eta_2}_{\mathbf{0}}\cdots h^{\eta_p}_{\mathbf{0}}\in{SU(2)^{\otimes p}},
  \end{align}
  i.e.,
  $U_{se}\mathfrak{C}^{\epsilon_{se},\epsilon_{mu}}_{[\widehat{\alpha}]_p}U^\dag_{se}
  =\mathfrak{C}^{\bar{\epsilon}_{se},\epsilon_{mu}}_{[\hat{\alpha}]_p}$,
  where $h^{\eta_i}_{\mathbf{0}}
  =\frac{1}{\sqrt{2}}({\cal S}^{\bf 0}_{\hspace{.01cm}{\bf 0}}+i{\cal S}^{\eta_i}_{\mathbf{0}})$
  is a one-qubit rotation,
  $\epsilon_{se}=\epsilon_{11}\epsilon_{22}\cdots\epsilon_{pp}$,
  $\bar{\epsilon}_{se}=\bar{\epsilon}_{11}\bar{\epsilon}_{22}\cdots\bar{\epsilon}_{pp}$,
  $\eta_i=\rho_1\rho_2\cdots\rho_p$ is a $p$-digit string with $\rho_{j\neq i}=0$ and
  $\rho_i=\epsilon_{ii}+\bar{\epsilon}_{ii}$,
  $1\leq i,j\leq p$ and $\epsilon_{ii},\bar{\epsilon}_{ii},\rho_j\in{Z_2}$.
 \end{lemma}
 \vspace{6pt} \noindent
  This lemma is easily affirmed by a plain calculation.
  The local action 
  $h^{\eta_i}_{\mathbf{0}}
  =I\otimes I\cdots\otimes I\otimes e^{i\frac{\pi}{4}{\cal S}^1_0}\otimes I\cdots\otimes I$ 
  prescribes a $SU(2)$ rotation $e^{i\frac{\pi}{4}{\cal S}^1_0}$
  on the $i$-th qubit.
  Whereupon the mutual-parity strings of $p$-th-kind Cartan subalgebras
  are legitimate labellings to distinguish equivalence classes of Cartan
  subalgebras under local transformations.
 \vspace{6pt}
 \begin{cor}\label{coronumeqclassC}
  The set of Cartan subalgebras of the Lie algebra $su(2^p)$
  can be partitioned into a number $2^{\frac{p(p-1)}{2}}$ of local equivalence classes
  $\{Cl_{\epsilon_{mu}};\epsilon_{mu}\in{Z^{\frac{p(p-1)}{2}}_2}\}$, 
  and every Cartan subalgebra of the $k$-th kind $\mathfrak{C}_k$ in the class
  $Cl_{\epsilon_{mu}}$, $0\leq k<p$,
  is related to a Cartan subalgebra of the $p$-th kind
  $\mathfrak{C}^{\epsilon_{se},\epsilon_{mu}}_{[\widehat{\alpha}]_p}$ with the mutual-parity string
  $\epsilon_{mu}$ 
  by the local transformation
  \begin{align}\label{eqpthCsameclass}
   U=U_{se}U_{k\rightarrow p}\in{SU(2)^{\otimes p}},
  \end{align}
  here  $U\mathfrak{C}_k U^{\dag}=\mathfrak{C}^{\epsilon_{se},\epsilon_{mu}}_{[\widehat{\alpha}]_p}$
  and $U$ being a composition of local transformations of Eqs.~\ref{eqloctransktop} and~\ref{eqpthCsamemuU}.
 \end{cor}
 \vspace{3pt}
 \begin{proof}
  This directly results from Corollary~\ref{corolocequiC} and
  Lemma~\ref{lempthkindC} by examining the number of $p$-th-kind
  Cartan subalgebras having the same self parities but different
  mutual parities.
 \end{proof}
 \vspace{6pt} \noindent
  On the other hand, two Cartan subalgebras of the $p$-th kind
  differing in mutual-parity strings are linked via
  some numbers of local and two-qubit nonlocal transformations.
 \vspace{6pt}
 \begin{cor}\label{coropthCnonlocal}
  For two local equivalence classes $Cl_{\epsilon_{mu}}$ and $Cl_{\bar{\epsilon}_{mu}}$
  of Cartan subalgebras in the Lie algebra $su(2^p)$,
  a Cartan subalgebra of the $p$-th kind
  $\mathfrak{C}^{\epsilon_{se},\epsilon_{mu}}_{[\widehat{\alpha}]_p}$ in
  $Cl_{\epsilon_{mu}}$ is mapped to one Cartan subalgebra of the
  $p$-th kind
  $\mathfrak{C}^{\bar{\epsilon}_{se},\bar{\epsilon}_{mu}}_{[\hat{\alpha}]_p}$
  in $Cl_{\bar{\epsilon}_{mu}}$
  by the transformation
 \begin{align}\label{eqpthCnonlo}
  U=\prod^{p}_{i,j=1,\hspace{1pt} i\leq j}h^{\eta_{ij}}_{\mathbf{0}}
  =h^{\eta_{11}}_{\mathbf{0}}h^{\eta_{12}}_{\mathbf{0}}\cdots h^{\eta_{pp}}_{\mathbf{0}}\notin{SU(2)^{\otimes p}},
 \end{align}
  where
  $U\mathfrak{C}^{\epsilon_{se},\epsilon_{mu}}_{[\widehat{\alpha}]_p}U^\dag
  =\mathfrak{C}^{\bar{\epsilon}_{se},\bar{\epsilon}_{mu}}_{[\hat{\alpha}]_p}$,
  $h^{\eta_{ij}}_{\mathbf{0}}
  =\frac{1}{\sqrt{2}}({\cal S}^{\bf 0}_{\hspace{.01cm}{\bf 0}}+i{\cal S}^{\eta_{ij}}_{\mathbf{0}})$
  is a one-qubit rotation if $i=j$ or a bipartite action if $i<j$,
  $\epsilon_{se}=\epsilon_{11}\epsilon_{22}\cdots\epsilon_{pp}$,
  $\epsilon_{mu}=\epsilon_{12}\epsilon_{13}\cdots\epsilon_{p-1p}$,
  $\bar{\epsilon}_{se}=\bar{\epsilon}_{11}\bar{\epsilon}_{22}\cdots\bar{\epsilon}_{pp}$,
  $\bar{\epsilon}_{mu}=\bar{\epsilon}_{12}\bar{\epsilon}_{13}\cdots\bar{\epsilon}_{p-1p}$,
  $\eta_{ij}=\rho_1\rho_2\cdots\rho_i\cdots\rho_j\cdots\rho_p$ is the $p$-digit string with
  $\rho_i=\sum^p_{s=1}(\epsilon_{is}+\bar{\epsilon}_{is})$,
  $\rho_j=\sum^p_{s=1}(\epsilon_{js}+\bar{\epsilon}_{js})$
  and $\rho_{r\neq i,j}=0$, $1\leq i,j,r\leq p$ and $\epsilon_{ij},\bar{\epsilon}_{ij},\rho_r\in{Z_2}$.
 \end{cor}
 \vspace{3pt}
 \begin{proof}
  The validation of the corollary is through a simple calculation.
  The transformation $h^{\eta_{ij}}_{\mathbf{0}}
  =I\otimes I\otimes\cdots \otimes I\otimes e^{i\frac{\pi}{4}{\cal S}^1_0}\otimes  I\otimes \cdots \otimes I$
  is a $SU(2)$ rotation
  on the $i$-th qubit if $i=j$,
  or a two-qubit operator
  acting on the $i$-th and $j$-th qubits otherwise.
 \end{proof}
 \vspace{6pt} \noindent
  A {\em hypercube-topology} structure of dimension $2^{\frac{p(p-1)}{2}}$ helps
  envision the conjugations among the $2^{\frac{p(p-1)}{2}}$ local equivalence classes
  of the Lie algebra $su(2^p)$,
  in which each vertex of the structure represents one equivalence class and
  the edge joining two vertices marks a nonlocal bipartite
  transformation bridging the two equivalence classes.

\section{Transformation Connecting Decomposition Paths\label{sectrans}}
\renewcommand{\theequation}{\arabic{section}.\arabic{equation}}
\setcounter{equation}{0} \noindent
 As described in \cite{Su}, a unitary action $U\in SU(N)$, $2^{p-1}<N\leq 2^p$,
 admits a recursive factorization according to a decomposition
 sequence $seq_{dec}=\{\hspace{2pt}{\cal A}_{[r]};\hspace{.1cm}r=0,1,2,\cdots,p,\hspace{.2cm} {\cal A}_{[0]}={\cal A}\
  {\rm and}\ {\cal A}_{[p]}=\mathfrak{t}_{[p]}\hspace{2pt}\}$ within 
 $\{{\cal Q}(\mathcal{A};2^p-1)\}$.
 Except the center subalgebra as $r=0$, the member ${\cal A}_{[r]}$ is a
 conditioned subspace in the quotient algebra.
 In other words, a decomposition sequence decides a decomposition path in the group $SU(N)$
 that leads to a factorization of $U$.
 For example, a unitary action $U\in SU(8)$ can be
 factorized into the following form,
  \begin{align*}
 U=e^{ia_{[3]1}}e^{ia_{[2]1}}e^{ia_{[3]2}}e^{ia_{[1]1}}
 e^{ia_{[3]3}}e^{ia_{[2]2}}e^{ia_{[3]4}}e^{ia_{[0]1}}
 e^{ia_{[3]5}}e^{ia_{[2]3}}e^{ia_{[3]6}}e^{ia_{[1]2}}
 e^{ia_{[3]7}}e^{ia_{[2]4}}e^{ia_{[3]8}}\text{,}
 \end{align*}
 where $a_{[i]j}$ is a vector in the subspace ${\cal A}_{[i]}$,
 $i=0,\cdots,3$ and $j=1,2,\cdots,2^i$.
 To determine these vectors in the corresponding subspace, specifically
 calculating the parameter associated to each generator in the subspace,
 it requires a recursive application of
  the SVD (the singular value decomposition) to the action $U$.
 Note that only the type-{\bf AI} decomposition
 is concerned at present and the continuation to the other types
 will be expounded in the sequels~\cite{SuTsai2,SuTsai3} .
 Even being systematic and recursive, the process of calculating these
 parameters is cumbersome in particular when the dimension $N$ is large.
 Nevertheless, since a decomposition sequence can be transformed
 to any another one,
 in practice the calculation is necessary
 for only one decomposition sequence.

 A choice is made for convenience and termed as the {\em referential decomposition sequence},
 \begin{align}\label{eqdecomseqintrAI}
 \hat{seq}_{dec}=\{\hspace{2pt}{\mathcal{C}}_{[r]};\hspace{1pt}0\leq r\leq p,\hspace{2pt}
 {\mathcal{C}}_{[0]}={\mathfrak{C}}_{[\mathbf 0]}\hspace{2pt}\text{ and }
 \hspace{2pt}{\mathcal{C}}_{[r]}={W}^0_{\beta_r}\text{ for }r\neq 0 \hspace{2pt}\},
 \end{align}
 which is selected from the intrinsic
 quotient algebra of rank zero $\{{\cal Q}(\mathfrak{C}_{[\mathbf{0}]};2^p-1)\}$.
 Every designated component 
 ${W}^0_{\beta_r}={W}^0(\mathfrak{B}_{\beta_r})
 =\{{\cal S}^{\eta_r}_{\beta_r}:\forall\hspace{2pt}\eta_r\in{Z^p_2},\hspace{2pt}\eta_r\cdot\beta_r=1\}\in\{{\cal Q}(\mathfrak{C}_{[\mathbf{0}]})\}$
 is a conditioned subspace of the maximal bi-subalgebra
 $\mathfrak{B}_{\beta_r}
 =\{{\cal S}^{\eta_r}_{\beta_r}:\forall\hspace{2pt}\eta_r\in{Z^p_2},
 \hspace{2pt}\eta_r\cdot\beta_r=0\}\subset\mathfrak{C}_{[\mathbf{0}]} $
 and
 exclusively consists of
 spinor generators of the odd self parity and of the identical binary partitioning
 $\beta_r=b_1b_2\cdots b_{p}$
 with $b_i=\delta_{i,r}$ for $i=1,2,\cdots,p$.
 Let $\Gamma_{\{\mathcal{A}_{[r]}\}}(U)$ denote the set of determined
 parameters in factorizing $U$ according to the decomposition sequence
 ${\{\mathcal{A}_{[r]};0\leq r\leq p\}}$.
 The factorizations of $U$ according to the two decomposition sequences
 are related via the conjugate formula
 $\Gamma_{\{\mathcal{A}_{[r]}\}}(Q^{\dagger}UQ)=\Gamma_{\{Q\mathcal{A}_{[r]}Q^{\dagger}\}}(U)$,
 where the transformation $Q$ {\em connects} the two sequences, {\em i.e.},
 $Q\mathcal{A}_{[r]}Q^{\dagger}={\mathcal{C}}_{[r]}$.
 This connecting mapping will be constructed in three steps:
 the first diagonalizing the center subalgebra $\mathcal{A}_{[0]}$,
 the $2$nd correcting parities and the $3$rd appropriately exchanging
 conditioned subspaces.
 All these transformations can be formed
 in compositions of {\em basic transformations}
 \begin{align}\label{eqbasic-trans}
 h^{\zeta}_{\alpha}=
 \frac{1}{\sqrt{2}}({\cal S}^{\nu}_{\mathbf 0}+i\cdot(-i)^{\zeta\cdot\alpha}{\cal S}^{\zeta}_{\alpha})
 \hspace{3pt}\text{ with }\hspace{2pt}
 \hspace{3pt}\nu\cdot\alpha=0,
 \end{align}
 $\zeta\cdot\alpha\in{Z_2}$, which are spinor-to-spinor $s$-rotations in~Appendix~B of~\cite{Su}.
 For the simplicity, it is set $\nu={\mathbf 0}$.
 Such transformations make possible the {\em spinor-to-spinor} mapping.
 \vspace{6pt} \noindent
 \begin{lemma}\label{lemh-trans}
 A basic transformation $h^{\zeta}_{\alpha}$ maps a spinor generator
 ${\cal S}^{\eta}_{\beta}$ to 
 itself if\hspace{2pt}
 $[{\cal S}^{\zeta}_{\alpha},{\cal S}^{\eta}_{\beta}]=0$ or to
 ${\cal S}^{\zeta+\eta}_{\alpha+\beta}$ if\hspace{2pt}
 $[{\cal S}^{\zeta}_{\alpha},{\cal S}^{\eta}_{\beta}]\neq 0$.
 \end{lemma}
 \vspace{3pt}
 \begin{proof}
 The RHS of the straightforward derivation
 $h^{\zeta}_{\alpha}{\cal S}^{\eta}_{\beta}{h^{\zeta}_{\alpha}}^{\dagger}
 =\frac{1}{2}(1+(-1)^{\eta\cdot\alpha+\zeta\cdot\beta}){\cal S}^{\eta}_{\beta}
 +\frac{1}{2}i^{\zeta\cdot\alpha}\cdot(-1)^{\eta\cdot\alpha}(1-(-1)^{\eta\cdot\alpha+\zeta\cdot\beta})
 {\cal S}^{\zeta+\eta}_{\alpha+\beta}$,
 reduces to ${\cal S}^{\eta}_{\beta}$
 if $\eta\cdot\alpha+\zeta\cdot\beta=0$ or, up to a phase, to ${\cal S}^{\zeta+\eta}_{\alpha+\beta}$
 if $\eta\cdot\alpha+\zeta\cdot\beta=1$.
 \end{proof}
 \vspace{6pt}\noindent
 Of particular note is that the transforming scheme introduced here
 is not only applicable to a group action of dimension a power of $2$,
 but also
 to an action in $SU(N)$ of dimension $2^{p-1}<N<2^p$ by first {\em embedding} its algebra $su(N)$
 to the space of $su(2^p)$.
 Refer to Appendix~B in~\cite{Su} for the simple formulation of embedding.

 \vspace{6pt}\noindent
 {\bf Diagonalization Transformation $R$}\noindent

 The first step of connecting the two decompositions is to have
 the center subalgebra $\mathcal{A}_{[0]}$ {\em return} to the intrinsic
 subspace, {\em i.e.}, diagonalizing $\mathcal{A}_{[0]}$.
 Suppose the center subalgebra is a Cartan subalgebra of the $k$-th kind
 $\mathcal{A}_{[0]}=\mathfrak{C}^{\hspace{1pt}\{\epsilon\}}_{[\alpha]_k}=\{{\cal S}^{\xi_l}_{\alpha_l}:0\leq l<2^k\}$
 and let $\{\hspace{2pt}\alpha_1,\alpha_2,\ldots,\alpha_k\hspace{2pt}\}$ be a generating set
 of the subgroup $[\alpha]_k$.
 Since a basic transformation $h^{\zeta}_{\alpha}$ maps ${\cal S}^{\zeta'}_{\alpha}$
 to ${\cal S}^{\zeta+\zeta'}_{\mathbf 0}$ and leaves ${\cal S}^{\eta}_{\beta}$ invariant
 if $(\zeta+\zeta')\cdot\alpha=1$ and $\zeta\cdot\beta+\eta\cdot\alpha=0$,
 the operator $R$ of the form
 \begin{align}\label{eqdial-trans}
 R=\prod^{k}_{i=1}h^{\zeta_i}_{\alpha_i}
       =h^{\zeta_k}_{\alpha_k}h^{\zeta_{k-1}}_{\alpha_{k-1}}\cdots h^{\zeta_1}_{\alpha_1}
       \text{}
 \end{align}
 diagonalizes $\mathcal{A}_{[0]}$ into $\mathfrak{C}_{[\mathbf 0]}$,
 respecting the condition $\xi_i\cdot\alpha_j+\zeta_j\cdot\alpha_i=\delta_{ij}$ and
 $\{{\cal S}^{\xi_i}_{\alpha_i}\}\subset\mathcal{A}_{[0]}$ for $1\leq i,j\leq k$.
 This product has no order, because the basic transformations so chosen commute with each other.
 A note is that 
 the operator has an alternative
\begin{align}\label{eqalterdialR}
 R=\prod^{p}_{i=1}h^{\zeta_i}_{\alpha_i}
 =h^{\zeta_p}_{\alpha_p}h^{\zeta_{p-1}}_{\alpha_{p-1}}\cdots h^{\zeta_1}_{\alpha_1}
\end{align}
 additionally multiplied by $p-k$ transformations
 $h^{\zeta_t}_{\alpha_t=\mathbf{0}}$, here $k< t\leq p$
 and $\zeta_t\cdot\alpha_i=0$ for $1\leq i\leq k$.

 \vspace{6pt}\noindent
 {\bf Parity Transformation $P$}\noindent

 Once the center subalgebra $\mathcal{A}_{[0]}$ is diagonalized
 as shown in the $2$nd column of Fig.~\ref{EPR-process},
 each $\mathcal{A}_{[r]}$ is mapped to
 some conditioned subspace in
 $\{{\cal Q}(\mathfrak{C}_{[{\mathbf 0}]};2^p-1)\}$, {\em i.e.},
 $R{\cal A}_{[0]}R^{\dagger}=\mathfrak{C}_{[\mathbf 0]}$ and
 $R{\cal A}_{[r]}R^{\dagger}=W^{\sigma_r}_{\alpha_r}=W^{\sigma_r}(\mathfrak{B}_{\alpha_r})$,
 here $\sigma_r\in{Z_2}$, $\alpha_r\in Z^p_2$, $1\leq r\leq p$
 and $\mathfrak{B}_{\alpha_r}$ being a maximal bi-subalgebra of $\mathfrak{C}_{[\mathbf 0]}$.
 All spinor generators in the conditioned subspace
 $W^{\sigma_r}_{\alpha_r}
 =\{{\cal S}^{\zeta_r}_{\alpha_r}:\forall\hspace{2pt}\zeta_r\in{Z^p_2},\hspace{2pt}\zeta_r\cdot\alpha_r=1+\sigma_r\}$
 are associated with the identical binary partitioning $\alpha_r$
 and self parity $1+\sigma_r$ yet which is not necessarily odd.
 It demands a transformation of parity correction that fixes the self parity of each conditioned subspace
 to $1$ and keeps the binary partitioning unchanged.
 The operator
 \begin{align}\label{eqphase-trans}
 P=h^{\eta}_{\mathbf 0}\text{}
 \end{align}
 serves the purpose, where $\eta\cdot\alpha_r=\sigma_r\text{ and }1\leq r\leq p$.

 \begin{figure}[!ht]
 \begin{center}
 \[\begin{array}{ccccccc}
 \begin{array}{c}
 {\cal A}_{[0]}={\cal A}\\
 {\cal A}_{[1]}=W_{[1]}\\
 {\cal A}_{[2]}=W_{[2]}\\
 \vdots\\
 {\cal A}_{[p]}=W_{[p]}
 \end{array}
 &
 \begin{array}{c}
 R\\
 \longrightarrow
 \end{array}
 &
 \begin{array}{c}
 \mathfrak{C}_{[\mathbf 0]}\\
 {W}^{\sigma_1}_{\alpha_1}\\
 {W}^{\sigma_2}_{\alpha_2}\\
 \vdots\\
 {W}^{\sigma_p}_{\alpha_{p}}
 \end{array}
 &
 \begin{array}{c}
 P\\
 \longrightarrow
 \end{array}
 &
 \begin{array}{c}
 \mathfrak{C}_{[\mathbf 0]}\\
 {W}^{0}_{\alpha_1}\\
 {W}^{0}_{\alpha_2}\\
 \vdots\\
 {W}^{0}_{\alpha_{p}}
 \end{array}
 &
 \begin{array}{c}
 E\\
 \longrightarrow
 \end{array}
 &
 \begin{array}{c}
 \mathfrak{C}_{[\mathbf 0]}={\cal C}_{[0]}\\
 {W}^{0}_{\beta_1}={\cal C}_{[1]}\\
 {W}^{0}_{\beta_2}={\cal C}_{[2]}\\
 \vdots\\
 {W}^{0}_{\beta_{p}}={\cal C}_{[p]}
 \end{array}
 \end{array}\]\\
 \fcaption{The transformations connecting two decomposition sequences.\label{EPR-process}}
 \end{center}
 \end{figure}

 \vspace{6pt}\noindent
 {\bf Exchange Transformation $E$}\noindent

 With the application of $R$ and $P$,
 the original decomposition sequence is transformed to the sequence
 in the third column of Fig.~\ref{EPR-process}.
 The $3$rd operator $E$ then comes into play that
 replaces each subspace ${W}^{0}_{\alpha_r}$ by ${W}^{0}_{\beta_r}$, $1\leq r\leq p$.
 In contrast to the simple forms of the operators $R$ and $P$,
 the construction of $E$ requires an iterative composition of the mapping $e_{\alpha+\beta}$.
 \vspace{6pt}
  \begin{lemma}\label{leme-transrank0}
  Two conditioned subspaces of the same self parity
  ${W}^\epsilon_\alpha$ and ${W}^\epsilon_\beta$
  in the quotient algebra
  $\{{\cal Q}(\mathfrak{C}_{[{\mathbf 0}]};2^p-1)\}$
  are mapped to each other by the transformation
  $e_{\alpha+\beta}=h^{\zeta}_{\alpha+\beta}h^{\eta}_{\alpha+\beta}$, i.e.,
  $e_{\alpha+\beta}W^{\epsilon}_{\alpha}e^{\dagger}_{\alpha+\beta}=W^{\epsilon}_{\beta}$,
  if the parity rules are satisfied
  $\zeta\cdot(\alpha+\beta)=\eta\cdot(\alpha+\beta)=(\zeta+\eta)\cdot\alpha=(\zeta+\eta)\cdot\beta=1$.
  \end{lemma}
  \vspace{3pt}
  \begin{proof}
  Given a spinor ${\cal S}^{\eta}_{\alpha+\beta}$,
  the condition subspace $W^\epsilon_{\alpha}$ admits the bisection
  $W^{\epsilon}_{\alpha}
  =\{{\cal S}^{\xi_0}_{\alpha}:\forall\hspace{2pt}\xi_0\in{Z^p_2},\hspace{2pt}\xi_0\cdot\alpha=1+\epsilon
   \text{ and }\xi_0\cdot(\alpha+\beta)+\eta\cdot\alpha=0\}\cup
   \{{\cal S}^{\xi_1}_{\alpha}:\forall\hspace{2pt}\xi_1\in{Z^p_2},\hspace{2pt}\xi_1\cdot\alpha=1+\epsilon
   \text{ and }\xi_1\cdot(\alpha+\beta)+\eta\cdot\alpha=1\}$,
  such that ${\cal S}^{\eta}_{\alpha+\beta}$ is commuting with
  the subspace of the $1$st half $\{{\cal S}^{\xi_0}_{\alpha}\}$
  but not with the $2$nd $\{{\cal S}^{\xi_1}_{\alpha}\}$, {\em cf.} Lemma~\ref{lemconparscomm}.
  By Lemma~\ref{lemh-trans},
  the $1$st half is left invariant
  and the $2$nd is mapped to $\{{\cal S}^{\xi_1+\eta}_{\beta}\}$
  when the basic transformation $h^{\eta}_{\alpha+\beta}$ is applied.
  With the application of the $2$nd action $h^{\zeta}_{\alpha+\beta}$,
  the $1$st half subspace $\{{\cal S}^{\xi_0}_{\alpha}\}$ is mapped to
  $\{{\cal S}^{\xi_0+\zeta}_{\beta}\}$ for $\xi_0\cdot(\alpha+\beta)+\zeta\cdot\alpha=1$
  due to the parity condition $\xi_0\cdot(\alpha+\beta)+\eta\cdot\alpha=0$
  and the parity rule $(\zeta+\eta)\cdot\alpha=1$, and however the consequent subspace of the $2$nd half
  $\{{\cal S}^{\xi_1+\eta}_{\beta}\}$ keeps invariant
  for $(\xi_1+\eta)\cdot(\alpha+\beta)+\zeta\cdot\beta=0$
  due to the condition $\xi_1\cdot(\alpha+\beta)+\eta\cdot\alpha=1$
  and the rule $(\zeta+\eta)\cdot\beta=1$.
  Accordingly, the condtioned subspace $W^\epsilon_\alpha$ is transformed to
  $\{{\cal S}^{\xi_0+\zeta}_{\beta}\}\cup\{{\cal S}^{\xi_1+\eta}_{\beta}\}=W^\epsilon_\beta$
  by the mapping
  $e_{\alpha+\beta}=h^{\zeta}_{\alpha+\beta}h^{\eta}_{\alpha+\beta}$.
  Finally, the invariance of the self parity of the subspaces, {\em i.e.},
  $(\xi_0+\zeta)\cdot\beta=(\xi_1+\eta)\cdot\beta=1+\epsilon$, is easily
  verified by adding the parity conditions to the parity rules.
  \end{proof}
  \vspace{6pt}\noindent
  In spite of four parity rules imposed for the phase-string selection,
  only three of them are independent.
  Within the quotient algebra $\{{\cal Q}(\mathfrak{C}_{[{\mathbf 0}]};2^p-1)\}$,
  a basic transformation always maps one half of spinor generators of
  a conditioned subspace to another subspace and leaves the other half unchanged.
  Taking advantage of this fact, the transformation $e_{\alpha+\beta}$ is constructed by composing
  two basic transformations in the way that each of them respectively keeps
  one half of the operated subspace invariant and delivers the other half to
  the same tageted subspace.

  It is worthy to note that, among the others, the conjugate pair
  ${W}^0_{\alpha+\beta}$ and ${W}^1_{\alpha+\beta}$
  and the center subalgebra $\mathfrak{C}_{[{\mathbf 0}]}$
  in $\{{\cal Q}(\mathfrak{C}_{[{\mathbf 0}]};2^p-1)\}$
  are three invariant subspaces of the
  transformation $e_{\alpha+\beta}$.
  The transformation
  $e_{\alpha+\beta}=h^{\zeta}_{\alpha+\beta}h^{\eta}_{\alpha+\beta}$
  earns more invariant conditioned subspaces
  $W^\epsilon_\gamma\in\{{\cal Q}(\mathfrak{C}_{[{\mathbf 0}]};2^p-1)\}$,
  with either the odd or the even self parity,
  as long as the phase strings $\zeta$ and $\eta$
  are chosen realizing an additional parity rule
  $(\zeta+\eta)\cdot\gamma=0$.
  While its {\em conjugate} version is followed $(\zeta+\eta)\cdot\gamma'=1$,
  a conditioned subspace $W^\epsilon_{\gamma'}$ is mapped by $e_{\alpha+\beta}$
  to another subspace $W^\epsilon_{\alpha+\beta+\gamma'}$ of the identical self parity.
  The exchange transformation $E$ is derived
  recursively making use of the mapping $e_{\alpha+\beta}$.

  To transform the decomposition  in
  the $3$rd column of Fig.~\ref{EPR-process} to the referential sequence in the final
  column, the first mapping to exercise is 
  $e_{\alpha_1+\beta_1}$ that obeys the parity
  rules in Lemma~\ref{leme-transrank0}. After this action, each subspace
  ${W}^0_{\alpha_j}$ is mapped to ${W}^0_{\alpha^{(1)}_j}$, $2\leq j\leq p$, here
  either $\alpha^{(1)}_j=\alpha_j+\alpha_1+\beta_1$ or $\alpha^{(1)}_j=\alpha_j$.
  The $2$nd action is to apply $e_{\alpha^{(1)}_2+\beta_2}$ to the current
  decomposition sequence complying with the additional parity rule such that
  the already exchanged subspace ${W}^0_{\beta_1}$ remains invariant.
  Iterating to the $r$-th action, the application of
  $e_{\alpha^{(r-1)}_r+\beta_r}=h^{\zeta_r}_{\alpha^{(r-1)}_r+\beta_r}h^{\eta_r}_{\alpha^{(r-1)}_r+\beta_r}$,
  $1\leq i<r<j\leq p$,
  transforms the subspace ${W}^0_{\alpha^{(r-1)}_{r}}$ to ${W}^0_{\beta_{r}}$ and
  the subspaces ${W}^0_{\alpha^{(r-1)}_{j}}$ to ${W}^0_{\alpha^{(r)}_{j}}$
  with either $\alpha^{(r)}_{j}=\alpha^{(r-1)}_{j}+\alpha^{(r-1)}_r+\beta_r$
       or $\alpha^{(r)}_{j}=\alpha^{(r-1)}_{j}$,
  and meanwhile preserves the invariance of the already exchanged subspaces ${W}^0_{\beta_i}$ 
  in obedience to the additional parity rule $(\zeta_r+\eta_r)\cdot\beta_i=0$.
 The exchange operator is thus cast into the composition form
 \begin{align}\label{eqexchange-trans}
 E=\prod^{p}_{r=1}e_{\alpha^{(r-1)}_r+\beta_r}
  =e_{\alpha^{(p-1)}_{p}+\beta_{p}}e_{\alpha^{(p-2)}_{p-1}+\beta_{p-1}}\ldots e_{\alpha^{(0)}_1+\beta_1}\text{, }
 \end{align}
 where
 $\alpha^{(0)}_1\equiv\alpha_1$ and the $r$-th action
 $e_{\alpha^{(r-1)}_r+\beta_r}=h^{\zeta_r}_{\alpha^{(r-1)}_r+\beta_r}h^{\eta_r}_{\alpha^{(r-1)}_r+\beta_r}$
 satisfies a set of  parity rules
 $(\zeta_r+\eta_r)\cdot\alpha^{(r-1)}_r=(\zeta_r+\eta_r)\cdot\beta_r
 =\zeta_r\cdot(\alpha^{(r-1)}_r+\beta_r)=\eta_r\cdot(\alpha^{(r-1)}_r+\beta_r)=1$ and
 $(\zeta_r+\eta_r)\cdot\beta_i=0$ for $1\leq i<r\leq p$.
 Notice that every step of the iteration is permitted, for
 the string pair $(\zeta_r,\eta_r)$ having as many as
 $2^{2p-r-1}$ choices in the $r$-th iteration.
 The exposition is closed with
 the transformation connecting the two decomposition sequences.
 \vspace{6pt} \noindent
  \begin{thm}\label{thmEPR}
  The transformation Q mapping a decomposition to the referential sequence  
  is a composition of the operators of diagonalization R, parity correction P and
  exchange E, i.e., Q\hspace{1.5pt}=\hspace{1.5pt}E\hspace{1.5pt}P\hspace{1.5pt}R.
  \end{thm}
  \vspace{6pt} \noindent

\nonumsection{References}

\appendix{~~Coset of Phase Strings\label{appcosets}}

 To generate a maximal bi-subalgebra of a Cartan subalgebra, it is legitimate to
 {\em bisect} the subalgebra by a maximal subgroup over either the binary-partitioning
 or the phase strings.
 Given a Cartan subalgebra
 $\mathfrak{C}^{\hspace{1.pt}\{\epsilon\}}_{[\alpha]_k}
 =\{{\cal S}^{\zeta_i}_{\alpha_i}:0\leq i<2^k, \alpha_i\in [\alpha]_k\text{ and }\zeta_i\in [\zeta]_q \}$
 as in Fig.~\ref{phasecoset},
 the $k$ independent binary-partitioning strings of $[\alpha]_k$ and their corresponding
 sets of phase strings are respectively displayed in the $1$st and $2$nd columns.
 The set $\{\zeta_0\}=[\zeta_0]_{p-k}$ pertaining to $\alpha_0={\mathbf 0}$
 forms a subgroup of $p-k$ generators in the subgroup $[\zeta]_q$, $q=p-k+h$.
 In the partition of $[\zeta]_q$ generated by
 $[\zeta_0]_{p-k}$, each phase-string set $\{\zeta_i\}$, $i=1,2,\cdots,k$, is a coset of
 $[\zeta_0]_{p-k}$ with a string $\zeta_i$ randomly picked in the set being the coset leader, and
 a number $h$ of cosets among them are independent, $0\leq h\leq k$.
 A bit-type maximal bi-subalgebra is created by bisecting
 $\mathfrak{C}^{\hspace{1.pt}\{\epsilon\}}_{[\alpha]_k}$ according to a maximal subgroup
 in $[\alpha]_k$.
 In other words, a such bi-subalgebra
 $\mathfrak{B}^{\hspace{1.pt}[\zeta_0]_{p-k}}_{[\alpha]_{k-1}}=\{{\cal S}^{\zeta_l}_{\alpha_l}: \forall\hspace{2pt}\alpha_l\in[\alpha]_{k-1}\}$
 is obtained when all spinor
 generators with binary-partitioning strings belonging to a chosen maximal subgroup
 $[\alpha]_{k-1}$ are collected. As a result, the set of phase strings
 $\{\zeta_l:\forall\hspace{2pt}{\cal S}^{\zeta_l}_{\alpha_l}\in
 \mathfrak{C}^{\hspace{1.pt}\{\epsilon\}}_{[\alpha]_k},\hspace{2pt}
 \alpha_l\in[\alpha]_{k-1}\}$, which are attached to spinor generators
 of the binary partitioning $\alpha_l\in[\alpha]_{k-1}$, is either a ($1$st) maximal subgroup in
 $[\zeta]_q$ as $h=k$ or $[\zeta]_q$ itself as $h<k$.
 This set is then denoted by $[\zeta_l]_{q'}$ with $q'=q-1\text{ or }q$.
 Since there are $2^k-1$ options of maximal subgroups $[\alpha]_{k-1}$ to be chosen from a given
 subgroup $[\alpha]_k$, a Cartan subalgebra of the $k$-th kind has a total number
 $2^k-1$ of bit-type maximal bi-subalgebras.
 \begin{figure}[!hbp]
 \begin{center}
 \begin{align*}
 \hspace{-138pt}\mathfrak{C}^{\hspace{1.pt}\{\epsilon\}}_{[\alpha]_k}
 =\{{\cal S}^{\zeta_i}_{\alpha_i}:0\leq i<2^k, \hspace{2pt}\alpha_i\in [\alpha]_k\text{ and }\zeta_i\in [\zeta]_q \}
 \end{align*}
 \[\begin{array}{ccccc}
   \begin{array}{cccc}
   \alpha_0&&& [\zeta_0]_{p-k}\\
   \alpha_1&&& \zeta_1+[\zeta_0]_{p-k}\\
   \alpha_2&&& \zeta_2+[\zeta_0]_{p-k}\\
   \vdots&&&\vdots\\
   \hspace{10pt}\alpha_{k-1}&&&\hspace{10pt} \zeta_{k-1}+[\zeta_0]_{p-k}\\
   \alpha_k&&& \zeta_k+[\zeta_0]_{p-k}
   \end{array}
 &&&&
 \begin{array}{cccc}
   [\zeta_0]_{p-k-1}&&& [\zeta_0]^c_{p-k-1}\\
   \zeta_1+[\zeta_0]_{p-k-1}&&& \zeta_1+[\zeta_0]^c_{p-k-1}\\
   \zeta_2+[\zeta_0]_{p-k-1}&&& \zeta_2+[\zeta_0]^c_{p-k-1}\\
   \vdots&&&\vdots\\
   \hspace{10pt}\zeta_{k-1}+[\zeta_0]_{p-k-1}&&&\hspace{10pt}\zeta_{k-1}+[\zeta_0]^c_{p-k-1}\\
   \zeta_k+[\zeta_0]_{p-k-1}&&& \zeta_k+[\zeta_0]^c_{p-k-1}
   \end{array}
 \end{array}\]

 \end{center}
 \fcaption{In the $2$nd column, the phase cosets of $[\zeta_0]_{p-k}$ pertaining to $k$ independent binary-partitioning strings
 $\alpha_i$, $i=1,2,\cdots,k$ and $\alpha_0={\mathbf 0}$ are listed; each coset is bisected into two subcosets as shown in
 the $3$rd and the final columns respectively according to the subgroup
 $[\zeta_0]_{p-k-1}\subset [\zeta_0]_{p-k}$ and its complementary subset
 $[\zeta_0]^c_{p-k-1}=[\zeta_0]_{p-k}-[\zeta_0]_{p-k-1}$.
 \label{phasecoset}}
 \end{figure}

 While producing a phase-type maximal bi-subalgebra of
 $\mathfrak{C}^{\hspace{1.pt}\{\epsilon\}}_{[\alpha]_k}$,
 denoted as $[\breve{\zeta}_0]_{q'}$ is a maximal subgroup
 in $[\zeta]_q$ to be assigned first. The
 components of this subgroup are demonstrated in the last two columns of Fig.~\ref{phasecoset}.
 The subgroup $[\zeta_0]_{p-k-1}$ on the top is a maximal subgroup arbitrarily taken in
 $[\zeta_0]_{p-k}$ that divides $[\zeta_0]_{p-k}$  into two subsets
 $\{\hat{\zeta}_0\}=[\zeta_0]_{p-k-1}$ and
 $\{\hat{\zeta}_0\}^c=[\zeta_0]_{p-k}-[\zeta_0]_{p-k-1}$.
 Each coset $\{\zeta_i\}$ associated with the binary-partitioning
 $\alpha_i$, $i=1,2,\cdots,k$, is likewise divided into two subsets
 $\{\hat{\zeta}_i\}=\zeta_i+[\zeta_0]_{p-k-1}$ and
 $\{\hat{\zeta}_i\}^c=\zeta_i+[\zeta_0]^c_{p-k-1}$ with the coset leader $\zeta_i$
 being any string in $\{\hat{\zeta}_i\}$.
 A maximal subgroup $[\breve{\zeta}_0]_{q'}$ of $[\zeta]_q$ is yielded from the generating set
 that is formed by taking the union of
 $\{\hat{\zeta}_0\}$ and either component $\{\hat{\zeta}_i\}$ or $\{\hat{\zeta}_i\}^c$
 of each coset $\{\zeta_i\}$, $i=1,2,\cdots,k$.
 Only the selection $\bigcup^k_{l=0}\{\hat{\zeta}_l\}$ has $h$ independent cosets and the
 other selections of the union have $h+1$ ones instead.
 The maximal subgroup
 $[\breve{\zeta}_0]_{q'}$ is thus tagged with the subscript $q'=q\text{ or }q-1$.
 Collecting all spinor generators with phase strings in $[\breve{\zeta}_0]_{q'}$,
 a phase-type maximal bi-subalgebra
 $\mathfrak{B}^{\hspace{1.pt}[\zeta_0]_{p-k-1}}_{[\alpha]_{k}}=\{{\cal S}^{\zeta_r}_{\alpha_r}: \forall\hspace{2pt}\zeta_r\in[\breve{\zeta}_0]_{q'}\}$
 is constructed.
 Through this coset-selecting process, the number of phase-type maximal bi-subalgebras in
 a Cartan subalgebra of the $k$-th kind amounts to $(2^{p-k}-1)2^k$, noting that there
 are $2^{p-k}-1$ choices of $[\zeta_0]_{p-k-1}$ in a given $[\zeta_0]_{p-k}$.
 Therefore as asserted in Corollary~\ref{lemnummaxbi}, a Cartan subalgebra of any kind in the Lie algebra $su(2^p)$
 has a total number $2^p-1$ of maximal bi-subalgebras, which exactly suffice the need to
 {\em define} the same number of conjugate pairs in a rank-zero quotient algebra.
 Owing to the bit-phase duality allowing to treat
 $[\alpha]_k$ as phase strings and $[\zeta]_q$ as binary-partitioning strings,
 there have other ways to generate
 maximal bi-subalgebras, which however lead to the identical complete set.
 Explicit examples of complete sets of maximal bi-subalgebras for some Cartan subalgebras
 are given in Appendix~\ref{appQAsu8}.

\appendix{~~Unique Maximal Bi-subalgebra for a Spinor\label{appBconstruct}}
 This is an alternative proof for Lemma~\ref{lemunimaxBcomm}. 
 Given a Cartan subalgebra $\mathfrak{C}^{\hspace{1.pt}\{\epsilon\}}_{[\alpha]_k}=
 \{{\cal S}^{\zeta_i}_{\alpha_i}:
 0\leq i<2^k, \alpha_0={\mathbf 0},\hspace{2pt} \alpha_i\in[\alpha]_k\text{ and }\zeta_i\in[\zeta]_q\}$
 of $su(2^p)$ and a spinor generator
 ${\cal S}^{\zeta}_{\alpha}\in su(2^p)$,
 the unique maximal bi-subalgebra $\mathfrak{B}$ such that
 $[{\cal S}^{\zeta}_{\alpha},\mathfrak{B}]=0$ will be constructed.
 As ${\cal S}^{\zeta}_{\alpha}\in\mathfrak{C}^{\hspace{1.pt}\{\epsilon\}}_{[\alpha]_k}$,
 the bi-subalgebra $\mathfrak{B}$ is obviously the
 Cartan subalgebra $\mathfrak{C}^{\hspace{1.pt}\{\epsilon\}}_{[\alpha]_k}$ itself.
 Thus in the following, only the circumstance
 ${\cal S}^{\zeta}_{\alpha}\in su(2^p)-\mathfrak{C}^{\hspace{1.pt}\{\epsilon\}}_{[\alpha]_k}$
 is assumed.
 It will be immediately clear that $\mathfrak{B}$ is a bit-type if $\alpha\in[\alpha]_k$
 or a phase-type otherwise.

 Suppose the string $\alpha$ is in $[\alpha]_k$.
 Since $\zeta_0\cdot\alpha=0$ for any $\alpha\in[\alpha]_k$,
 it holds that
 $\forall\hspace{2pt}\zeta_0\in\{\zeta_0\}=[\zeta_0]_{p-k}$,
 $[{\cal S}^{\zeta}_{\alpha},{\cal S}^{\zeta_0}_{\mathbf 0}]=0$.
 Moreover, if there is a string $\zeta_i$ in the coset $\{\zeta_i\}$ associated with the
 binary-partitioning $\alpha_i$ such that
 $[{\cal S}^{\zeta}_{\alpha},{\cal S}^{\zeta_i}_{\alpha_i}]=0$,
 also
 $[{\cal S}^{\zeta}_{\alpha},{\cal S}^{\zeta_i}_{\alpha_i}]=0,
 \forall\hspace{2pt}\zeta_i\in\{\zeta_i\}$.
 This is owing to the coset rule that $\forall\hspace{2pt}\zeta_i,\zeta'_i\in\{\zeta_i\}$,
 $\zeta_i+\zeta'_i\in[\zeta_0]_{p-k}$, $0<i<2^k$.
 Now consider the generating set
 $\mathcal{G}=\{{\cal S}^{\zeta_0}_{\alpha_0},{\cal S}^{\zeta_1}_{\alpha_1},
 {\cal S}^{\zeta_2}_{\alpha_2},\ldots,{\cal S}^{\zeta_k}_{\alpha_k}\}$ consisting of
 $k+1$ spinors from $\mathfrak{C}^{\hspace{1.pt}\{\epsilon\}}_{[\alpha]_k}$, where
 ${\alpha_j}$, $j=1,2,\cdots,k$, are $k$
 independent generators randomly chosen from $[\alpha]_k$.
 It is impossible for ${\cal S}^{\zeta}_{\alpha}$ to commute with $\mathcal{G}$ entirely
 because of
 the assumption 
 ${\cal S}^{\zeta}_{\alpha}\notin\mathfrak{C}^{\hspace{1.pt}\{\epsilon\}}_{[\alpha]_k}$.
 Then there must exist a cut for
 $\mathcal{G}=\mathcal{M}\cup(\mathcal{G}-\mathcal{M})$ at an integer $0<r\leq k$ with
 $\mathcal{M}=\{{\cal S}^{\zeta_1}_{\alpha_1},{\cal S}^{\zeta_2}_{\alpha_2},\ldots,
   {\cal S}^{\zeta_r}_{\alpha_r}\}$ and
 $\mathcal{G}-\mathcal{M}=\{{\cal S}^{\zeta_0}_{\alpha_0},{\cal S}^{\zeta_{r+1}}_{\alpha_{r+1}},\ldots,
   {\cal S}^{\zeta_k}_{\alpha_k}\}$, such that 
 $[{\cal S}^{\zeta}_{\alpha},\mathcal{G}-\mathcal{M} ]=0$ and
 $[{\cal S}^{\zeta}_{\alpha},{\cal S}^{\zeta_l}_{\alpha_l}]\neq 0$
 for every $1\leq l\leq r$.
 As a consequence of the cut, the  subset
 $\mathcal{M'}=\{{\cal S}^{\zeta_1+\zeta_2}_{\alpha_1+\alpha_2},
   {\cal S}^{\zeta_1+\zeta_3}_{\alpha_1+\alpha_3},\ldots,
   {\cal S}^{\zeta_1+\zeta_r}_{\alpha_1+\alpha_r}\}$ commuting with ${\cal S}^{\zeta}_{\alpha}$,
 {\em i.e.,} $[{\cal S}^{\zeta}_{\alpha},\mathcal{M'} ]=0$, is derived.
 It attains the purpose that $[{\cal S}^{\zeta}_{\alpha},\mathcal{G'} ]=0$,
 where the modified generating set $\mathcal{G'}=\mathcal{M'}\cup(\mathcal{G}-\mathcal{M})$
 has only $k$ spinors and the set of its binary-partitioning strings
 $\{\alpha_0={\mathbf 0},\alpha_1+\alpha_2,\alpha_1+\alpha_3,\ldots,\alpha_1+\alpha_r,
 \alpha_{r+1},\alpha_{r+2}\ldots,\alpha_{k}\}$ generates the subgroup $[\alpha]_{k-1}$
 uniquely determined by the given ${\cal S}^{\zeta}_{\alpha}$.
 So the bit-type maximal bi-subalgebra $\mathfrak{B}^{[\zeta_0]_{p-k}}_{[\alpha]_{k-1}}$
 commuting with ${\cal S}^{\zeta}_{\alpha}$ is constructed when all spinor generators
 of binary-partitioning strings in $[\alpha]_{k-1}$ are assembled.

 In the other case as $\alpha\notin[\alpha]_k$
 for the given spinor  ${\cal S}^{\zeta}_{\alpha}$,
 the unique maximal subgroup
 $[\zeta_0]_{p-k-1}\subset[\zeta_0]_{p-k}$ such that
 $[{\cal S}^{\zeta}_{\alpha},{\cal S}^{\nu}_{\mathbf 0}]=0$
 for all $\nu\in[\zeta_0]_{p-k-1}$ should be firstly decided.
 Since
 $[{\cal S}^{\zeta}_{\alpha},{\cal S}^{\nu_1+\nu_2}_{\mathbf 0}]=0$ for
 any $\nu_1\text{ and }\nu_2\in[\zeta_0]_{p-k}$ but
 $[{\cal S}^{\zeta}_{\alpha},{\cal S}^{\nu_1}_{\mathbf 0}]\neq 0\text{ and }
  [{\cal S}^{\zeta}_{\alpha},{\cal S}^{\nu_2}_{\mathbf 0}]\neq 0$,
 a such  subgroup must exist.
 Actually it is easy to locate $[\zeta_0]_{p-k-1}$
 by pondering the $k+1$ conditions over $p$-digit strings, that is,
 $\zeta_0\cdot\alpha=0$ and $\zeta_0\cdot\alpha_j=0$ for $\zeta_0\in Z^p_2$ and
 $\alpha_j, j=1,2,\dots,k$, being any $k$ independent generators of $[\alpha]_k$.
 Let the subgroup $[\zeta_0]_{p-k-1}$ generate an aforesaid partition of
 the whole group of phase strings $[\zeta]_q$
 as illustated in Fig.~\ref{phasecoset}.
 In this partition, the group $[\zeta]_q$ is divided into $k+1$ coset pairs where
 the $0$th pair is composed of $[\zeta_0]_{p-k-1}$ and
 $[\zeta_0]^c_{p-k-1}=[\zeta_0]_{p-k}-[\zeta_0]_{p-k-1}$, and the $i$-th pair
 composed of
 $\{\hat{\zeta}_i\}=\nu_i+[\zeta_0]_{p-k-1}$ and
 $\{\hat{\zeta}_i\}^c=\nu_i+[\zeta_0]^c_{p-k-1}$ with
 the coset leader $\nu_i\in\{\hat{\zeta}_i\}$, $i=1,2,\cdots,k$.
 The spinor  ${\cal S}^{\zeta}_{\alpha}$ should commute with either one coset
 in each pair due to the relation that
 $\forall\hspace{2pt}\zeta_i\in\{\hat{\zeta}_i\}, 
  \hspace{2pt}\zeta'_i\in\{\hat{\zeta}_i\}^c, i=1,2,\cdots,k$,
 $[{\cal S}^{\zeta}_{\alpha},{\cal S}^{\zeta_i}_{\alpha_i}]=0$ and
 $[{\cal S}^{\zeta}_{\alpha},{\cal S}^{\zeta'_i}_{\alpha_i}]\neq 0$ if
 $[{\cal S}^{\zeta}_{\alpha},{\cal S}^{\nu_i}_{\alpha_i}]=0$, or
 $[{\cal S}^{\zeta}_{\alpha},{\cal S}^{\zeta_i}_{\alpha_i}]\neq 0$ and
 $[{\cal S}^{\zeta}_{\alpha},{\cal S}^{\zeta'_i}_{\alpha_i}]=0$ if
 $[{\cal S}^{\zeta}_{\alpha},{\cal S}^{\nu_i}_{\alpha_i}]\neq 0$.
 The maximal subgroup $[\breve{\zeta}_0]_{q'}\subset[\zeta]_q$ demanded to bisect
 the given Cartan subalgebra is then produced from the generating set that is
 formed by taking the union of $[\zeta_0]_{p-k-1}$ and either
 $\{\hat{\zeta}_i\}$ or $\{\hat{\zeta}_i\}^c, i=1,2,\cdots,k$, whose associated
 spinor generators commuting with ${\cal S}^{\zeta}_{\alpha}$.
 Including all the spinor generators of phase strings in $[\breve{\zeta}_0]_{q'}$,
 the phase-type maximal bi-subalgebra $\mathfrak{B}^{[\zeta_0]_{p-k-1}}_{[\alpha]_k}$
 commuting with ${\cal S}^{\zeta}_{\alpha}$ is acquired.

\newpage

\appendix{~~Quotient Algebras of $su$(8)\label{appQAsu8}}
 \begin{figure}[!hbp]
 \begin{center}
 \begin{align*}
 &\mathfrak{C}_{[\bf 0]}\\
 \hspace{40pt}{\cal S}^{000}_{000},\hspace{2pt}{\cal S}^{001}_{000},\hspace{2pt}{\cal S}^{010}_{000},\hspace{2pt}{\cal S}^{011}_{000}&,
 \hspace{2pt}{\cal S}^{100}_{000},\hspace{2pt}{\cal S}^{101}_{000},\hspace{2pt}{\cal S}^{110}_{000},\hspace{2pt}{\cal S}^{111}_{000}
 \end{align*}
 \[\begin{array}{ccccccc}
 \mathfrak{B}_{1}&&W_{001}
 &{\cal S}^{000}_{001}\text{, }{\cal S}^{010}_{001}\text{, }{\cal S}^{100}_{001}\text{, }{\cal S}^{110}_{001}
 &\hspace{24pt}
 &{\cal S}^{001}_{001}\text{, }{\cal S}^{011}_{001}\text{, } {\cal S}^{101}_{001}\text{, }{\cal S}^{111}_{001}
 &\hat{W}_{001}\\
 &&&&&&\\
 \mathfrak{B}_{2}&&W_{010}
 &{\cal S}^{000}_{010}\text{, }{\cal S}^{001}_{010}\text{, } {\cal S}^{100}_{010}\text{, }{\cal S}^{101}_{010}
 &
 &{\cal S}^{010}_{010}\text{, }{\cal S}^{011}_{010}\text{, } {\cal S}^{110}_{010}\text{, }{\cal S}^{111}_{010}
 &\hat{W}_{010}\\
 &&&&&&\\
 \mathfrak{B}_{3}&&W_{011}
 &{\cal S}^{000}_{011}\text{, }{\cal S}^{011}_{011}\text{, } {\cal S}^{100}_{011}\text{, }{\cal S}^{111}_{011}
 &
 &{\cal S}^{001}_{011}\text{, }{\cal S}^{010}_{011}\text{, } {\cal S}^{101}_{011}\text{, }{\cal S}^{110}_{011}
 &\hat{W}_{011}\\
 &&&&&&\\
 \mathfrak{B}_{4}&&W_{100}
 &{\cal S}^{000}_{100}\text{, }{\cal S}^{001}_{100}\text{, } {\cal S}^{010}_{100}\text{, }{\cal S}^{011}_{100}
 &
 &{\cal S}^{100}_{100}\text{, }{\cal S}^{101}_{100}\text{, } {\cal S}^{110}_{100}\text{, }{\cal S}^{111}_{100}
 &\hat{W}_{100}\\
 &&&&&&\\
 \mathfrak{B}_{5}&&W_{101}
 &{\cal S}^{000}_{101}\text{, }{\cal S}^{010}_{101}\text{, } {\cal S}^{101}_{101}\text{, }{\cal S}^{111}_{101}
 &
 &{\cal S}^{001}_{101}\text{, }{\cal S}^{011}_{101}\text{, } {\cal S}^{100}_{101}\text{, }{\cal S}^{110}_{101}
 &\hat{W}_{101}\\
 &&&&&&\\
 \mathfrak{B}_{6}&&W_{110}
 &{\cal S}^{000}_{110}\text{, }{\cal S}^{001}_{110}\text{, } {\cal S}^{110}_{110}\text{, }{\cal S}^{111}_{110}
 &
 &{\cal S}^{010}_{110}\text{, }{\cal S}^{011}_{110}\text{, } {\cal S}^{100}_{110}\text{, }{\cal S}^{101}_{110}
 &\hat{W}_{110}\\
 &&&&&&\\
 \mathfrak{B}_{7}&&W_{111}
 &{\cal S}^{000}_{111}\text{, }{\cal S}^{011}_{111}\text{, } {\cal S}^{101}_{111}\text{, }{\cal S}^{110}_{111}
 &
 &{\cal S}^{001}_{111}\text{, }{\cal S}^{010}_{111}\text{, } {\cal S}^{100}_{111}\text{, }{\cal S}^{111}_{111}
 &\hat{W}_{111}
 \end{array} \]
 \vspace{30pt}
 \[ \begin{array}{ccc}
 \mathfrak{B}_{1}=
 \{\hspace{2pt}{\cal S}^{000}_{000},{\cal S}^{010}_{000},{\cal S}^{100}_{000},{\cal S}^{110}_{000}\hspace{2pt}\},
 &\mathfrak{B}_{2}=
 \{\hspace{2pt}{\cal S}^{000}_{000},{\cal S}^{001}_{000},{\cal S}^{100}_{000},{\cal S}^{101}_{000}\hspace{2pt}\},
 &\mathfrak{B}_{3}=
 \{\hspace{2pt}{\cal S}^{000}_{000},{\cal S}^{011}_{000},{\cal S}^{100}_{000},{\cal S}^{111}_{000}\hspace{2pt}\},\\
 &&\\
 \mathfrak{B}_{4}=
 \{\hspace{2pt}{\cal S}^{000}_{000},{\cal S}^{001}_{000},{\cal S}^{010}_{000},{\cal S}^{011}_{000}\hspace{2pt}\},
 &\mathfrak{B}_{5}=
 \{\hspace{2pt}{\cal S}^{000}_{000},{\cal S}^{010}_{000},{\cal S}^{101}_{000},{\cal S}^{111}_{000}\hspace{2pt}\},
 &\mathfrak{B}_{6}=
 \{\hspace{2pt}{\cal S}^{000}_{000},{\cal S}^{001}_{000},{\cal S}^{110}_{000},{\cal S}^{111}_{000}\hspace{2pt}\},\\
 &&\\
 \mathfrak{B}_{7}=
 \{\hspace{2pt}{\cal S}^{000}_{000},{\cal S}^{011}_{000},{\cal S}^{101}_{000},{\cal S}^{110}_{000}\hspace{2pt}\}.&&
 \end{array} \]\\
 \end{center}
 \fcaption{The quotient algebra given by the $0$th-kind Cartan subalgebra $\mathfrak{C}_{[\mathbf{0}]}$ and
 the complete set of corresponding maximal bi-subalgebras listed at the bottom;
 the relation holds $\forall\hspace{2pt}\alpha,\beta\in Z^3_2$,
 $\mathfrak{B}_{\alpha}\sqcap\mathfrak{B}_{\beta}=\mathfrak{B}_{\alpha+\beta}$
 for maximal bi-subalgebras redenoted as
 $\mathfrak{B}_{\alpha}=\{{\cal S}^{\nu}_{000}:\forall\hspace{2pt}\nu\in Z^3_2,\hspace{2pt}\nu\cdot\alpha=0\}$,
 {\em cf.} Theorem~\ref{thmGofC}.\label{1stshellFig}}
 \end{figure}

 \begin{figure}
 \begin{center}
 \begin{align*}
 &\mathfrak{C}^{0}_{[100]}\\
 \hspace{40pt}{\cal S}^{000}_{000},\hspace{2pt}{\cal S}^{001}_{000},\hspace{2pt}{\cal S}^{010}_{000},\hspace{2pt}{\cal S}^{011}_{000}&,\hspace{2pt}
 {\cal S}^{000}_{100},\hspace{2pt}{\cal S}^{001}_{100},\hspace{2pt}{\cal S}^{010}_{100},\hspace{2pt}{\cal S}^{011}_{100}
 \end{align*}
 \hspace{1pt}\[\begin{array}{ccccccc}
 \mathfrak{B}_{1}&&W_{1}
 &{\cal S}^{100}_{000}\text{, }{\cal S}^{101}_{000}\text{, } {\cal S}^{110}_{000}\text{, }{\cal S}^{111}_{000}
 &\hspace{24pt}
 &{\cal S}^{100}_{100}\text{, }{\cal S}^{101}_{100}\text{, } {\cal S}^{110}_{100}\text{, }{\cal S}^{111}_{100}
 &\hat{W}_{1}\\
 &&&&&&\\
 \mathfrak{B}_{2}&&W_{2}
 &{\cal S}^{000}_{001}\text{, }{\cal S}^{010}_{001}\text{, } {\cal S}^{000}_{101}\text{, }{\cal S}^{010}_{101}
 &
 &{\cal S}^{001}_{001}\text{, }{\cal S}^{011}_{001}\text{, } {\cal S}^{001}_{101}\text{, }{\cal S}^{011}_{101}
 &\hat{W}_{2}\\
 &&&&&&\\
 \mathfrak{B}_{3}&&W_{3}
 &{\cal S}^{100}_{001}\text{, }{\cal S}^{110}_{001}\text{, } {\cal S}^{101}_{101}\text{, }{\cal S}^{111}_{101}
 &
 &{\cal S}^{101}_{001}\text{, }{\cal S}^{111}_{001}\text{, } {\cal S}^{100}_{101}\text{, }{\cal S}^{110}_{101}
 &\hat{W}_{3}\\
 &&&&&&\\
 \mathfrak{B}_{4}&&W_{4}
 &{\cal S}^{000}_{010}\text{, }{\cal S}^{001}_{010}\text{, } {\cal S}^{000}_{110}\text{, }{\cal S}^{001}_{110}
 &
 &{\cal S}^{010}_{010}\text{, }{\cal S}^{011}_{010}\text{, } {\cal S}^{010}_{110}\text{, }{\cal S}^{011}_{110}
 &\hat{W}_{4}\\
 &&&&&&\\
 \mathfrak{B}_{5}&&W_{5}
 &{\cal S}^{100}_{010}\text{, }{\cal S}^{101}_{010}\text{, } {\cal S}^{110}_{110}\text{, }{\cal S}^{111}_{110}
 &
 &{\cal S}^{110}_{010}\text{, }{\cal S}^{111}_{010}\text{, } {\cal S}^{100}_{110}\text{, }{\cal S}^{101}_{110}
 &\hat{W}_{5}\\
 &&&&&&\\
 \mathfrak{B}_{6}&&W_{6}
 &{\cal S}^{000}_{011}\text{, }{\cal S}^{011}_{011}\text{, } {\cal S}^{000}_{111}\text{, }{\cal S}^{011}_{111}
 &
 &{\cal S}^{001}_{011}\text{, }{\cal S}^{010}_{011}\text{, } {\cal S}^{001}_{111}\text{, }{\cal S}^{010}_{111}
 &\hat{W}_{6}\\
 &&&&&&\\
 \mathfrak{B}_{7}&&W_{7}
 &{\cal S}^{100}_{011}\text{, }{\cal S}^{111}_{011}\text{, } {\cal S}^{101}_{111}\text{, }{\cal S}^{110}_{111}
 &
 &{\cal S}^{101}_{011}\text{, }{\cal S}^{110}_{011}\text{, } {\cal S}^{100}_{111}\text{, }{\cal S}^{111}_{111}
 &\hat{W}_{7}\\
 \end{array} \]\\
 \vspace{20pt}
 \[ \begin{array}{ccc}
 \mathfrak{B}_{1}=
 \{\hspace{2pt}{\cal S}^{000}_{000},{\cal S}^{001}_{000},{\cal S}^{010}_{000},{\cal S}^{011}_{000}\hspace{2pt}\},
 &\mathfrak{B}_{2}=
 \{\hspace{2pt}{\cal S}^{000}_{000},{\cal S}^{010}_{000},{\cal S}^{000}_{100},{\cal S}^{010}_{100}\hspace{2pt}\},
 &\mathfrak{B}_{3}=
 \{\hspace{2pt}{\cal S}^{000}_{000},{\cal S}^{010}_{000},{\cal S}^{001}_{100},{\cal S}^{011}_{100}\hspace{2pt}\},\\
 &&\\
 \mathfrak{B}_{4}=
 \{\hspace{2pt}{\cal S}^{000}_{000},{\cal S}^{001}_{000},{\cal S}^{000}_{100},{\cal S}^{001}_{100}\hspace{2pt}\},
 &\mathfrak{B}_{5}=
 \{\hspace{2pt}{\cal S}^{000}_{000},{\cal S}^{001}_{000},{\cal S}^{001}_{100},{\cal S}^{010}_{100}\hspace{2pt}\},
 &\mathfrak{B}_{6}=
 \{\hspace{2pt}{\cal S}^{000}_{000},{\cal S}^{011}_{000},{\cal S}^{000}_{100},{\cal S}^{011}_{100}\hspace{2pt}\},\\
 &&\\
 \mathfrak{B}_{7}=
 \{\hspace{2pt}{\cal S}^{000}_{000},{\cal S}^{011}_{000},{\cal S}^{001}_{100},{\cal S}^{010}_{100}\hspace{2pt}\}.&&
 \end{array} \]\\
 \end{center}
 \fcaption{The quotient algebra given by a $1$st-kind Catan subalgebra $\mathfrak{C}^{0}_{[100]}$ and
 the complete set of corresponding maximal bi-subalgebras.\label{2ndshellFig}}
 \end{figure}

 \begin{figure}
 \begin{center}
 \begin{align*}
 &\mathfrak{C}^{10}_{[001,100]}\\
 \hspace{30pt}{\cal S}^{000}_{000},\hspace{2pt}{\cal S}^{010}_{000},\hspace{2pt}{\cal S}^{101}_{001},\hspace{2pt}{\cal S}^{111}_{001}&,
 \hspace{2pt}{\cal S}^{001}_{100},\hspace{2pt}{\cal S}^{011}_{100},\hspace{2pt}{\cal S}^{100}_{101},\hspace{2pt}{\cal S}^{110}_{101}
 \end{align*}
 \[\begin{array}{ccccccc}
 \mathfrak{B}_{1}&&W_{1}
 &{\cal S}^{100}_{000}\text{, }{\cal S}^{110}_{000}\text{, } {\cal S}^{001}_{001}\text{, }{\cal S}^{011}_{001}
 &\hspace{24pt}
 &{\cal S}^{101}_{100}\text{, }{\cal S}^{111}_{100}\text{, } {\cal S}^{000}_{101}\text{, }{\cal S}^{010}_{101}
 &\hat{W}_{1}\\
 &&&&&&\\
  \mathfrak{B}_{2}&&W_{2}
 &{\cal S}^{001}_{000}\text{, }{\cal S}^{011}_{000}\text{, } {\cal S}^{000}_{100}\text{, }{\cal S}^{010}_{100}
 &
 &{\cal S}^{100}_{001}\text{, }{\cal S}^{110}_{001}\text{, } {\cal S}^{101}_{101}\text{, }{\cal S}^{111}_{101}
 &\hat{W}_{2}\\
 &&&&&&\\
 \mathfrak{B}_{3}&&W_{3}
 &{\cal S}^{101}_{000}\text{, }{\cal S}^{111}_{000}\text{, } {\cal S}^{001}_{101}\text{, }{\cal S}^{011}_{101}
 &
 &{\cal S}^{000}_{001}\text{, }{\cal S}^{010}_{001}\text{, } {\cal S}^{100}_{100}\text{, }{\cal S}^{110}_{100}
 &\hat{W}_{3}\\
 &&&&&&\\
 \mathfrak{B}_{4}&&W_{4}
 &{\cal S}^{000}_{010}\text{, }{\cal S}^{101}_{011}\text{, } {\cal S}^{001}_{110}\text{, }{\cal S}^{100}_{111}
 &
 &{\cal S}^{010}_{010}\text{, }{\cal S}^{111}_{011}\text{, } {\cal S}^{011}_{110}\text{, }{\cal S}^{110}_{111}
 &\hat{W}_{4}\\
 &&&&&&\\
 \mathfrak{B}_{5}&&W_{5}
 &{\cal S}^{100}_{010}\text{, }{\cal S}^{001}_{011}\text{, } {\cal S}^{111}_{110}\text{, }{\cal S}^{010}_{111}
 &
 &{\cal S}^{110}_{010}\text{, }{\cal S}^{011}_{011}\text{, } {\cal S}^{101}_{110}\text{, }{\cal S}^{000}_{111}
 &\hat{W}_{5}\\
 &&&&&&\\
 \mathfrak{B}_{6}&&W_{6}
 &{\cal S}^{001}_{010}\text{, }{\cal S}^{110}_{011}\text{, } {\cal S}^{000}_{110}\text{, }{\cal S}^{111}_{111}
 &
 &{\cal S}^{011}_{010}\text{, }{\cal S}^{100}_{011}\text{, } {\cal S}^{010}_{110}\text{, }{\cal S}^{101}_{111}
 &\hat{W}_{6}\\
 &&&&&&\\
 \mathfrak{B}_{7}&&W_{7}
 &{\cal S}^{101}_{010}\text{, }{\cal S}^{010}_{011}\text{, } {\cal S}^{110}_{110}\text{, }{\cal S}^{001}_{111}
 &
 &{\cal S}^{111}_{010}\text{, }{\cal S}^{000}_{011}\text{, } {\cal S}^{100}_{110}\text{, }{\cal S}^{011}_{111}
 &\hat{W}_{7}\\
 \end{array} \]\\
 \vspace{20pt}
 \[ \begin{array}{ccc}
 \mathfrak{B}_{1}=
 \{\hspace{2pt}{\cal S}^{000}_{000},{\cal S}^{010}_{000},{\cal S}^{101}_{001},{\cal S}^{111}_{001}\hspace{2pt}\},
 &\mathfrak{B}_{2}=
 \{\hspace{2pt}{\cal S}^{000}_{000},{\cal S}^{010}_{000},{\cal S}^{001}_{100},{\cal S}^{011}_{100}\hspace{2pt}\},
 &\mathfrak{B}_{3}=
 \{\hspace{2pt}{\cal S}^{000}_{000},{\cal S}^{010}_{000},{\cal S}^{100}_{101},{\cal S}^{110}_{101}\hspace{2pt}\},\\
 &&\\
 \mathfrak{B}_{4}=
 \{\hspace{2pt}{\cal S}^{000}_{000},{\cal S}^{101}_{001},{\cal S}^{001}_{100},{\cal S}^{100}_{101}\hspace{2pt}\},
 &\mathfrak{B}_{5}=
 \{\hspace{2pt}{\cal S}^{000}_{000},{\cal S}^{101}_{001},{\cal S}^{011}_{100},{\cal S}^{110}_{101}\hspace{2pt}\},
 &\mathfrak{B}_{6}=
 \{\hspace{2pt}{\cal S}^{000}_{000},{\cal S}^{111}_{001},{\cal S}^{001}_{100},{\cal S}^{110}_{101}\hspace{2pt}\},\\
 &&\\
 \mathfrak{B}_{7}=
 \{\hspace{2pt}{\cal S}^{000}_{000},{\cal S}^{111}_{001},{\cal S}^{011}_{100},{\cal S}^{100}_{101}\hspace{2pt}\}.&&
 \end{array} \]\\
 \end{center}
 \fcaption{The quotient algebra given by a $2$nd-kind Cartan subalgebra $\mathfrak{C}^{10}_{[001,100]}$ and
 the complete set of corresponding maximal bi-subalgebras.\label{3rdshellFig}}
 \end{figure}

 \begin{figure}
 \begin{center}
 \begin{align*}
 &\mathfrak{C}^{100}_{[001,010,100]}\\
 \hspace{30pt}{\cal S}^{000}_{000},\hspace{2pt}{\cal S}^{101}_{001},\hspace{2pt}{\cal S}^{000}_{010},\hspace{2pt}{\cal S}^{101}_{011}&,
 \hspace{2pt}{\cal S}^{001}_{100},\hspace{2pt}{\cal S}^{100}_{101},\hspace{2pt}{\cal S}^{001}_{110},\hspace{2pt}{\cal S}^{100}_{111}
 \end{align*}
 \[\begin{array}{ccccccc}
 \mathfrak{B}_{1}&&W_{1}
 &{\cal S}^{001}_{000}\text{, }{\cal S}^{001}_{010}\text{, } {\cal S}^{000}_{100}\text{, }{\cal S}^{000}_{110}
 &\hspace{24pt}
 &{\cal S}^{101}_{001}\text{, }{\cal S}^{101}_{011}\text{, } {\cal S}^{100}_{101}\text{, }{\cal S}^{100}_{111}
 &\hat{W}_{1}\\
 &&&&&&\\
 \mathfrak{B}_{2}&&W_{2}
 &{\cal S}^{010}_{000}\text{, }{\cal S}^{111}_{001}\text{, } {\cal S}^{011}_{100}\text{, }{\cal S}^{110}_{101}
 &
 &{\cal S}^{010}_{010}\text{, }{\cal S}^{111}_{011}\text{, } {\cal S}^{011}_{110}\text{, }{\cal S}^{110}_{111}
 &\hat{W}_{2}\\
 &&&&&&\\
 \mathfrak{B}_{3}&&W_{3}
 &{\cal S}^{011}_{000}\text{, }{\cal S}^{110}_{011}\text{, } {\cal S}^{010}_{100}\text{, }{\cal S}^{111}_{111}
 &
 &{\cal S}^{110}_{001}\text{, }{\cal S}^{011}_{010}\text{, } {\cal S}^{111}_{101}\text{, }{\cal S}^{010}_{110}
 &\hat{W}_{3}\\
 &&&&&&\\
 \mathfrak{B}_{4}&&W_{4}
 &{\cal S}^{100}_{000}\text{, }{\cal S}^{001}_{001}\text{, } {\cal S}^{100}_{010}\text{, }{\cal S}^{001}_{011}
 &
 &{\cal S}^{101}_{100}\text{, }{\cal S}^{000}_{101}\text{, } {\cal S}^{101}_{110}\text{, }{\cal S}^{000}_{111}
 &\hat{W}_{4}\\
 &&&&&&\\
 \mathfrak{B}_{5}&&W_{5}
 &{\cal S}^{101}_{000}\text{, }{\cal S}^{101}_{010}\text{, } {\cal S}^{001}_{101}\text{, }{\cal S}^{001}_{111}
 &
 &{\cal S}^{000}_{001}\text{, }{\cal S}^{000}_{011}\text{, } {\cal S}^{100}_{100}\text{, }{\cal S}^{100}_{110}
 &\hat{W}_{5}\\
 &&&&&&\\
 \mathfrak{B}_{6}&&W_{6}
 &{\cal S}^{110}_{000}\text{, }{\cal S}^{011}_{001}\text{, } {\cal S}^{111}_{110}\text{, }{\cal S}^{010}_{111}
 &
 &{\cal S}^{110}_{010}\text{, }{\cal S}^{011}_{011}\text{, } {\cal S}^{111}_{100}\text{, }{\cal S}^{010}_{101}
 &\hat{W}_{6}\\
 &&&&&&\\
 \mathfrak{B}_{7}&&W_{7}
 &{\cal S}^{111}_{000}\text{, }{\cal S}^{010}_{011}\text{, } {\cal S}^{011}_{101}\text{, }{\cal S}^{110}_{110}
 &
 &{\cal S}^{010}_{001}\text{, }{\cal S}^{111}_{010}\text{, } {\cal S}^{110}_{100}\text{, }{\cal S}^{011}_{111}
 &\hat{W}_{7}\\
 \end{array} \]\\
 \vspace{20pt}
 \[ \begin{array}{ccc}
 \mathfrak{B}_{1}=
 \{\hspace{2pt}{\cal S}^{000}_{000},{\cal S}^{000}_{010},{\cal S}^{001}_{100},{\cal S}^{001}_{110}\hspace{2pt}\},
 &\mathfrak{B}_{2}=
 \{\hspace{2pt}{\cal S}^{000}_{000},{\cal S}^{101}_{001},{\cal S}^{001}_{100},{\cal S}^{100}_{101}\hspace{2pt}\},
 &\mathfrak{B}_{3}=
 \{\hspace{2pt}{\cal S}^{000}_{000},{\cal S}^{101}_{011},{\cal S}^{001}_{100},{\cal S}^{100}_{111}\hspace{2pt}\},\\
 &&\\
 \mathfrak{B}_{4}=
 \{\hspace{2pt}{\cal S}^{000}_{000},{\cal S}^{101}_{001},{\cal S}^{000}_{010},{\cal S}^{101}_{011}\hspace{2pt}\},
 &\mathfrak{B}_{5}=
 \{\hspace{2pt}{\cal S}^{000}_{000},{\cal S}^{000}_{010},{\cal S}^{100}_{101},{\cal S}^{100}_{111}\hspace{2pt}\},
 &\mathfrak{B}_{6}=
 \{\hspace{2pt}{\cal S}^{000}_{000},{\cal S}^{111}_{001},{\cal S}^{001}_{100},{\cal S}^{110}_{101}\hspace{2pt}\},\\
 &&\\
 \mathfrak{B}_{7}=
 \{\hspace{2pt}{\cal S}^{000}_{000},{\cal S}^{101}_{011},{\cal S}^{001}_{110},{\cal S}^{100}_{111}\hspace{2pt}\}.&&
 \end{array} \]\\
 \end{center}
 \fcaption{The quotient algebra given by a $3$rd-kind Cartan subalgebra $\mathfrak{C}^{100}_{[001,010,100]}$ and
 the complete set of corresponding maximal bi-subalgebras.\label{4thshellFig}}
 \end{figure}
 \end{document}